\documentclass[twocolumn]{aastex6}

\usepackage{amsmath,amssymb,amstext}

\usepackage[all]{hypcap} 

\usepackage{aas_macros}
\newcommand{\myemail}{ibanez@ph1.uni-koeln.de \\ 
mordecai@amnh.org \\
klessen@uni-heidelberg.de}

\defcitealias{Ibanez-Mejia2016GravitationalClouds}{Paper I}	

\shorttitle{Accreting molecular clouds in a Turbulent ISM.}
\shortauthors{Ib\'a\~nez-Mej\'{\i}a et al.}

\begin{document}

\title{Feeding vs.\ Falling: \\ The growth and collapse of molecular clouds in a turbulent interstellar medium.}

\author{Juan C. Ib\'a\~nez-Mej\'{\i}a \altaffilmark{1,2,3,4}, 
Mordecai-Mark Mac Low \altaffilmark{2,1}, 
Ralf S. Klessen \altaffilmark{1,5} and   
Christian Baczynski \altaffilmark{1}}

\affil{$^{1}$Universit\"at Heidelberg, Zentrum f\"ur Astronomie Heidelberg, Institut f\"ur Theoretische Astrophysik, Albert-Ueberle-Str. 2,D-69120 Heidelberg, Germany}
\affil{$^{2}$Department of Astrophysics, American Museum of Natural History, New York, NY 10024, USA}
\affil{$^{3}$I. Physikalisches Institut, Universit\"at zu K\"oln, Z\"ulpicher Str. 77, D-50937 K\"oln, Germany}
\affil{$^{4}$Max Planck Institute for Extraterrestrial Physics, Giessenbachstrasse 1, D-85748 Garching}
\affil{$^{5}$Universit\"at Heidelberg, Interdisziplin\"are Zentrum f\"ur Wissenschaftliches Rechnen, Im Neuenheimer Feld 205, D-69120 Heidelberg, Germany}

\email{\myemail}

\begin{abstract}
In order to understand the origin of observed molecular cloud properties, it is critical to understand how clouds interact with their environments during their formation, growth and collapse. 
It has been suggested that accretion-driven turbulence can maintain clouds in a highly turbulent state, preventing runaway collapse, and explaining the observed non-thermal velocity dispersions. 
We present 3D, adaptive-mesh-refinement, magnetohydrodynamical simulations of a kiloparsec-scale, stratified, supernova-driven, self-gravitating, interstellar medium, including diffuse heating and radiative cooling. 
These simulations model the formation and evolution of a molecular cloud population in the turbulent interstellar medium. 
We use zoom-in techniques to focus on the dynamics of the mass accretion and its history for individual molecular clouds.
We find that mass accretion onto molecular clouds proceeds as a combination of turbulent flow and near free-fall accretion of a gravitationally bound envelope. 
Nearby supernova explosions have a dual role, compressing the envelope, increasing mass accretion rates, but also disrupting parts of the envelope and eroding mass from the cloud's surface. 
It appears that the inflow rate of kinetic energy onto clouds from supernova explosions is insufficient to explain the net rate of change of the cloud kinetic energy. 
In the absence of self-consistent star formation, conversion of gravitational potential into kinetic energy during contraction seems to be the main driver of non-thermal motions within clouds.
We conclude that although clouds interact strongly with their environments, bound clouds are always in a state of gravitational contraction, close to runaway, and their properties are a natural result of this collapse.  
\end{abstract}

\maketitle

\section{Introduction}
Molecular clouds (MCs) are complex, dynamically evolving systems hosting all observed star formation in our own and other galaxies \citep[for reviews on star formation see][]{Williams1999TheIMF, Larson2003TheFormation, MacLow2004ControlTurbulence, Lada2005StarOverview, McKee2007TheoryFormation, Dobbs2014FormationFormation, Klessen2014PhysicalMedium}. 
MCs form out of the turbulent interstellar environment, and evolve in constant interaction with it, exchanging mass and energy through accretion flows and surface forces.
Determining how MCs evolve and interact with their turbulent environment is of critical importance to understanding what controls their properties and determines their future evolution.

The non-linear interplay between gravity, turbulence, and magnetic fields determines the evolution and collapse of MCs.
Observations show that molecular clouds have supersonic internal velocity dispersions \citep{Zuckerman1974RadioMolecules}, with kinetic energies comparable to the gravitational potential energy of the clouds \citep{Larson1981, Solomon1987MassClouds, Heyer2009RE-EXAMININGCLOUDS}. 
However, the origin of the observed motions inside dense clouds and whether they are strong and isotropic enough to prevent the clouds from collapsing remains poorly understood. 

It has been proposed that the process of mass growth in molecular clouds can drive sufficiently strong internal motions to explain the observations \citep[hereafter KH10]{Klessen2010Accretion-drivenDisks}.
However, few observations quantify mass accretion rates, while theoretical studies to date have relied on semi-analytic methods using efficiency parameters tuned to reproduce the observations \citep{Goldbaum2011THEACCRETION} or parameters motivated by numerical simulations of colliding flows to develop semi-empirical models for the evolution of gravitationally contracting clouds \citep{Zamora-Aviles2012ANACTIVITY}.
Some of the best evidence for mass growth in MCs comes from the observations by \citet{Fukui2009MOLECULARI} and \citet{Kawamura2009GlobularGalaxies}, who observed giant molecular clouds (GMC) in the Large Magellanic Cloud (LMC), matching their internal level of star-formation with their masses, deriving an evolutionary sequence for GMCs, and recovering average mass accretion rates.

Until recently, most simulations following the detailed
evolution of MCs providing the framework for star-formation models
were run in isolated, periodic boxes, with artificially driven
turbulence \citep{Klessen2000GravitationalTurbulence,
  Heitsch2001GravitationalTurbulence, Krumholz2005AGalaxies,
  Padoan2012AFormation, Federrath2012THEOBSERVATIONS}.
It remains unclear whether these idealized setups accurately capture the processes influencing the properties of MC in the Galaxy, and consequently represent real MCs and star forming environments.
More complex models of MC formation and evolution correspond to
colliding flow simulations \citep{Audit2005Thermalflow,
  Audit2010Onflows, Heitsch2005FormationStudy, Heitsch2006TheFlows,
  Heitsch2008RapidMovies, Heitsch2011FlowSimulations,
  Banerjee2009ClumpFormation, Vazquez-Semadeni2007MolecularConditions,
  Vazquez-Semadeni2009HIGH-EFFICIENCIES}.  These show that accretion
is an essential process in a cloud's life, and although accretion
driven turbulence can delay local and global gravitational collapse,
it is not an efficient mechanism to drive turbulence within clouds
that agrees with observations
\citep{Vazquez-Semadeni2007MolecularConditions,
  Vazquez-Semadeni2010MOLECULARFEEDBACK, Colin2013MolecularFeedback}

In this paper we use a kiloparsec-scale, magnetized, supernova (SN) driven, turbulent, stratified,  interstellar medium (ISM) simulation \citep[hereafter Paper I]{Ibanez-Mejia2016GravitationalClouds} to study mass accretion histories and rates.  
We first measure them in the entire simulated cloud population, and then in more detail in zoom-in, high-resolution simulations of three clouds. 
We derive an approximate relation for the mass accretion rates expected from gas self-gravity and from the background turbulent environment in Section~\ref{sec:mass_and_energy_influx}; introduce the numerical setup in Section~\ref{sec:numerical_model}; in Section~\ref{sec:results} compare the predicted and measured mass accretion rates for the full cloud population and follow in detail the accretion flows for three high-resolution clouds. 
We examine the energy budget in these clouds in Section~\ref{sec:discussion} and draw conclusions in Section~\ref{sec:conclusions}.

\section{Mass and Kinetic Energy Influx}
\label{sec:mass_and_energy_influx}
In order to provide baseline estimates for mass accretion rates, we analytically calculate the mass accretion rate expected for a cloud embedded in a uniform density environment.  
We consider two extreme cases. 
The first case is an initially stationary, gravitationally bound envelope falling onto the cloud due to the cloud's gravitational attraction, while the second case is a turbulent environment that deposits mass onto the cloud by a turbulent advective flux.

\subsection{Gravitationally Driven Accretion}
Assuming spherical symmetry for the cloud, we can express the mass accretion rate as 
\begin{eqnarray}
	\dot{M} = 4 \pi R^{2} v_{in} \rho_{ism}.
    \label{eq:macc}
\end{eqnarray}
where R is the cloud's radius, $\rho_{ism}$ is the background density and $v_{in}$ is the infall velocity. For the gravitationally driven case, the infall velocity corresponds to the free falling gas velocity at the surface of the cloud given by:
\begin{eqnarray}
	v_{in}^{2} = \frac{2 G M}{5 R}.
	\label{eq:vinf}
\end{eqnarray}

We adopt the observed empirical relation between the mass and size of molecular clouds \citep{Falgarone2004StructureClouds}
\begin{eqnarray}
	M = 40~\mathrm{M}_{\odot} \left(\frac{R}{1~\mathrm{pc}} \right)^{2}.
    \label{eq:Falgarone}
\end{eqnarray}
which implies that clouds are fractal structures with a fractal dimension of 2.
Combining Equations~(\ref{eq:macc}), (\ref{eq:vinf})
and~(\ref{eq:Falgarone}), we obtain a mass accretion relation as a
function of the density of the environment, and the cloud mass, given
by  
\begin{multline}
	\dot{M} = 6\times10^{-6} \mathrm{~M_{\odot}~yr^{-1}}  \\ 
    \left( \frac{n_{ism}}{1 \mbox{ cm}^{-3}}\right) \left( \frac{M}{10^3~\mathrm{M}_{\odot}} \right)^{1.25}
    \label{eq:macc_grav}
\end{multline}
where we have assumed that the number density $n_{ism} = \rho_{ism}/\mu$, with $\mu=1.3~ m_{\text{H}}$.

\subsection{Turbulence-Driven Accretion}
Turbulent motions carry a net mass flux of $\phi_{turb}=v_{ism} \rho_{ism}$, where $v_{ism}$ is the mean turbulent velocity and $\rho_{ism}$, the density of the environment.  
Multiplying by the cloud area yields an estimate of the total accretion rate.
Furthermore, we assume that the mean turbulent velocity of the environment is constant with
\begin{eqnarray}
	v_{ism} = 10~\mathrm{km~s}^{-1}.
	\label{eq:vconst}
\end{eqnarray}
Combining Equations~(\ref{eq:macc}), (\ref{eq:Falgarone})
and~(\ref{eq:vconst}), we obtain a relation for the estimated mass
accretion rate driven by the turbulence as
\begin{multline}
	\dot{M} = 1.026 \times 10^{-4} \mathrm{~M_{\odot}~yr^{-1}} \\
    \left( \frac{n_{ism}}{1~\mbox{cm}^{-3}}\right) \left( \frac{v_{ism}}{10~\mbox{km~s}^{-1}}\right) \left( \frac{M}{10^{3}~\mathrm{M}_{\odot}} \right).
    \label{eq:macc_turb}
\end{multline}
Although both Equations~(\ref{eq:macc_grav}) and~(\ref{eq:macc_turb}) depend on the cloud density, gravitationally driven turbulence has a steeper slope, as the infall velocity is expected to increase with increasing cloud mass, while in the turbulently driven accretion scenario, the increasing mass accretion rate depends on the growth of the surface area of the cloud with mass. 

\subsection{Energy Inflow in Accretion Flows}
We now calculate the amount of kinetic energy accreted by the cloud for these two extreme cases.
The influx of kinetic energy is given by:
\begin{eqnarray}
	\dot{E}_{in} = \frac{1}{2} \dot{M} v_{in}^2,
\end{eqnarray}
where $v_{in}$ is the velocity of the incoming material.
Substituting the gravitational mass accretion rate, Equation~(\ref{eq:macc_grav}), and the gravitational infall velocity, Equation~(\ref{eq:vinf}), we obtain the kinetic energy influx driven by self-gravity
%
\begin{multline}
	\dot{E}_{grav,in} = 6.53 \times 10^{29} \mathrm{~erg~s^{-1}} \\
    \left( \frac{n_{ism}}{1~\mathrm{cm}^{-3}} \right) \left(\frac{M}{10^{3}~\mathrm{M}_{\odot}} \right)^{1.75}.
	\label{eq:ekin_in_grav}
\end{multline}
%
For the cloud embedded in a turbulent environment, the influx of kinetic energy is given by
\begin{multline}
	\dot{E}_{turb, in} = 3.24\times 10^{33} \mathrm{~erg~s^{-1}} \\
    \left( \frac{n_{ism}}{1~\mathrm{cm}^{-3}} \right) \left( \frac{v_{ism}}{10~\mathrm{km~s^{-1}}}\right)^{3} \left(\frac{M}{10^3~\mathrm{M}_{\odot}} \right).
	\label{eq:ekin_in_turb}
\end{multline}

\subsection{Accretion Turbulence Driving Efficiency}
We now ask whether the incoming kinetic energy estimated in the previous section are sufficient to maintain the non-thermal motions observed in molecular clouds, either separately or together.
To do this, we derive the efficiency parameter introduced by KH10,
\begin{eqnarray}
	\epsilon = \left|\frac{\dot{E}_{decay}}{\dot{E}_{in}}\right|.
	\label{eq:epsilon_def}
\end{eqnarray}
This parameter compares the decay rate of turbulence in a molecular cloud to the influx of energy through gravitational accretion or turbulent advection. 
The turbulent kinetic energy, $E = (1/2) M \sigma^{2}$, for a cloud of mass $M$, and size $R$, decays at a rate \citep{MacLow1999TheClouds}
\begin{eqnarray}
	\dot{E}_{decay} = - \frac{1}{2} \frac{M \sigma^{3}}{R}.
	\label{eq:ekin_decay}
\end{eqnarray}
We estimate the velocity dispersion of a cloud to be related to the mass and size of the cloud by the virial relation 
\begin{eqnarray}
	\sigma = \left( \frac{G M}{5 R} \right)^{1/2}.
	\label{eq:sigma_virial}
\end{eqnarray}
Finally, we assume that clouds are fractal structures with fractal dimension of 2 as introduced in Equation~(\ref{eq:Falgarone}).  

Comparing the energy decay rate with the energy influx rate for
self-gravity driven turbulence, Equation~(\ref{eq:ekin_in_grav}), we
obtain the gravitational efficiency
\begin{eqnarray}
	\epsilon_{g} = 7.07\left(\frac{n_{ism}}{1~\mbox{cm}^{-3}}\right)^{-1} \left( \frac{M}{\mathrm{10^{3}~M_{\odot}}} \right)^{-0.5}.
	\label{eq:epsilon_grav}
\end{eqnarray}
This equation suggests that turbulence-driven accretion from the
gravitational collapse of the envelope, for clouds below
  $5\times10^{4}$~M$_{\odot}$ with an environment density of $n \leq
$1~cm$^{-3}$, would be insufficient to compensate for the energy decay
of the system ($\epsilon_g < 1$).  
However, clouds more massive than $5\times10^{4}$~M$_{\odot}$ or accreting from a denser environment, could drive sufficient turbulence to maintain the observed non-thermal motions.
Equation~(\ref{eq:epsilon_grav}) suggests that accretion driving efficiency scales inversely proportional to the environmental density, as argued by KH10. 

Now, comparing the energy decay rate with the energy accretion rate by the capture of gas from the turbulent environment, we obtain
\begin{multline}
	\epsilon_{t} = 1.42 \times10^{-3}  \\
    \left(\frac{n_{ism}}{1~\mbox{cm}^{-3}}\right)^{-1} \left(\frac{v_{ism}}{10~\mbox{km~s}^{-1}}\right)^{-3} \left( \frac{M}{\mathrm{10^{3}~M_{\odot}}} \right)^{0.25}.
	\label{eq:epsilon_turb}
\end{multline}
This relation suggests that the influx of kinetic energy from a turbulent background is sufficient to maintain the non-thermal motions observed in MCs, with a strong dependence on the velocity of the background turbulence and the density of the environment.
However, the accretion efficiency scales with the cloud mass to the power of $0.25$, suggesting that turbulence driven by accretion becomes less efficient with increasing cloud mass. 


\section{Numerical Model}
\label{sec:numerical_model}
%
We present models based on those originally presented in
\citetalias{Ibanez-Mejia2016GravitationalClouds}. These are three-dimensional numerical simulations of a self-gravitating, magnetized, SN-driven, turbulent, multiphase ISM using the FLASH v4.2.2 adaptive mesh refinement code \citep[AMR;][]{Fryxell2000FLASHFlashes}.
They use a directionally split, density, entropy and internal
  energy positivity-preserving, HLL3R approximate,  magnetohydrodynamical Riemann solver implemented in FLASH by \citet{Waagan2011ATests}.
The solver shows high efficiency and stability, particularly for the
high Mach number flows and low plasma $\beta$ values commonly found in astrophysics.
The code solves the ideal equations of magnetohydrodynamics, preserving $\nabla \cdot B = 0$ using divergence cleaning.
The energy equation is modified by including the discrete injection of SN explosions, as well as diffuse heating and radiative cooling.
The latter is included assuming an optically thin plasma with solar metallicity. 
The cooling curve is a piecewise power law approximating the results
of \citet{Dalgarno1972HeatingRegions}, with an electron fraction of
$n_{e}/n_{H} =10^{-2}$ for temperatures below $10^{4}$~K, and
collisional ionization equilibrium resonant line cooling \citet{Sutherland1993CoolingPlasmas} for temperatures above $2\times10^{4}$~K.
Photoelectric heating from irradiated dust grains is the dominant heating mechanism for the cold and warm neutral medium \citep{Bakes1994TheHydrocarbons}. 
Heating rates are assumed to be independent of gas density \citep{Wolfire1995TheMedium}, with a heating efficiency of $\epsilon= 0.05$, for an interstellar radiation field of G0 = 1.7 \citep{Draine1978PhotoelectricGas}. 
We assume the heating rate declines exponentially with height $\Gamma_{pe}(z) = \Gamma_{pe,0}e^{-z/h_{pe}}$, with a scale height $h_{pe} = 300$~pc.

The simulations use a stratified box consisting of a 1~kpc$^{2} \times 40$~kpc vertical section of the ISM of a disk galaxy, that captures the dynamics of the gas at the midplane, the vertical stratification in a background galactic potential, and the circulation of gas in a galactic fountain \citep{Joung2006,Joung2009,Hill2012VerticalMedium, Walch2015TheISM, Girichidis2016TheOutflows, Ibanez-Mejia2016GravitationalClouds, Gatto2017SILCCIII, Peters2017SILCCIV}.
The galactic disk and vertical stratification is maintained by a static gravitational potential, accounting for the gravitational influence of dark matter, stars
and gas.
Near the disk, the potential follows a modified version of the solar neighborhood potential derived by \citet{Kuijken1989TheSun},
transitioning to the inner halo potential of \citet{Dehnen1998MassWay}  at $∣z∣ \approx 4$~kpc. The potential asympotes to a \citet{Navarro1996TheHalos} potential at heights $∣z∣ >7.5$~kpc.
Gas self-gravity is included using the optimized multigrid hybrid
scheme in FLASH \citep{Ricker2008AMeshes, Daley2012HybridGravity},
based on the multigrid algorithm of \citet{Huang00Fastmeshes}, which
interfaces parallel FFTs with the multigrid solver to maintain the
scalability of the multigrid solver on oct-tree AMR meshes. 

Turbulence is driven by discrete SN explosions.
The SN rates are normalized to the galactic SN rate \citep{Tammann1994TheRate}: Type Ia and core-collapse SN have rates of 6.58 and 27.4~Myr$^{-1}$~kpc$^{-2}$, respectively. 
SN explosions are randomly positioned in the horizontal direction in the disk, and have an exponentially decaying vertical distribution with scale heights of 90~pc for core-collapse SNe and 325~pc for SNe Ia.
SNe are treated as thermal explosions, adding $10^{51}$~erg of energy to a sphere encompassing 60~M$_{\odot}$ of mass. 
No gas mass is added to the SN explosion. 
Clustered SNe are taken into account by assuming three-fifths of the core-collapse SN to be correlated in space and time. 
Clustered SN are assigned to a massless particle moving in a straight line, with velocity given by the local gas velocity at the time of cluster formation, with a maximum velocity of $20$~km~s$^{-1}$.

The simulations in \citetalias{Ibanez-Mejia2016GravitationalClouds} had a maximum resolution of 0.95 pc in a static nested grid, covering the Galactic midplane, $1$~kpc$^2 \times 100$~pc, and successively lower resolution at greater altitudes above and below the plane.
During the initial evolution of the simulation we systematically increase the resolution at the midplane in order to populate the turbulent cascade down to the smallest scales.
The simulations include an initially uniform magnetic field along the horizontal $\hat{x}$ direction that decays exponentially with height, with an initial midplane field strength of $B_{0} = 5 ~\mu$G.  
As the simulation proceeds, the magnetic field naturally evolves as it is advected and tangled by the fluid flow.   
Our initial value is chosen based on observations of radio synchrotron emissivities and the assumption of energy equipartition between magnetic fields and cosmic rays that suggest an average total magnetic field strength in the solar neighborhood of $6\pm2~\mu$G \citep{Beck2001GalacticFields}. 
Similarly, observations of Zeeman splitting towards HI regions find a (well defined) median magnetic field of $6~\mu$G, \citep{Heiles2005TheTurbulence, Crutcher2010MAGNETICANALYSIS} suggesting approximate equipartition between turbulent and magnetic energies.

In this paper, we use as an initial condition the simulation from \citetalias{Ibanez-Mejia2016GravitationalClouds} after running it for 230 Myr without gas self-gravity in order to fully develop a multi-phase turbulent ISM.  
By that time, 7,515 SNe have exploded and they continue to be randomly injected at the same rate during the subsequent evolution presented in this work. 
Further details about the setup and the dynamical evolution of the self-gravitating and non-self-gravitating runs are given in \citetalias{Ibanez-Mejia2016GravitationalClouds}.

All but one of the simulations presented here include self-gravity. 
The exception is used to compare the impact of turbulence alone. 
The analysis is performed from the moment self-gravity is turned
on, or is a coeval evolution for the simulation without self-gravity.

We re-simulated selected clouds with varying sizes and maximum resolutions using an adaptive mesh refinement zoom-in technique described below in Section~\ref{zoom-in_sec}.
The refinement condition in the clouds is chosen to require that the
local Jeans (\citeyear{Jeans1902TheNebula}) length be resolved with
four cells, satisfying the \citet{Truelove1997TheHydrodynamics}
criterion. 
Outside the zoom-in region, the nested, static refinement is reduced in resolution, to a maximum resolution of 1.9~pc at the midplane, as described in Table~\ref{tab:resolution_table}.
Our goal is to resolve the dynamics of the clouds and their environments, concentrating on gravitationally unstable gas, while simultaneously resolving the background dynamics of the midplane gas and its interaction with each cloud. 
\begin{table}
	\begin{center}
		\begin{tabular}{clc} 
			\hline
			resolution [pc]	& \multicolumn{1}{c}{height} & ref. type \\ 
    		\hline
			0.06--0.12 & Zoom-in box L$_{x,y,z} $ & AMR     \\
			1.90   &$\qquad\qquad\;\;$ z $\leq \lvert $300$\lvert$~pc& static  \\
			3.80   &$\lvert$300$\lvert$~pc  $<$ z $<\lvert$1$\lvert$~kpc& static  \\
			7.60   &$\;\;\lvert$1$\lvert$~kpc $<$ z $<\lvert$3$\lvert$~kpc& static \\
			15.2   &$\;\,\lvert$3$\lvert$~kpc $<$ z $<\lvert$10$\lvert$~kpc& static\\
			30.4   &$\lvert$10$\lvert$~kpc $<$ z $<\lvert$20$\lvert$~kpc& static  \\
    		\hline
		\end{tabular}
    \end{center}
    \vspace{-5mm}
   \caption{Grid structure for a re-simulated zoom-in cloud. 
   The environment maintains the nested refinement structure while the selected region increases the resolution using adaptive mesh refinement down to a resolution of 0.06--0.12~pc, depending on the cloud. \label{tab:resolution_table}}   
\end{table}
%
We initially defined clouds in the three-dimensional, position-position-position space as topologically connected structures above a density threshold of $n_{thr} \ge 100$~cm$^{-3}$. 
However, due to the high resolution achieved in these new simulations,
parts of the cloud connect and disconnect as material around
  the bulk of the cloud fluctuates above and below the density threshold.
For this reason we identify all isodensity contours within the zoom-in region, and
compare their bulk kinetic energy with respect to the center of mass with the gravitational potential energy of the bulk of the cloud. 
Clumps with gravitational energy exceeding kinetic energy are then considered part of the cloud. 
This results in a smooth change in mass as the cloud evolves, so we use this diagnostic to measure mass accretion rates.

We follow some of the cloud properties as a function of time, such as the total velocity dispersion and cloud virial parameter.
To obtain an estimate of the velocity dispersion, we calculate the mass-weighted, one-dimensional, velocity dispersion for the resolved density range given by
\begin{eqnarray}
\sigma^2_{1D} = \frac{1}{3}\frac{\sum_{i}^{N} \rho_{i} ({\textbf{v}}-
  {\bar{\textbf{v}}} )^{2}}{ \sum_{i} \rho_{i} },
\end{eqnarray}
where $\bar{\textbf{v}}$ is the mass-weighted average velocity summed over all cells in the cloud. 
Since $\sigma_{1D}$ corresponds to the one-dimensional, non-thermal, velocity dispersion, we compute the total velocity dispersion including the average mass-weighted sound speed, $\bar{c}_{s}$,
\begin{eqnarray}
\sigma^2_{tot} = \sigma^2_{1D} + \bar{c}^2_{s}.
\end{eqnarray}
We use the approximate version of the virial parameter neglecting surface, magnetic, and time-dependent terms that is often used in studies of molecular cloud dynamics \citep{Bertoldi1992Pressure-confinedClouds,Ballesteros-Paredes2006SixClouds,Kauffmann2013LOWFIELDS}, given by
\begin{eqnarray}
\alpha_{vir} = \frac{5 \sigma^2_{tot} R}{G M_{\rho}},
\end{eqnarray} 
where $\sigma_{tot}$ and $M_{\rho}$ are the mass is the total velocity dispersion for the resolved density range, and $R$ is the radius of the cloud, if the total volume of the cloud is mapped to a sphere.

In order to follow the dynamics of the accreting gas we inject passive tracer particles in and around the cloud.
A total of 200$^{3}$ tracer particles are added in a lattice with a volume equal to the zoom-in region at $t_{sg}=0$, such that, initially, we have one tracer particle every $(0.5$~pc$)^{3}$.
These passive tracers are evolved using a second-order Runge-Kutta scheme.

We analyze a total of five simulations in this chapter, listed in Table \ref{tab:simulations_list}. 
This Table includes the names, maximum resolutions, the size of the the zoom-in box, and whether or not they include gas self-gravity.
\begin{table}
	\begin{center}
		\begin{tabular}{cllc} 
            \hline
			Name	& Refined region  & resolution & Self-Gravity \\ 
    		\hline
			StBx\_1pc\_NoSG  & 1~kpc$^2\times$100~pc & 0.95 & No     \\
			StBx\_1pc\_SG    & 1~kpc$^2\times$100~pc & 0.95 & Yes  \\
			M3    &  (100~pc)$^{3}$ & 0.06 &  Yes \\
			M4    &  (100~pc)$^{3}$ & 0.06 &  Yes \\ 
			M8    &  (100~pc)$^{3}$ & 0.12 &  Yes  \\
    		\hline
		\end{tabular}
    \end{center}
    \vspace{-5mm}
    \caption{List of the simulations analyzed in this work. The
      columns correspond to their names, sizes of the high-resolution
      box, maximum resolutions, and if self-gravity was included or
      not in the simulation. 
    \label{tab:simulations_list}}
\end{table}

\subsection{Zoom-in technique}
\label{zoom-in_sec}
In order to decide which clouds to zoom in to, we reran the simulation at multiple resolutions, in each case extracting and following a cloud population in time, and then comparing the catalogs between resolutions. 
To do this, we first restarted from the \citetalias{Ibanez-Mejia2016GravitationalClouds} simulation with maximum resolution of 0.95 pc, and injected a lattice of $200\times200\times50$ tracer particles covering a volume of $1000\times1000\times100$~pc$^{3}$ around the midplane. 
We ran the simulations for 15~Myr without self-gravity and extract a cloud catalog every megayear. 
We then connected the time evolution of the clouds using the tracer particles. 
We included an extra layer of nested refinement within $z\pm50$~pc, for a maximum resolution of $\Delta x = 0.47$~pc, and ran the same 15~Myr of evolution, again without gas self-gravity, extracting a cloud population at every megayear and connecting this cloud population in time. 
We compared these two cloud catalogs searching for new clouds formed during the final 5~Myr of these 15~Myr of evolution. 
We look for clouds that appear in the $0.47$~pc cloud catalog before they are found in the $0.95$~pc catalog, suggesting that we are capturing these clouds at earlier stages of their formation and ensuring that we capture their dynamics as they grow.
Once a cloud has been selected for zoom-in, we restart from the $0.47$~pc resolution simulation and position constrain the AMR to a box of size $(100$~pc$)^{3}$ centered on the cloud's center of mass. 
The zoom-in box has a uniform background resolution of $0.47$~pc and a maximum resolution in Jeans unstable regions of 0.06--0.12~pc depending on the cloud (see table \ref{tab:simulations_list}).  
Table \ref{tab:resolution_table} shows the grid structure once a cloud has been tagged for refinement. 

Our goal is to resolve the dynamics of the environment, as well as the internal dynamics of the cloud, in order to quantify the influence of mass accretion onto the clouds. 
We therefore need to restrict our analysis to the period of time while
the collapse and fragmentation of the cloud is resolved.

We consider a cloud to be resolved as long as more than half of the cloud mass has a density below the critical Jeans density which depends on the grid resolution. 
The critical Jeans density is defined as 
\begin{eqnarray}
    n_{crit}  = (1281 \text{ cm}^{-3}) \left( \frac{\lambda_{j}}{1
        \mbox{ pc}} \right)^{-2} \left( \frac{T}{10 \mbox{ K}} \right),
\end{eqnarray}
where $\lambda_{j}$ is the Jeans (\citeyear{Jeans1902TheNebula}) length, and $T$ is the average temperature of the cloud.
We use a Jeans length $\lambda_j = 4 \Delta x_{min}$, as our resolution limit.
We run the evolution of our zoom-in clouds for 10~Myr. 
However, they become unresolved as they enter runaway collapse.
We limit our analysis of the influence of the accreted material on the cloud properties to the period of the evolution during which the cloud is resolved by our definition.

The zoom-in method presented here is similar to the one described by \citet{Seifried2017SILCC-ZOOM}, who concentrated on the detailed evolution of the physical and chemical properties of two clouds from the simulations by \citet{Walch2015TheISM}. 
The main difference between our methods is the refinement timescale $\tau$, the time over which the code is allowed to go to the highest refinement level once the zoom-in region is activated.
\citet{Seifried2017SILCC-ZOOM} explores a range of refinement timescales of $\tau =0$ to 4.5~Myr, suggesting that an optimal timescale of $\tau \geq 1.5$~Myr is necessary to avoid spurious fragmentation. In this paper we adopt a refinement timescale of $\tau=0$~Myr, meaning that we allow the code to refine on Jeans unstable regions instantaneously after the zoom-in region is enabled. The reason why we do not see spurious, large-scale, rotating clumps, is because we already start with a uniform grid resolution of $\Delta x=0.5$~pc around the midplane, and also that we concentrated on low average density clouds. Therefore when we activate the zoom-in region, only a small fraction of mass is in Jeans unstable regions and the refinement follows the gravitational collapse of the cloud.

%
%

\section{Results}
\label{sec:results}
\subsection{Overview}
\begin{figure*}[!h]
\centering 
\includegraphics[width=1.\textwidth]{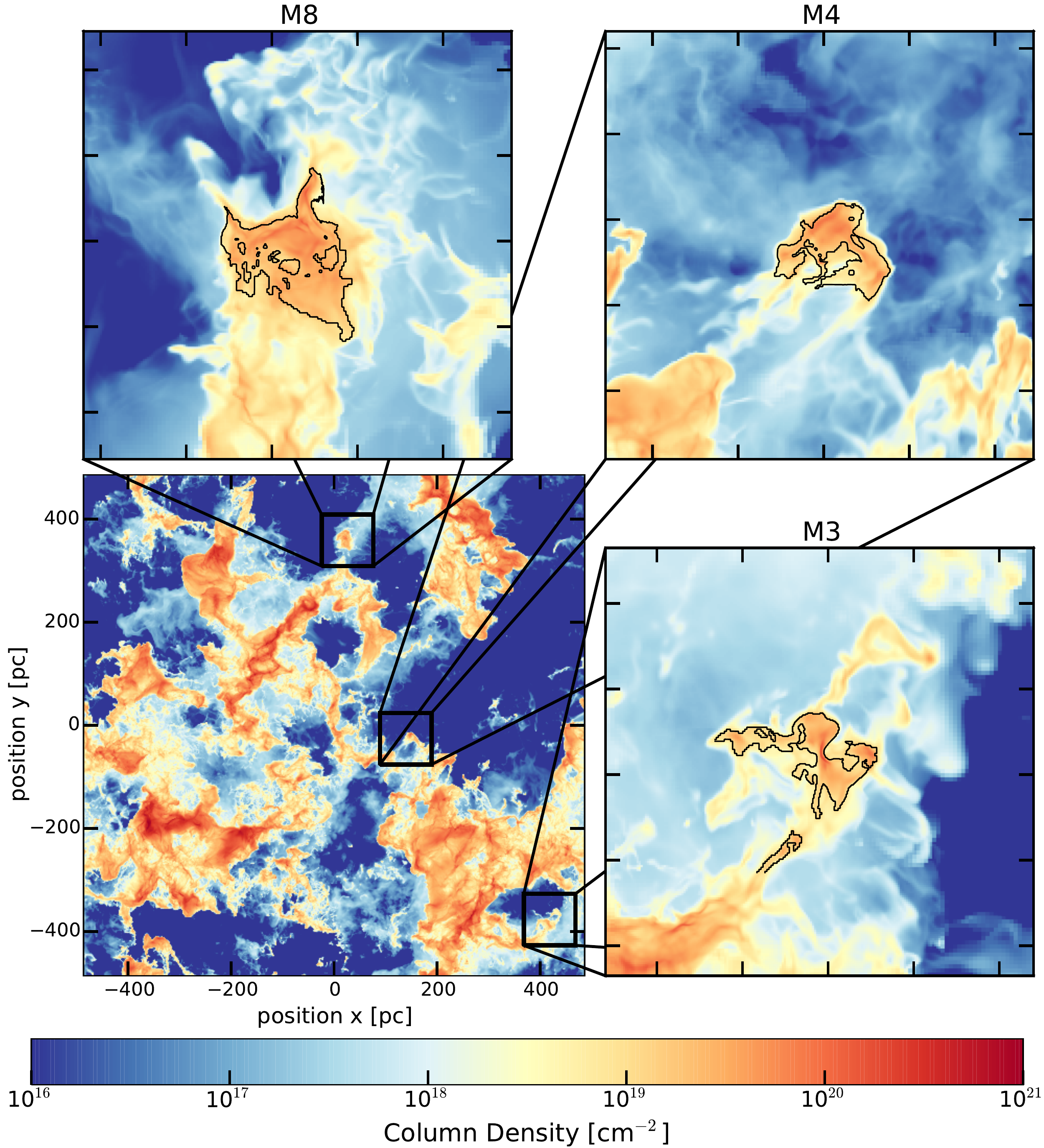} 
\caption{Column density of the simulation projected perpendicular to the Galactic midplane at the moment before self-gravity is turned on at $t_{SG} = 0$. 
The zoom-in boxes around the target clouds are superposed.  
Close-ups of the (100~pc)$^3$ zoom-in boxes around each of the clouds are also shown (but after 10~Myr of evolution at higher resolution), with a black contour outlining the $n = 100$~cm$^{-3}$ region. 
An animation of the subsequent 10 Myr of evolution can be found online.
\label{fig:midplane-and-clouds}} 
\end{figure*}

Figure~\ref{fig:midplane-and-clouds} shows a face-on projection of the box from the simulation with 0.47~pc resolution at the midplane, at the time when the target clouds were identified.
The three target clouds are also shown in close-up windows, and their evolution is shown in animations available on line.
These clouds formed in the turbulent ISM have complex density distributions and shapes, with predominantly elongated and filamentary structures. 
They evolve in a state of hierarchical, gravitational contraction, while simultaneously interacting with their environment, accreting and losing material as large scale turbulence and nearby SN blast waves interact with the clouds.  

We follow this interchange of material using tracer particles. 
The shape of the clouds continuously changes due to a combination of gas self-gravity and surface forces.  
In every case, a large fraction of the injected tracer particles quickly disperses throughout the simulation box, while a smaller fraction interacts with the clouds and their envelopes. 
Of the order of $10^5$ particles are accreted by each cloud by the end of the simulation. 

We stop the simulations $10$~Myr after including self-gravity. 
By this point massive stars should certainly be present and feeding back energy in the form of radiation, winds and SN explosions, influencing not only the clouds' properties but their environments as well.
As we do not include self-consistent star-formation and feedback in these simulations, running this setup for longer would certainly lead to unphysical results. 
Indeed, as noted at the end of Section~\ref{zoom-in_sec}, we already must cut off our analysis earlier, when too much mass has reached unresolved densities.

We want to determine the main process driving the accretion of mass onto clouds, and whether or not the kinetic energy carried by the accreted material can sustain the observed non-thermal linewidths in the interior of MCs.  
To answer these questions we focus on two aspects: First, we compare the mass accretion rate for the cloud population in the global simulations with and without self-gravity, and compare the results to the predictions of mass accretion rate estimated from the gravitational collapse of a uniform density, spherical envelope, and from turbulent accretion in a uniform density environment (Sect.~\ref{sec:cloud_population}).
Second, we follow the dynamics of the accreted material in the high-resolution zoom-in simulations, calculating the amount of kinetic energy entering the cloud boundary, in order to quantify the importance of this accretion to the dynamical evolution of the cloud in comparison to the kinetic energy provided by internal gravitational contraction.
The analysis scripts used to produce the figures and calculation in this paper can be found in \citet{AccPaperScripts}

\begin{figure*}[!t]
\centering 
\includegraphics[width=0.48\textwidth]{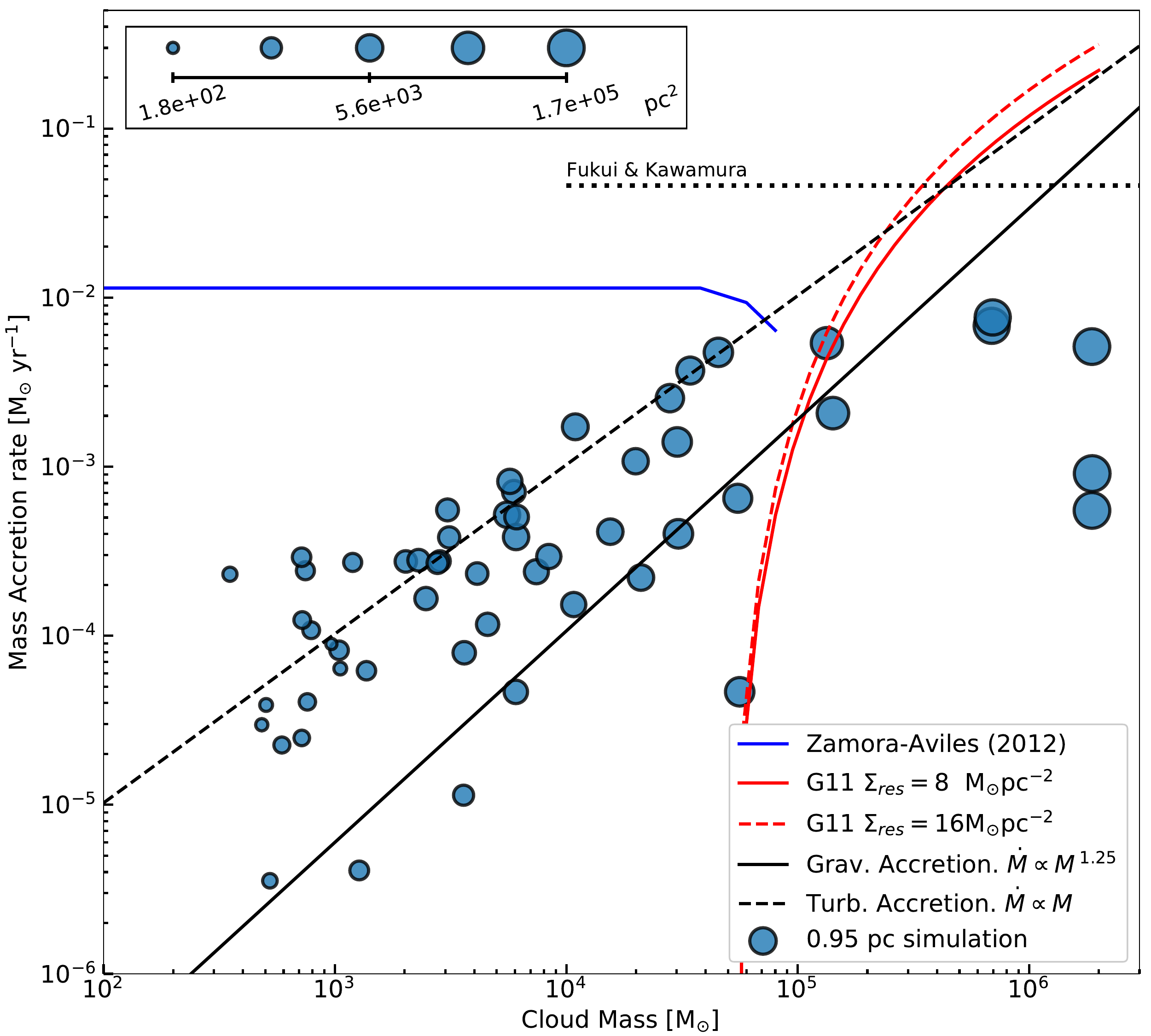}
\includegraphics[width=0.48\textwidth]{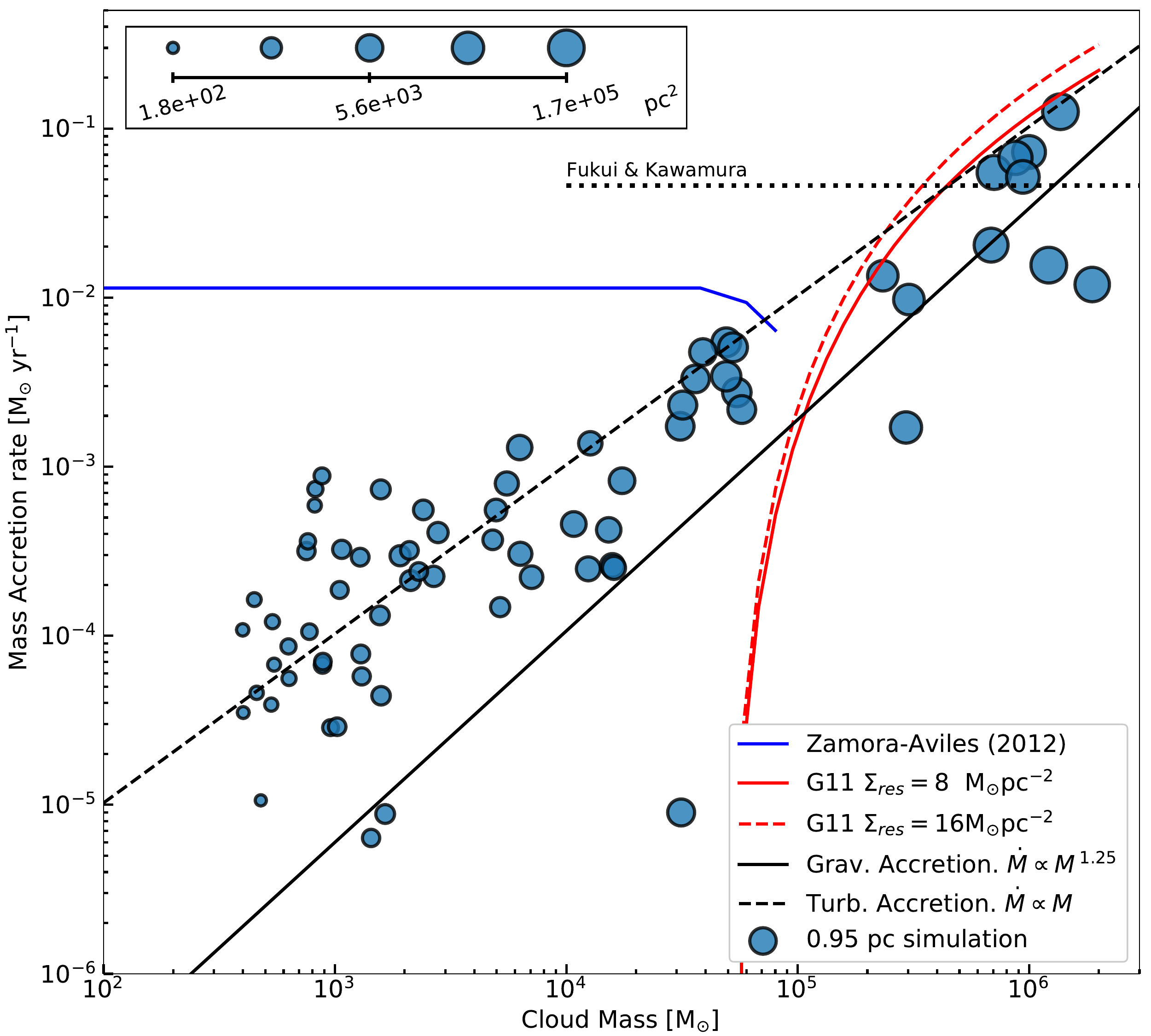} 
\caption{Mass accretion rates measured for a cloud population over
  5~Myr of evolution with self-gravity turned off (left, model
  StBx\_1pc\_NoSG) or on (right, model StBx\_1pc\_SG). The sizes of
  the circles correspond to the surface area of the clouds.  A
  horizontal black dotted line shows the mass accretion rate inferred for
  GMCs in the LMC by \citet{Fukui2009MOLECULARI, Kawamura2009THEFORMATION}, and
  \citet{Fukui2010MolecularGalaxies}. The solid black line corresponds
  to the gravitationally driven accretion case
  (Eq.~\ref{eq:macc_grav}), and the dashed line corresponds to the 
  turbulence driven accretion case (Eq.~\ref{eq:macc_turb}), for an ambient
  density of $n_{ism}=1$~cm$^{-3}$ and ambient velocity of
  $v_{ism}=10$~km~s$^{-1}$. For comparison, we include the mass accretion rates used in the semi-empirical model by \citet{Zamora-Aviles2012ANACTIVITY} (\emph{solid blue line}) and the semi-analytic model by \citet{Goldbaum2011THEACCRETION} (G11 in figure legend) for low surface density, $\Sigma_{res}=8$~M$_{\odot}$~pc$^{-2}$, (\emph{solid red line}) and high surface density, $\Sigma_{res}=16$~M$_{\odot}$~pc$^{-2}$, (\emph{dashed red line}) reservoirs.
\label{fig:MassAcc_CloudPop}} 
\end{figure*}

\subsection{Cloud Population}
\label{sec:cloud_population}
In this subsection we present the results of our simulations with 0.95~pc resolution at the midplane, with and without gas self-gravity. 
We concentrate on the influence of gas self-gravity on the total mass accretion rates. 

Figure \ref{fig:MassAcc_CloudPop} shows the measured mass accretion rates for the cloud population over 5~Myr of evolution with and without self-gravity. 
At first glance, it is striking how similar the mass accretion rates for clouds with and without self-gravity are for the mass range between $10^{3}$--$10^{5}$~M$_{\odot}$.
At the high mass end, however, there is a clear difference of up to two orders of magnitude in the measured mass accretion rates with and without self-gravity.

Both panels in Figure \ref{fig:MassAcc_CloudPop} show the mass accretion rates predicted for a gravitationally collapsing envelope, Equation~(\ref{eq:macc_grav}), and the mass flux rate across the cloud surface due to a turbulent background, Equation~(\ref{eq:macc_turb}). 
For the mass range between $10^{2}$--$10^{5}$~M$_{\odot}$, the mass accretion rates measured in the simulations closely follow the estimated accretion rate from a turbulent background, with little variation with the presence of self-gravity. 

At the high mass end, the clouds in the simulation without self-gravity do not accrete mass at rates estimated by either of the accretion models.
On the other hand, the simulations including self-gravity do accrete mass at the estimated rates. 
This likely occurs because in the presence of self-gravity clouds attract gas in their environments at rates similar to those estimated due to a free-falling envelope.  
Therefore, the accretion rates measured in this cloud population are determined by the combined action of turbulence and self-gravity.  

Figure~\ref{fig:MassAcc_CloudPop} shows the averaged mass accretion rates derived by \citet{Fukui2009MOLECULARI}, \citet{Kawamura2009THEFORMATION} and \citet{Fukui2010MolecularGalaxies} for the evolution of GMCs in the LMC. 
They measured an average mass accretion rate of $\dot{M} \approx 5 \times 10^{-2}$~M$_{\odot}$~yr$^{ -1}$ for GMCs with masses of $M > 10^{5}$~M$_{\odot}$. 
It is evident that without gas self-gravity, massive clouds above $10^{5}$~M$_{\odot}$ do not accrete mass at the rate estimated by observations, whereas in the presence of gas self-gravity, the accretion rates of the massive clouds lie close to the estimate from observations of massive clouds in the LMC.
This behavior suggest that the mass growth of MCs proceeds as a combination of turbulent driven and gravitational driven accretion.

For comparison, we include in Figure~\ref{fig:MassAcc_CloudPop} the mass accretion rates implemented in the semi-empirical model for the evolution of gravitationally contracting MCs by \citet{Zamora-Aviles2012ANACTIVITY}, and the semi-analytic model of clouds growing in virial equilibrium by \citet{Goldbaum2011THEACCRETION}.

In the \citet{Zamora-Aviles2012ANACTIVITY} model, the mass accretion rate is roughly constant for almost three decades in cloud mass. 
This is because their accretion rates are a direct outcome of their simulation setup, where two streams of uniform density gas collide in the middle of the box with a constant inflow velocity, in essence a similar concept to the turbulence driven accretion model we present. 
The difference in the resulting accretion rates between our two models lies in the different cloud shapes. 
Clouds in \citet{Zamora-Aviles2012ANACTIVITY} model are cylindrical and the surface area across which gas is accreted remains constant through most of the cloud's lifetime, in contrast to our model, where we assume spherically symmetric clouds where the mass and size scale according to Equation (\ref{eq:Falgarone}).

Mass accretion rates in the \citet{Goldbaum2011THEACCRETION} model
correspond to a gravitationally collapsing, constant surface density
reservoir. This model only provides mass accretion rates for cloud
masses above $5\times10^{4}$~M$_{\odot}$, given that this is their
initial cloud mass. In their model the accretion rate rises quickly
and then levels off to a magnitude similar to the one predicted in our
model in the mass range where comparisons with observational
estimations of accretion rates are available \citep{Fukui2010MolecularGalaxies}.

\subsection{High Resolution Clouds}
We now discuss the evolution of the individual zoomed-in clouds and their interactions with their environments.

Although the initial masses and virial parameters of the clouds are
relatively similar, the three clouds show substantial differences in
the evolution of their structural parameters and mass accretion
histories, giving us qualitative information on what aspects of the
behavior are historically dependent, and what are typical of clouds.

\subsection{Overview}
\label{subsec:evol}

The three clouds that we consider have properties listed in
Table~\ref{tab:zoom_results}.
We give brief descriptions of their
histories, including the location of SN explosions with respect to the
cloud center of mass ${\mathbf{d}}_{CM} = {\mathbf{x}}_{CM} -
{\mathbf{x}}_{SN}$. 

\begin{table*}
	\begin{center}
		\begin{tabular}{lccccccccc} 
            \hline
Name  & $M_0$ & $\langle\dot{M}\rangle$ & $\alpha_{vir,0}$ & $x_0$ &  $y_0$ & $z_0$ & $v_{x,0}$ &
                         $v_{y,0}$ &  $v_{z,0}$ \\
            & (M$_{\odot}$) & (M$_{\odot}$ yr$^{-1}$)  & &
                             \multicolumn{3}{c}{(pc)} & \multicolumn{3}{c}{ ( km s$^{-1}$)} \\ 
    		\hline
M3    &3600  & 2.1(-4) & 0.4   &  458 & -380 & 17 & 0   &   3 & -2  \\
M4    & 3200 & 3.9(-4) & 0.45 & 180  & -30   & 8   & -1 & -1 & -1   \\
M8    & 7500 & 4.8(-3) & 0.3   & 65    & 359   & 21 & 1   & -1 & -1   \\
    		\hline
		\end{tabular}
    \end{center}
    \vspace{-4mm}
    \caption{Parameters for the high resolution clouds, giving
      initial mass, average mass accretion rate, and initial virial parameter, position, and velocity. 
    \label{tab:zoom_results}}
\end{table*}

\begin{figure*}[!t]
\centering 
\includegraphics[width=1\textwidth]{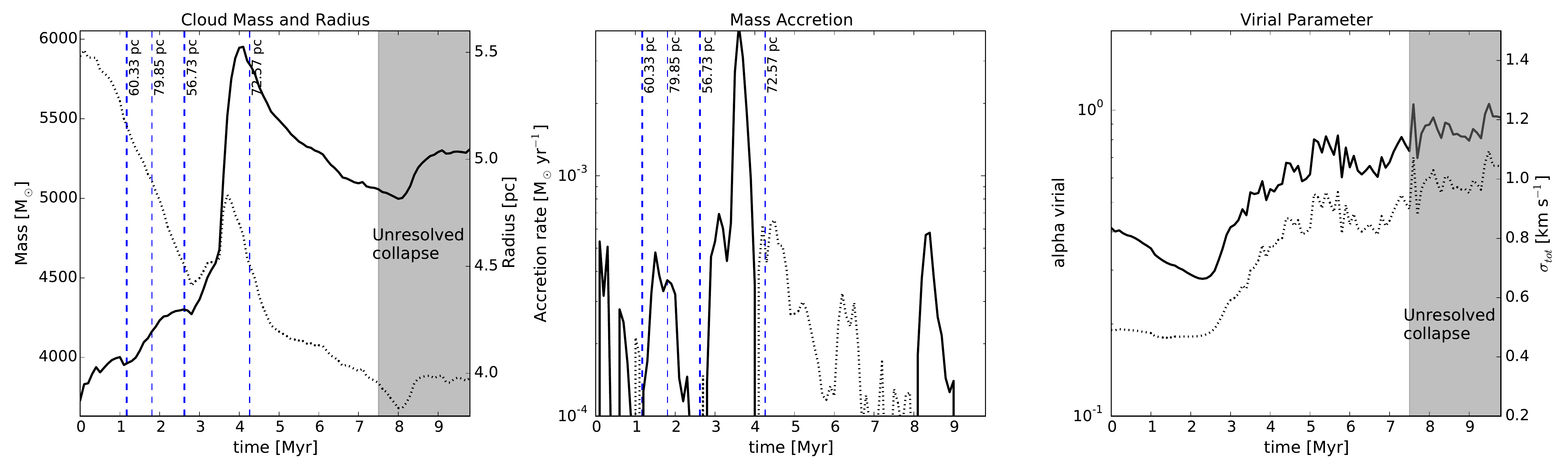} 
\vspace*{-6mm}
\caption{Evolution of the properties of cloud M3 starting at $t_{SG} =
  0$, when self-gravity is included. During the shaded period more than
  half the cloud's mass exceeds the Jeans resolution criterion
  (see Sect.~\ref{zoom-in_sec}).
{\em{Left:}} Evolution of cloud ({\em solid line}) mass and ({\em
  dotted line}) radius as a function of time. Vertical {\em dashed blue
lines} mark nearby SN explosions, with line thickness 
inversely related to the given center-of-mass distance.  
{\em Center:} Mass accretion rate as a function of time. The {\em solid
line} shows accretion, while the {\em dotted line} shows
mass loss.  
{\em Right:} Evolution of cloud ({\em solid line}) virial
parameter and ({\em dotted line}) total velocity dispersion. 
\label{fig:M3_evol} } 
\end{figure*}

\paragraph{Cloud M3} This cloud gained the least amount of mass during its evolution.
The cloud has an elongated structure that develops into a long, dense filament of $\approx 20$~pc in length as the cloud contracts. 
This filament then fragments as the cloud continues contracting as a whole.
The cloud is affected by four nearby SN explosions.
The first SN occurs at $t=1.17$~Myr, at a distance of ${\bf{d}}_{CM} = (30, -51, 11)$~pc compressing the cloud.
The second explosion occurs at $t=1.81$~Myr, at a distance of ${\bf{d}}_{CM} = (15, -78, -11)$~pc from the cloud.
This event does not have a major impact on the cloud, as it explodes in a large bubble of rarefied gas below the midplane and can expand freely in all directions. 
The third SN occurs at $t=2.62$~Myr at ${\bf{d}}_{CM} =(7, 51, -23)$~pc.
This explosion is not only closer, but also occurs in a more confined environment, sweeping a large amount of gas towards the cloud.
A final SN explosion occurs at $t=4.24$~Myr, at a distance of ${\bf{d}}_{CM} = (60, -13, 38)$~pc, contributing to the fragmentation of the cloud and clearing of the environment. 
Together, these explosions act to deliver gas and compress parts of
the cloud, while also ablating and fragmenting other parts of the
cloud. 
As a result, the cloud does not show a smooth accretion rate,
but rather a highly chaotic one, with sudden peaks of mass growth but
also periods of mass loss, as shown in Figure~\ref{fig:M3_evol}.  This rather low mass cloud is
significantly affected by the turbulence in the environment, and the collective influence of SN explosions.

The left panel of Figure~\ref{fig:M3_evol} shows that Cloud M3 grows in mass over the initial 4~Myr of evolution, but then loses roughly 15\% of its mass during the following $\sim4$~Myr.
As the cloud continues interacting with its environment, the densest parts of the cloud undergo gravitational collapse, with more than half of its mass reaching unresolved density by $t=7.1$~Myr, in regions that appear likely to undergo vigorous star formation.

Looking at the virial parameter in the right panel of Figure~\ref{fig:M3_evol}, we see that cloud M3 starts with a low value that continues dropping for the first 2~Myr. At the moment the blast wave of the nearby SN hits the cloud, the virial parameter jumps, and then begins to steadily climb to higher values as the cloud contracts gravitationally.

\paragraph{Cloud M4} This is the cloud with the most nearby SN
explosions, but it actually shows less chaotic behavior than the other
two. 

The first SN near the cloud occurs at $t=0.42$~Myr, at ${\bf{d}}_{CM} = (-30, 72, 14)$~pc.
This SN is behind a giant cloud that shields M4 from any influence.
A second SN occurs at $t=2.09$~Myr, at ${\bf{d}}_{CM} = (-19, 46,
33)$~pc. A third SN explodes far above the midplane at $t=3.75$, and
distance of ${\bf{d}}_{CM} =(-3, 23, 48) $~pc, where it can freely
expand, slightly disturbing the envelope. The fourth and fifth
supernovae, exploding at ${\bf{d}}_{CM} = (8, -3, 67) $~pc, and $(23,
-30, 82) $~pc, occur after the cloud has mostly collapsed below the
resolution limit and thus have little effect for numerical reasons.

\begin{figure*}
\centering 
\includegraphics[width=1\textwidth]{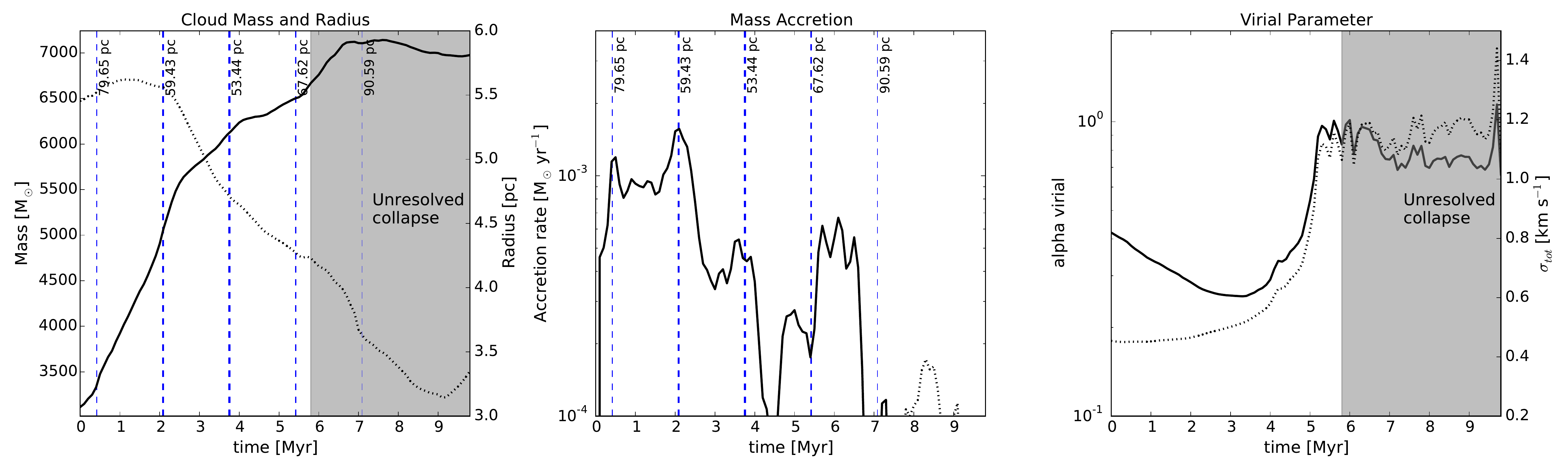} 
\caption{Evolution of properties of cloud M4 
using same diagnostics and notation as Figure~\ref{fig:M3_evol}.
\label{fig:M4_evol} } 
\end{figure*}

This cloud has a rather chaotic accretion history, affected by the turbulence of the
environment. 
After 0.5~Myr, M4 begins to accrete gas at a rate of $\dot{M}\approx
5\times 10^{-4}$~M$_{\odot}$~yr$^{-1}$, which monotonically increases
over time as expected from its increasing mass (see
Fig.~\ref{fig:M4_evol}). The second SN compresses the envelope and
transiently increases the mass accretion rate up to $\dot{M} \approx
1.3\times 10^{-3}$~M$_{\odot}$~yr$^{-1}$.  However, the blast wave also disrupted the cloud envelope, which 
results in a subsequent period of lowered mass accretion rate.

At the end of the first $2.5$~Myr of evolution, cloud M4 has
gained almost $70\%$ of its initial mass while roughly
preserving a constant size. 
At later times, cloud M4 continues accreting gas but now at a slightly lower rate of $\dot{M}\approx $2--3$ \times 10^{-4}$~M$_{\odot}$~yr$^{-1}$.
This lower accretion rate is a consequence of a reduced reservoir, as the cloud envelope has either been accreted onto the cloud or eroded by the turbulent environment.
The third SN has only minor impact on the
cloud properties as it freely expands upwards.

At a time close to when the cloud becomes unresolved, $\approx 6.5$~Myr, there is almost no gas left in the envelope, and the cloud continues contracting without further accretion. 
At this point vigorous star formation should be occurring in the centers of gravitational collapse, which would impact the cloud properties from within.

\begin{figure*}[t]
\centering 
\includegraphics[width=1\textwidth]{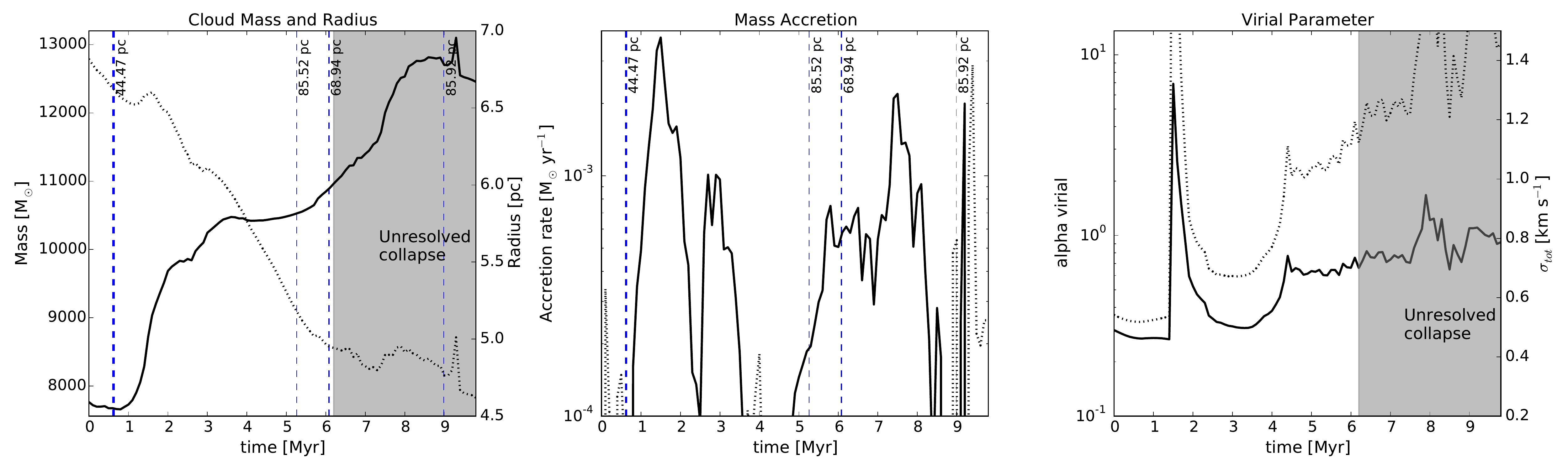} 
\caption{Evolution of properties of cloud M8
using same diagnostics and notation as Figure~\ref{fig:M3_evol}.
\label{fig:M8_evol} } 
\end{figure*}

\paragraph{Cloud M8} The most massive cloud has an interesting
history, as it is shocked by several nearby SN explosions that have a
major impact on the cloud and its envelope.  
The first nearby SN explodes at time $t=0.62$~Myr at ${\bf{d}}_{CM} = (-29, 14, -30) $~pc.
This SN shocks the cloud envelope $\approx 0.2$~Myr, later, and triggers an increased mass accretion rate that lasts for over a megayear (see Fig.~\ref{fig:M8_evol}).
There is a steep rise in the cloud mass from $t\approx 1.1$--2.0~Myr, with mass accretion rates climbing to $\dot{M} \approx 4 \times 10^{-3}$~M$_{\odot}$~yr$^{-1}$.  
Although the cloud continues accreting mass from $t\approx 2$--3.5~Myr, a lot
of mass is simultaneously torn off of the cloud on the side opposite
to the SN explosion. 

Because of the sudden accretion of fast moving material, the virial
parameter of the cloud briefly jumps to $\alpha_{vir}\approx 8$. 
Some of this gas becomes part of the cloud and shares its kinetic energy with the gas in the cloud. 
However, the large virial number results in the expected large amount of mass loss from the cloud, including erosion from the main body and breakup of large fragments. 

The turbulence induced in the cloud decays in a crossing time as expected \citep{MacLow1999TheClouds}, resulting in a fast drop of the virial parameter. 
This nearby explosion also plays a key role in perturbing the envelope, quickly pushing it onto the cloud or clearing it. 
This leads to a low accretion rate period from $t\approx 3.5$--5.5~Myr, a period during which the cloud just contracts gravitationally. 

At $t=5.26$~Myr, a second nearby SN blows up above the cloud, at ${\bf{d}}_{CM}= (-33, -69, \,38) $~pc.
While the SN shock front moves down towards the cloud, a third SN below the cloud explodes at  $t=6.02$~Myr, at ${\bf{d}}_{CM} =(32, 3, -61) $~pc.
These two subsequent SN explosions coming from opposite directions compress the cloud and its envelope, delivering another rush of mass to the cloud.
At times from $t\approx 5.5$--7~Myr, the cloud accretes material at a rate of
$\dot{M} \approx 5 \times 10^{-4}$~M$_{\odot}$~yr$^{-1}$.  
During its active history, the densest peaks of cloud M8 continuously contracting until they become unresolved in the simulation. 
At $t\approx 6.2$~Myr, half of the cloud mass has already collapsed to
unresolved structures, and the subsequent evolution of the cloud
dynamics is unresolved by our standard.

\subsection{Instantaneously Accreted Gas}
\label{subsec:inst}
\begin{figure*}[!h]
\centering 
\includegraphics[width=1\textwidth]{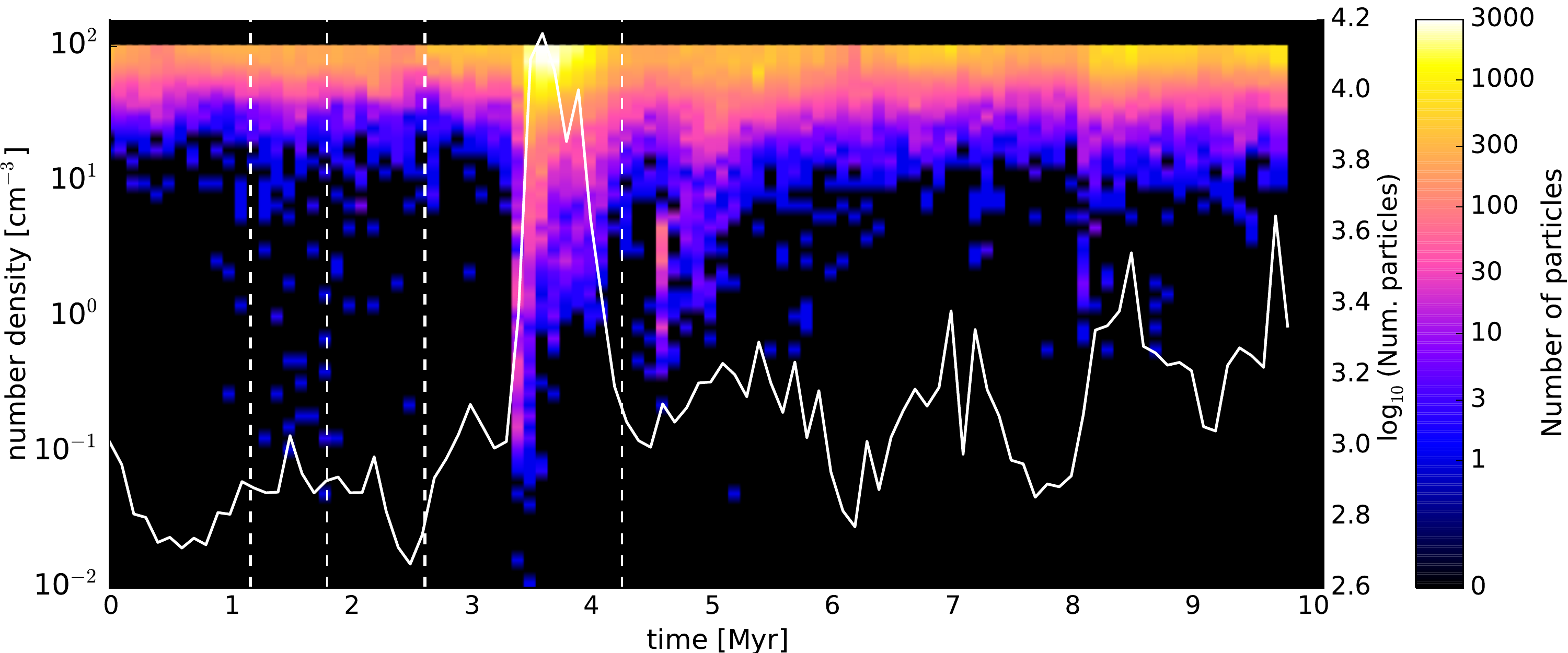}
\includegraphics[width=1\textwidth]{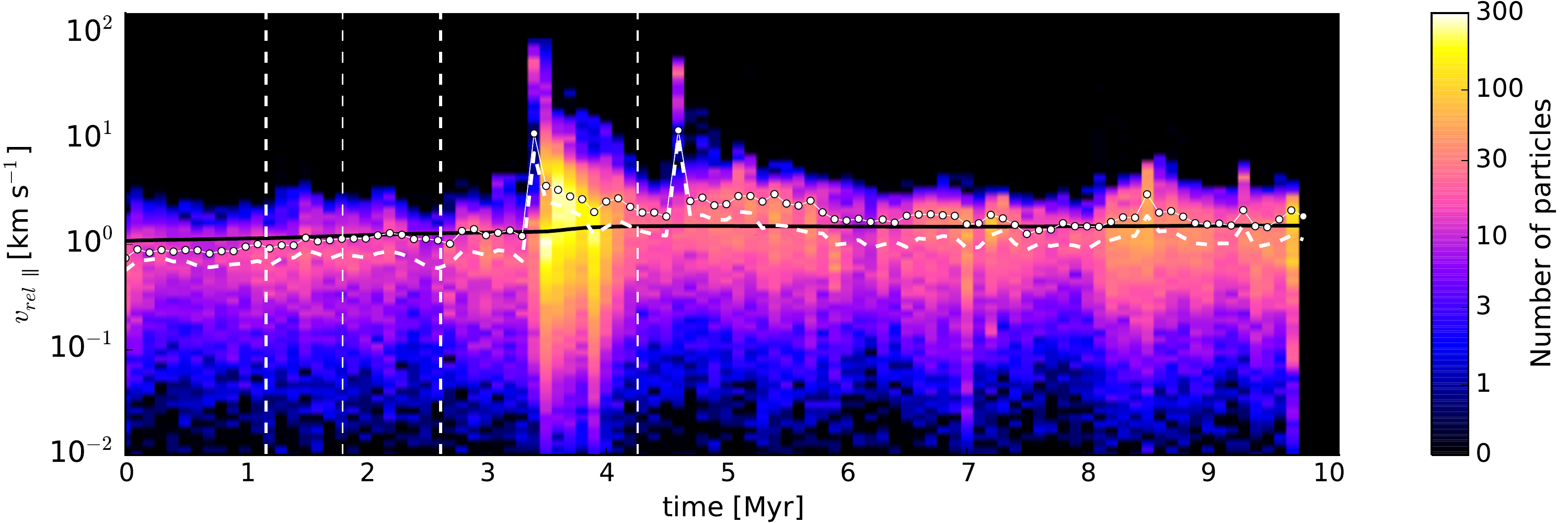} 
\includegraphics[width=1\textwidth]{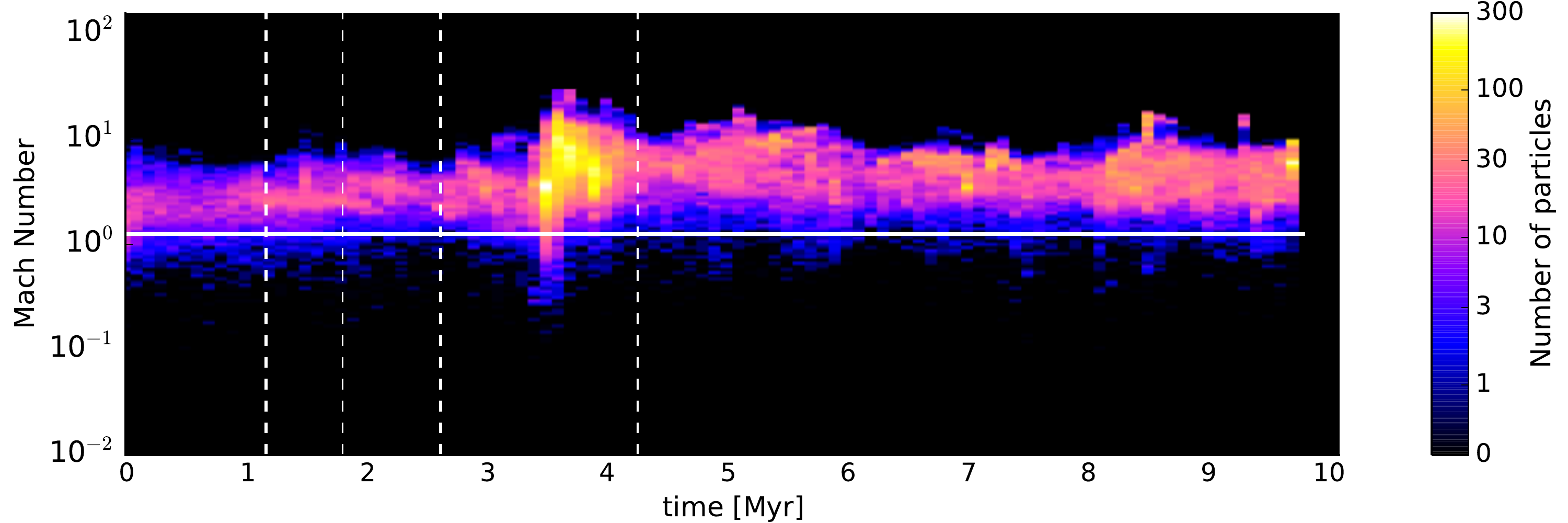} 
\caption{
    Properties of tracer particles accreted onto cloud M3 in the last sample before accretion (samples were taken every 0.1~Myr).  Vertical {\em dashed} lines show times of nearby SNe with thickness proportional to distance (see Fig.~\ref{fig:M3_evol}). 
  {\em (Top)} Number density at the location of the tracer particle.
  The solid white line shows the total number of tracer particles accreted in each sample.  
  {\em (Middle)} Local velocity component parallel to the density gradient relative to the cloud's center of mass velocity.  
The solid black line gives the free-fall velocity of the cloud, assuming the cloud mass is contained in a sphere of the same volume as the cloud (Eq.~\ref{eq:vinf}). 
The dashed and dotted white lines give the mean parallel and perpendicular velocities. 
  {\em (Bottom)} Local Mach number. Solid white line separates between subsonic and supersonic flow.
  \label{fig:M3_InstAcc} } 
\end{figure*}

\paragraph{Cloud M3} The number density distribution of accreted
tracer particles in this cloud (top panel of
Fig.~\ref{fig:M3_InstAcc}) shows that most of the particles entering
the cloud, unsurprisingly, have number densities slightly lower than
the number density threshold used to define the cloud.

This behavior is expected as the material has to climb up a density gradient to enter the cloud. 
However, there are significant variations of the mean number density
of the accreted material, revealing clues about the different
mechanisms of mass accretion onto the cloud. 

Cloud M3 accretes gas over the first $\approx 3$~Myr, as it is continuously shocked by SN shocks. 
Then, after the explosion at 2.6~Myr, a large amount of gas in the envelope becomes part of the cloud as seen in the middle panel of Figure~\ref{fig:M3_evol} and the top panel of \ref{fig:M3_InstAcc}. 
We identify two stages of the cloud's response to the shock.
During the short first stage, the cloud accretes low-density, fast-moving material. 
Figure~\ref{fig:M3_InstAcc} shows that there is an almost uniform
distribution of gas with densities between $n_{ISM}\approx 0.1$--90 that
climb up the density gradient to suddenly become part of the cloud,
with velocities parallel to the density gradient almost reaching
$v_{\parallel} \approx 100$~km~s$^{-1}$.  

In the second stage, this rapid accretion is followed by a slightly longer stage of enhanced accretion from the dense envelope compressed by the SN blast wave. 
This is seen as a clump of gas with enhanced density accreted over a timescale of $\approx 0.5$~Myr after the fast accretion stage. 
The accreted material in this second stage climbs the density gradient
with velocities ranging from $v_{\parallel}\approx 2$--5~km~s$^{-1}$,
faster than the free fall velocity (middle panel of Fig.~\ref{fig:M3_InstAcc}). 
This gas has high velocity both parallel and perpendicular to the density gradient.  
The perpendicular velocity corresponds to random gas motions that probably are not driven by gravity, and thus presumably come from the turbulent environment. 

The nearby SN explosion also causes mass loss, as it disrupts part of the envelope and even detaches fragments of the cloud. 
Although the cloud core has been compressed, some of its outermost material has been stirred and is now only marginally bound, so that turbulent motions in the environment and the final nearby SN explosion at 4.2~Myr can detach it. 
Thus, during the following $\approx$4~Myr of evolution the cloud loses mass (Fig.~\ref{fig:M3_evol}).
Note that Figure~\ref{fig:M3_InstAcc} only shows particles entering the cloud, but not particles leaving it.

Perpendicular velocities systematically exceed parallel velocities (dotted and dashed lines in the middle panel of Figure~\ref{fig:M3_InstAcc}). 
This suggests that the mass accretion rate depends strongly on environmental turbulence.  
However the importance of gravity is emphasized by the inability of nearby SN explosions to completely disrupt the cloud's envelope, and the amount of gas moving at the free-fall velocity. 

The Mach number distribution of the infalling gas uniformly exceeds unity.  
The accretion driven by the SN blast reaches Mach $\approx 20$.  
However, even at other times, the inflow remains supersonic,
suggesting freely-falling gas is far from any sort of hydrostatic
equilibrium.

\begin{figure*}
\centering 
\includegraphics[width=1\textwidth]{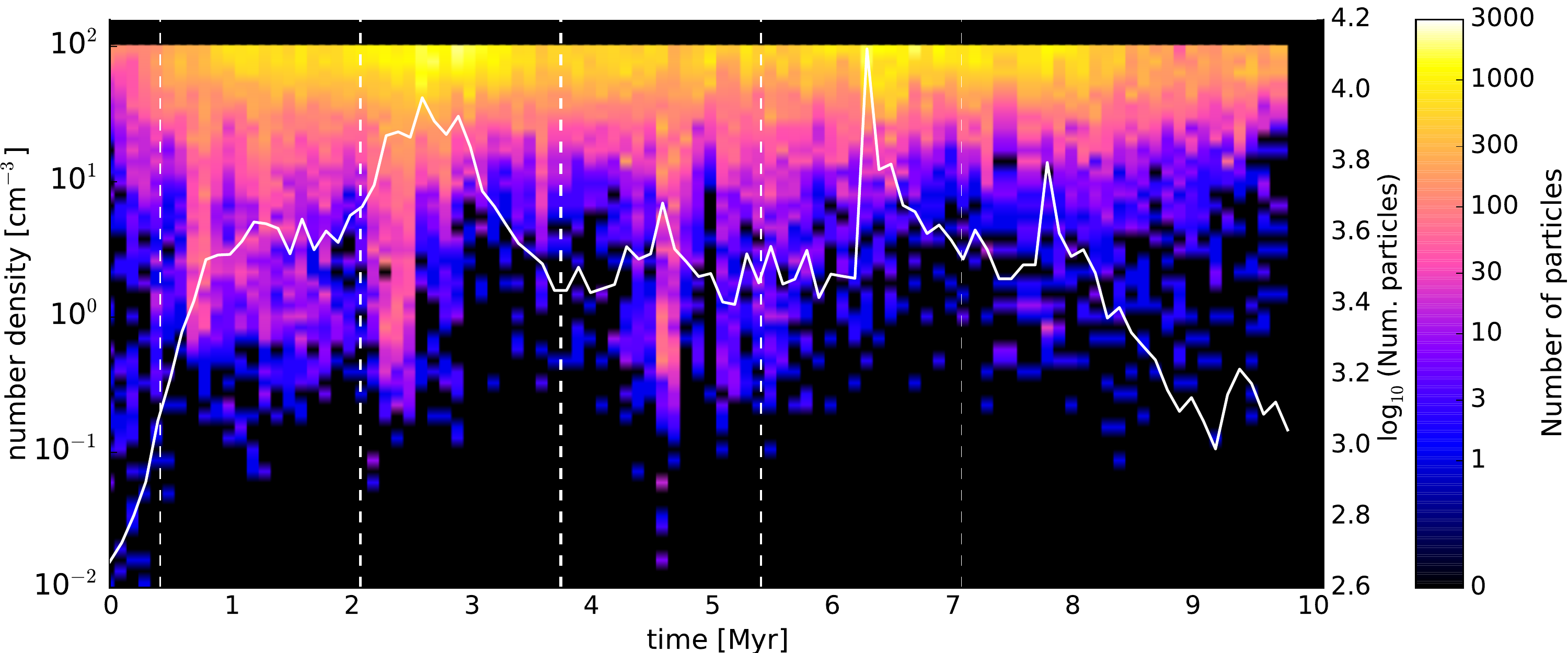} 
\includegraphics[width=1\textwidth]{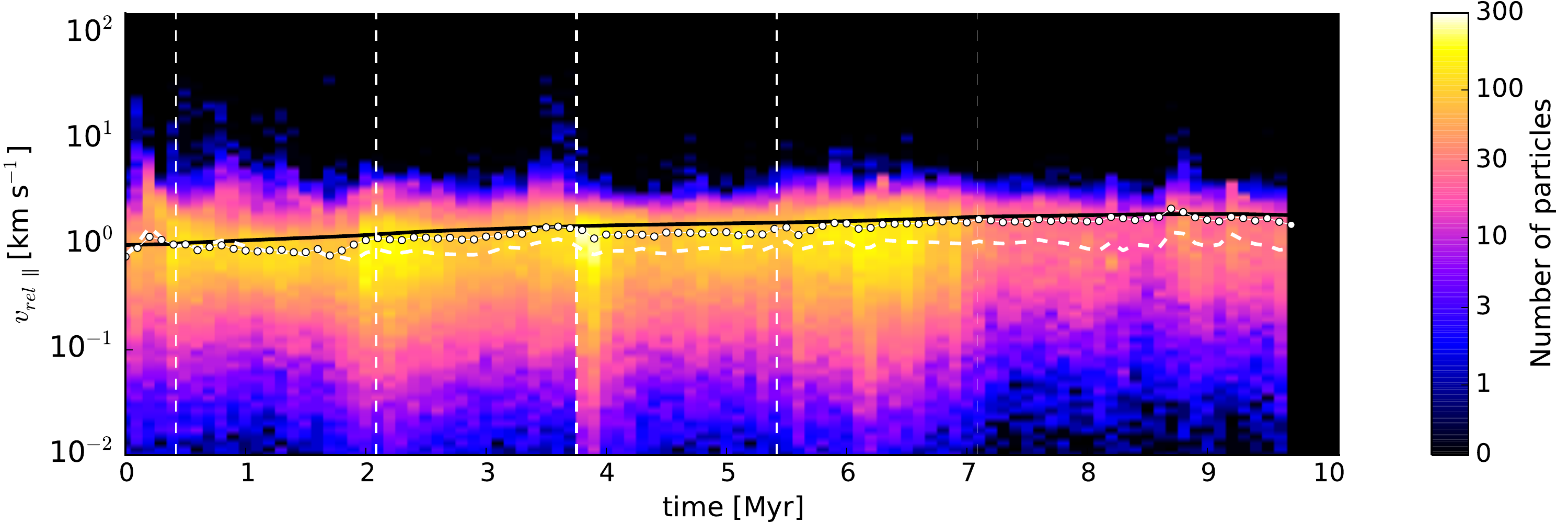} 
\includegraphics[width=1\textwidth]{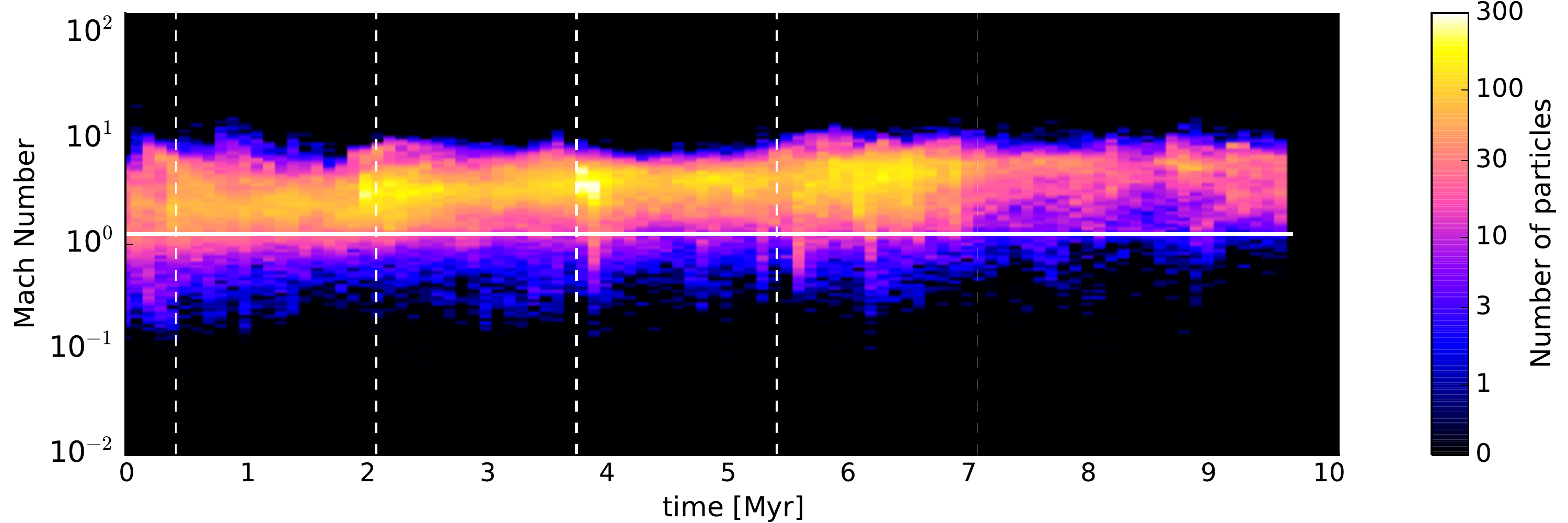} 
\caption{
    Properties of tracer particles accreted onto cloud M4, using same
    diagnostics and notation as Figure~\ref{fig:M3_InstAcc}. 
  \label{fig:M4_InstAcc}} 
\end{figure*}

\paragraph{Cloud M4} This shows the most uniform accretion history of the three clouds. 
Although for this cloud nearby SN events have little effect on the cloud, they do trigger some minor surges of accretion of low density gas, like the ones observed at $t \approx 0.7$~Myr, 2.3~Myr, and 4.5~Myr (top panel of Figure~\ref{fig:M4_InstAcc}). 
Some of these events also appear in the parallel velocity distribution (second panel, Figure \ref{fig:M4_InstAcc}) as sudden peaks of the velocity distribution of the particles. 
However, as discussed in Section~\ref{subsec:evol}, the
overall evolution of this cloud is not significantly affected by the
nearby SN explosions but rather is relatively uniform in time. 

For this cloud, we find that the mean parallel and perpendicular velocity components of the accreted gas are roughly equal during the whole evolution of the cloud.  
This suggests that the mass accretion into this cloud may be equally due to gravitationally infall of the envelope and capture of turbulent material from the cloud. 

Most of the mass inflow is supersonic, with ${\cal M} \approx $1--10.
Although there are no major SN events nearby, cloud M4 is embedded in a dynamic, turbulent environment, full of hot diffuse gas, that repeatedly shocks the cloud, sending bursts of accretion onto the cloud, with supersonic accretion velocities.

\begin{figure*}[t]
\centering 
\includegraphics[width=0.99\textwidth]{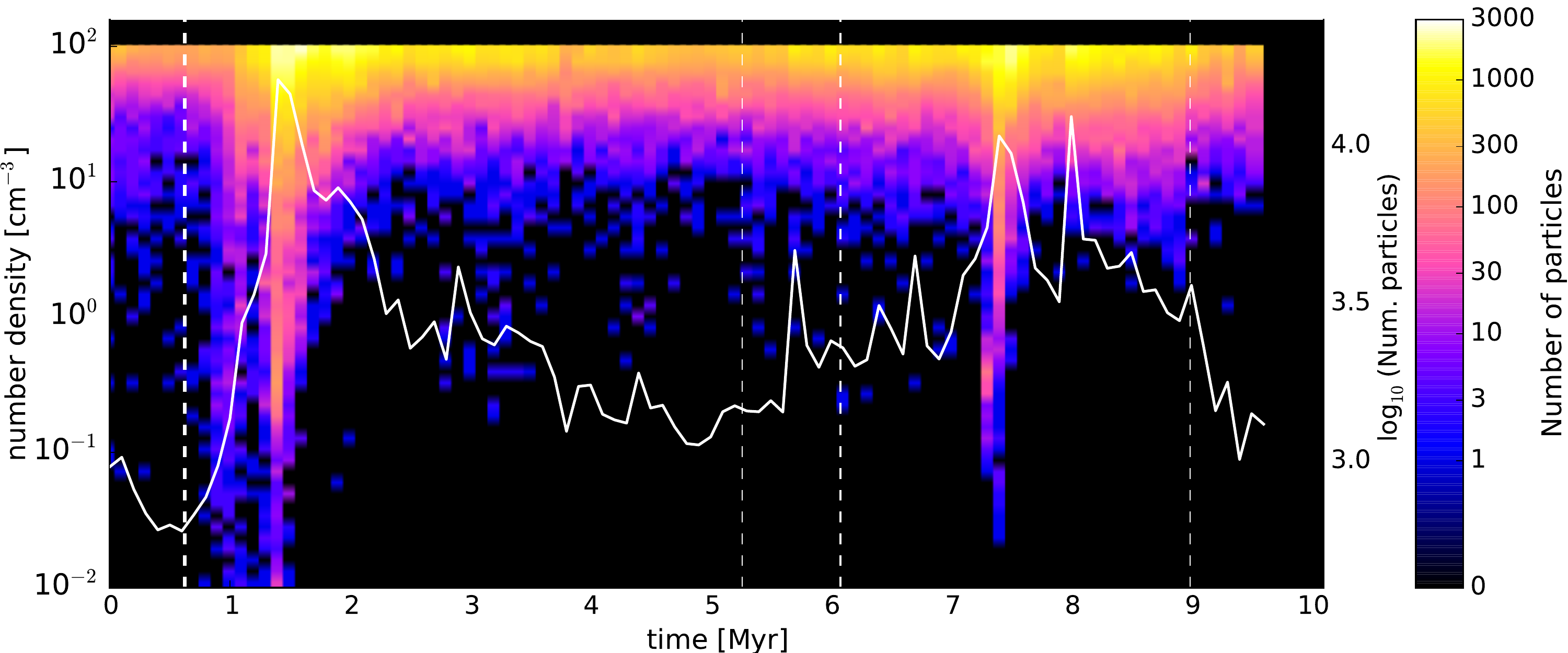} 
\includegraphics[width=0.99\textwidth]{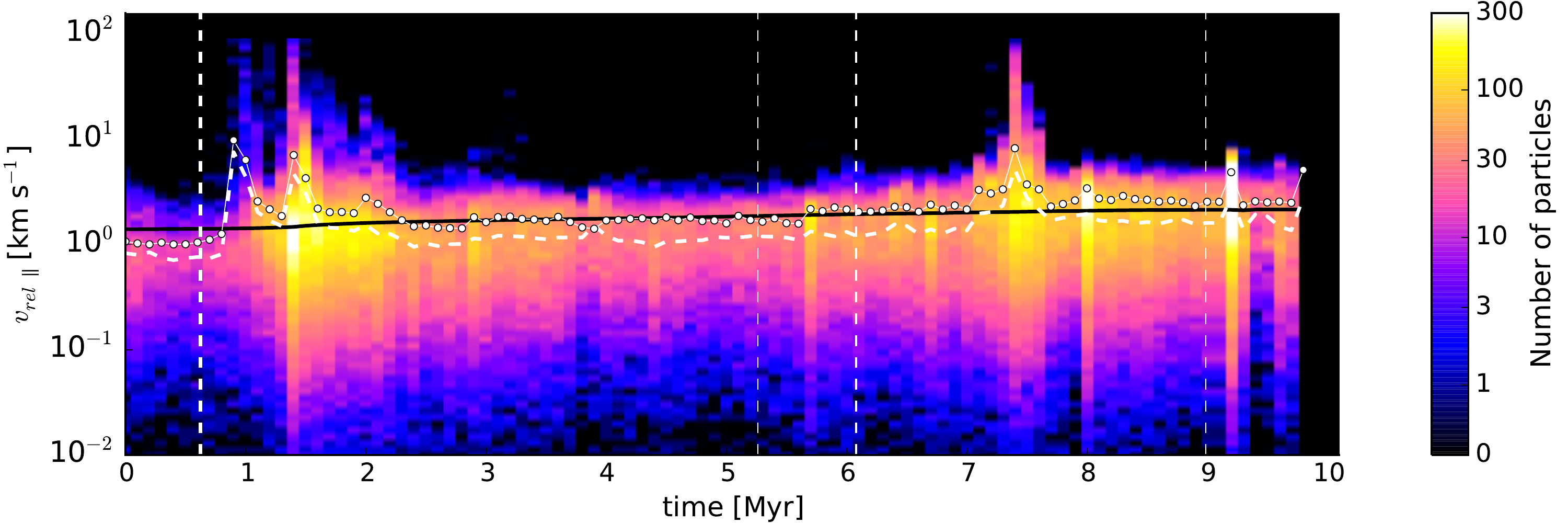} 
\includegraphics[width=0.99\textwidth]{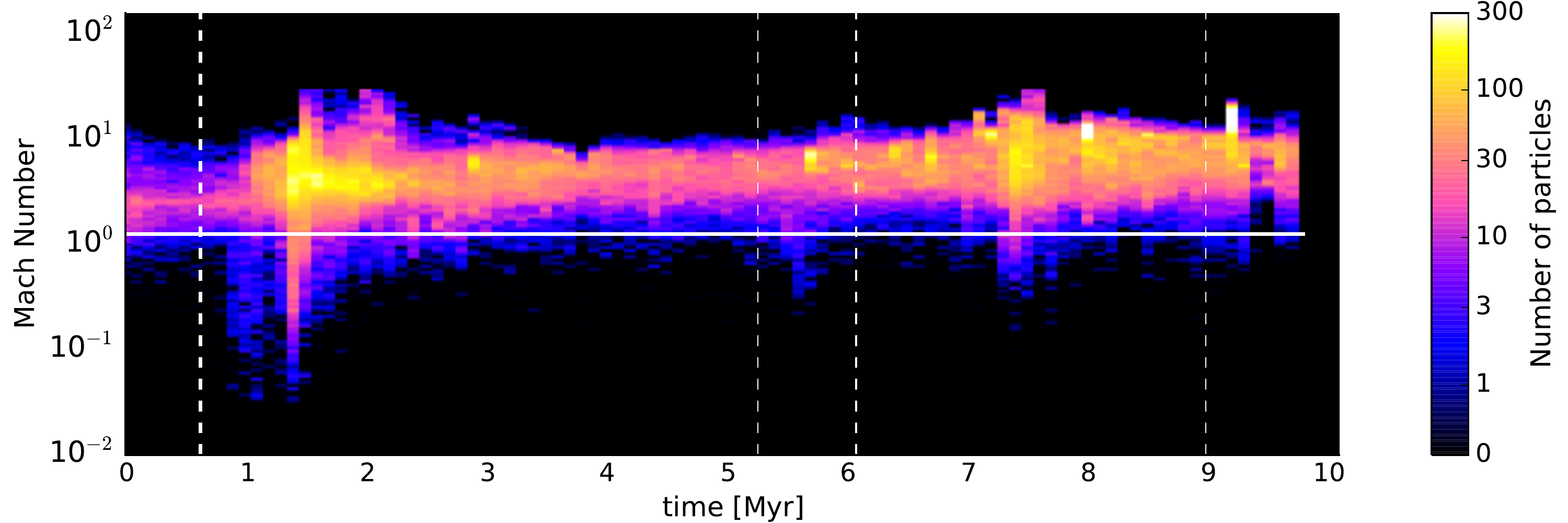} 
\caption{
    Properties of tracer particles accreted onto cloud M8, using same
    diagnostics and notation as Figure~\ref{fig:M3_InstAcc}. 
  \label{fig:M8_InstAcc}} 
\end{figure*}

\paragraph{Cloud M8} This cloud shows large fluctuations in the accretion rate, number of
accreted particles, and the densities and velocities of these accreted
particles (Fig.~\ref{fig:M8_InstAcc}).
The first blast wave shocking the cloud occurs at $t\approx 1$~Myr, when a burst of fast moving, low density gas is accreted onto the cloud. 
The mass accreted onto the cloud also shows two stages of accretion after the nearby explosion, as in cloud M3. 
There is a short stage during which the envelope is quickly compressed and fast moving material is accreted onto the cloud, followed by a longer-lived phase during which the now dense, compressed envelope, falls onto the cloud at slightly lower velocities.
However, for this cloud, in particular, the first explosion cleared the envelope on the side facing the SN, exposing the cloud surface to the low-density, turbulent environment.
This results in an extended stage of accretion of low-density gas, lasting until $t \approx 1.8$~Myr. 
Following this active period, the cloud goes through a period of low accretion that lasts roughly $4$~Myr.
During this time, it mostly accretes gas that slowly climbs up the density gradient and becomes part of the cloud.
This gas, however, is highly turbulent, with mean perpendicular velocities, similar to the velocity expected purely from gravitational infall and only a small dispersion around the mean. 

At the end of this period, two nearby SN explosions trigger another accretion burst, compressing the cloud and its envelope, as a large fragment of dense gas joins the cloud.
The captured fragment just became gravitationally bound, and moves at speeds close to the velocity expected from gravitational infall.
However this event is difficult to follow in this figure as there is no obvious signature of the parallel or perpendicular velocity components expected from such an event.  

Looking at the overall behavior of the parallel and perpendicular velocities of the accreted gas, we notice that the perpendicular velocity is systematically higher than the parallel velocity throughout the evolution.
This suggests that the accretion of gas onto this cloud is mostly due to gas delivered by the turbulent environment and not only due to the gravitational collapse of the envelope. 
This turbulent accretion rate shows large velocity fluctuations, particularly after the explosion of a nearby SN. 

\subsubsection{Global Evolution}

\begin{figure*}[!t]
\centering 
\includegraphics[width=1.0\textwidth]{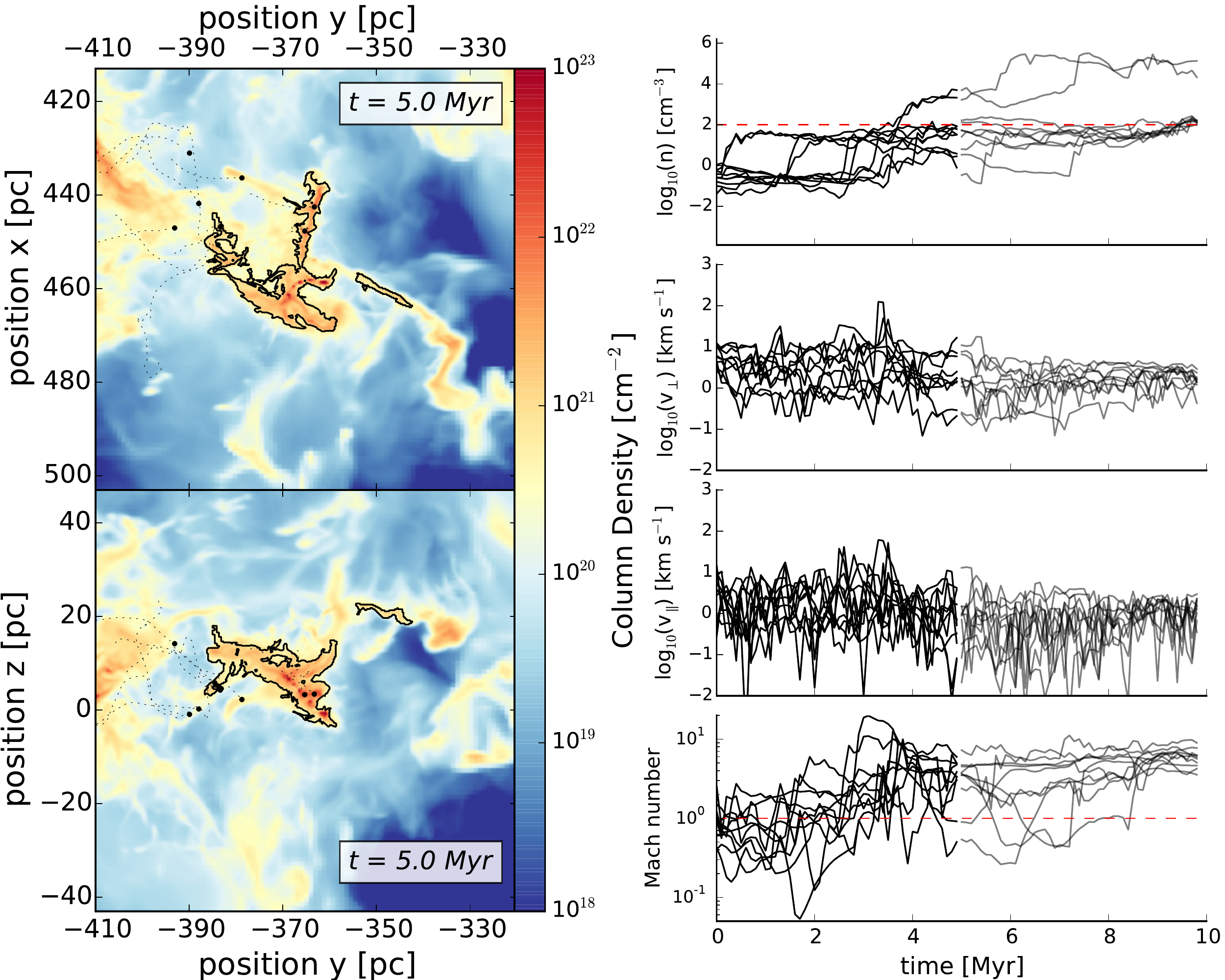}
\caption{Evolution of randomly selected tracer particles eventually accreted by cloud
  M3. 
    {\em{Left:}} Column density projection at $t=5.0$~Myr along
  lines of sight {\em (top)} perpendicular and {\em (bottom)} parallel to the
  midplane. The solid black contour follows the projected surface of the cloud.
	Both panels include the location of the ten particles at the
        current time (before many of the particles have been accreted)
        and their trajectories from the moment of injection up to the
        current time. 
	{\em{Right:}} Four panels showing the dynamical properties
        sampled by each of the ten particles along their trajectories.
	Panels show: {\em(top)} Local number density traced by the
        particles.  
    The horizontal dashed red line shows the cloud density threshold.
	{\em (second and third)} Velocity of the particles perpendicular and parallel to the local density gradient.
	{\em (bottom)} Mach number of the particles calculated using the local adiabatic sound speed of the gas and with respect to the center of mass of the cloud. 
	A dashed red line shows the sound speed, while the transition
        from black to grey shows the current time. 
	An animation of the full time evolution of this figure can be found online.
    \label{fig:M3_GlobalAcc} }  
\end{figure*}

\paragraph{Cloud M3} Figure~\ref{fig:M3_GlobalAcc} shows the evolution of ten randomly selected tracer particles that end up being accreted by this cloud. 
The number densities traced by these particles show that particles
tend to remain for relatively long times in the stable phases of the
ISM, and quickly jump between phases.  

The first group of particles lives in the dense phase, $n \approx 10$--90~cm$^{-3}$, sampling gas densities near the density threshold used to define the cloud. We call this group cold envelope particles. 
There is a second group moving at higher velocities and sampling lower densities, $n\approx 0.1$--1~cm$^{-3}$. We call this group warm environment particles.
The distinction between these two groups is seen in the top panel of Figure~\ref{fig:M3_GlobalAcc} between $t\approx 0$--3~Myr.
At $t\approx1.5$~Myr and $t\approx2.8$~Myr, some of the particles in the warm environment rapidly shift to the cold envelope as they get compressed towards the cloud and quickly climb up the density gradient.
This change of gas phase is preceded by small peaks in both the parallel and perpendicular velocities, followed by drops in the velocities because the particles conserve momentum as they move into a denser environment. 

The mean velocity of the particles perpendicular to the density gradient in the warm environment is of the order of $v_{\perp} \approx 1$--10~km~s$^{-1}$, and their mean parallel velocities are systematically lower almost by an order of magnitude. 
Once the particles become part of the cold envelope both their parallel and perpendicular velocities drop to $v_{\parallel, \perp} \approx 0.1$--1~km~s$^{-1}$.
Due to these low velocities, particles remain in the cold envelope for long times as they move towards the cloud. 
Sudden events, such as a nearby SN explosion, can compress the envelope, pushing the gas above the density threshold to become part of the cloud. 
However, there are some cases where particles already in the cold envelope return to the warm environment when the cold envelope is eroded by turbulence, as temporarily occurs at $t\approx 5$ and~6~Myr, for example.

While the warm environment particle population traces Mach numbers below unity, the cold envelope and cloud populations always move at locally supersonic speeds (bottom panel of Figure \ref{fig:M3_GlobalAcc}).  The sudden accretion of fast-moving material pushed by the nearby SN explosion causes the incoming gas to briefly reach Mach numbers of up to ${\cal M}\approx 20$. 

\begin{figure*}[!t]
\centering 
\includegraphics[width=1.0\textwidth]{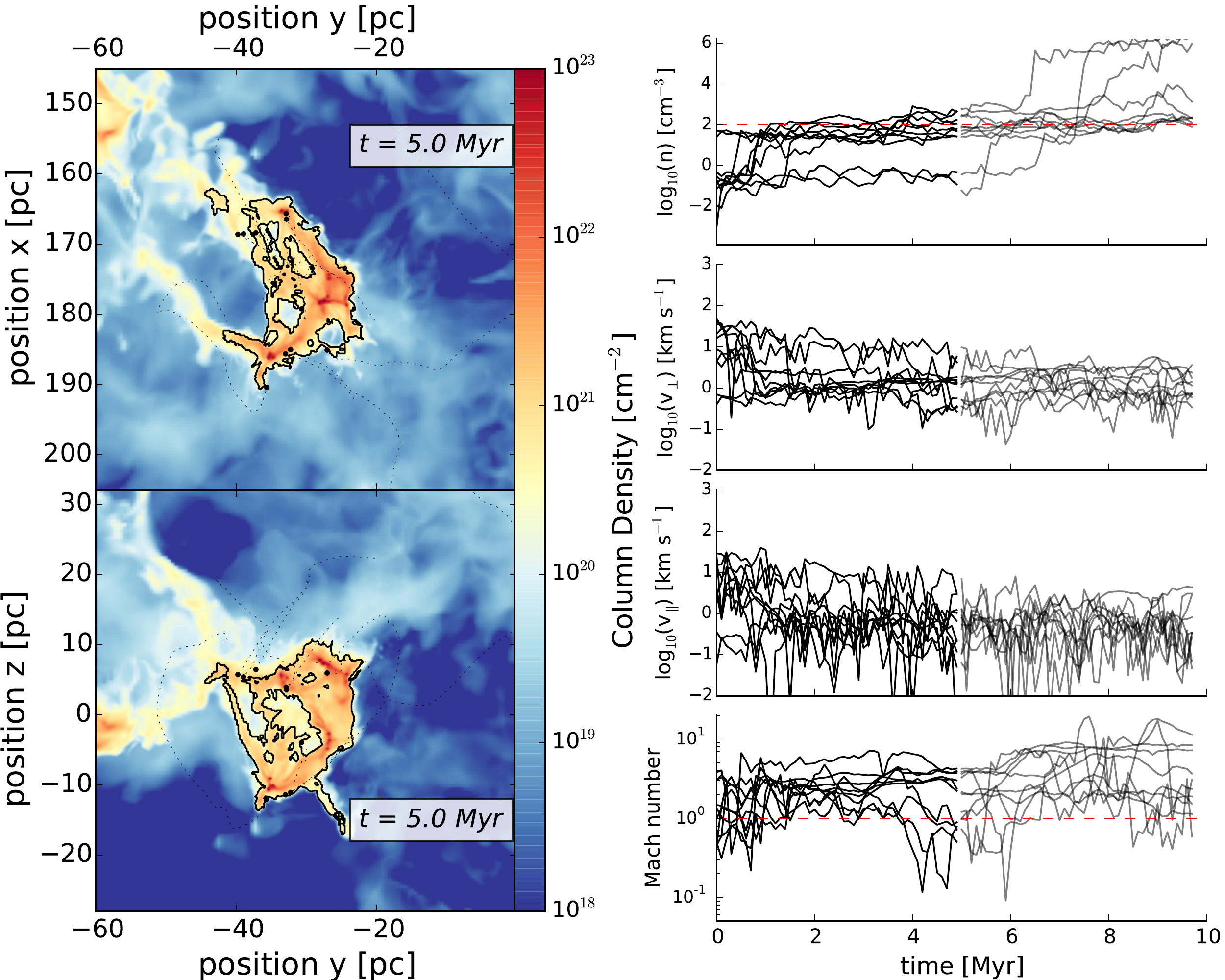} 
\caption{
Evolution of randomly selected tracer particles eventually accreted by cloud
  M4 using same diagnostics and notation as Figure~\ref{fig:M3_GlobalAcc}. 
    \label{fig:M4_GlobalAcc}} 
\end{figure*}

\paragraph{Cloud M4} The global evolution of the accreted gas onto cloud M4 shown in Figure~\ref{fig:M4_GlobalAcc} differs from the other two clouds because this cloud is embedded in a hotter, lower density environment.
In this energetic environment, cloud M4 is constantly shocked by hot,
rarefied gas that quickly becomes part of the cloud.

The accreted material arrives with velocities $\approx 10$--100~km~s$^{-1}$, but velocity drops quickly as gas climbs the density gradient.
During this time, warm environment particles occasionally pass through sharp transitions into the cold envelope, while particles in the cold envelope slowly move around on their way to joining the cloud.  
Three distinct populations appear in the top panel of Figure~\ref{fig:M4_GlobalAcc}, reflecting the three thermal phases of the ISM. 

Although this cloud is the one that experiences the most nearby SN explosions, they do not impact the cloud too severely, as it is already embedded in a hot environment.
For this reason, the accretion onto the cloud proceeds in a more uniform fashion, without strong fluctuations. 
As particles enter the cloud with velocities of $v_{\parallel} \approx 1$~km~s$^{-1}$, they produce only slow shocks with Mach numbers of ${\cal M}\approx$1--5.

Out of these ten randomly selected particles, only one particle shows two sudden jumps in density, passing through shocks with Mach numbers of ${\cal M} \approx 10$.
This particle is shocked by a SN at $t\approx4.9$~Myr, and is quickly pushed from the warm environment to the cold envelope.
Almost immediately the particle becomes part of the cloud and is
sucked in by a gravitationally collapsing density peak, maintaining
high Mach numbers in this cold, dense environment.

\begin{figure*}
\centering 
\includegraphics[width=1.0\textwidth]{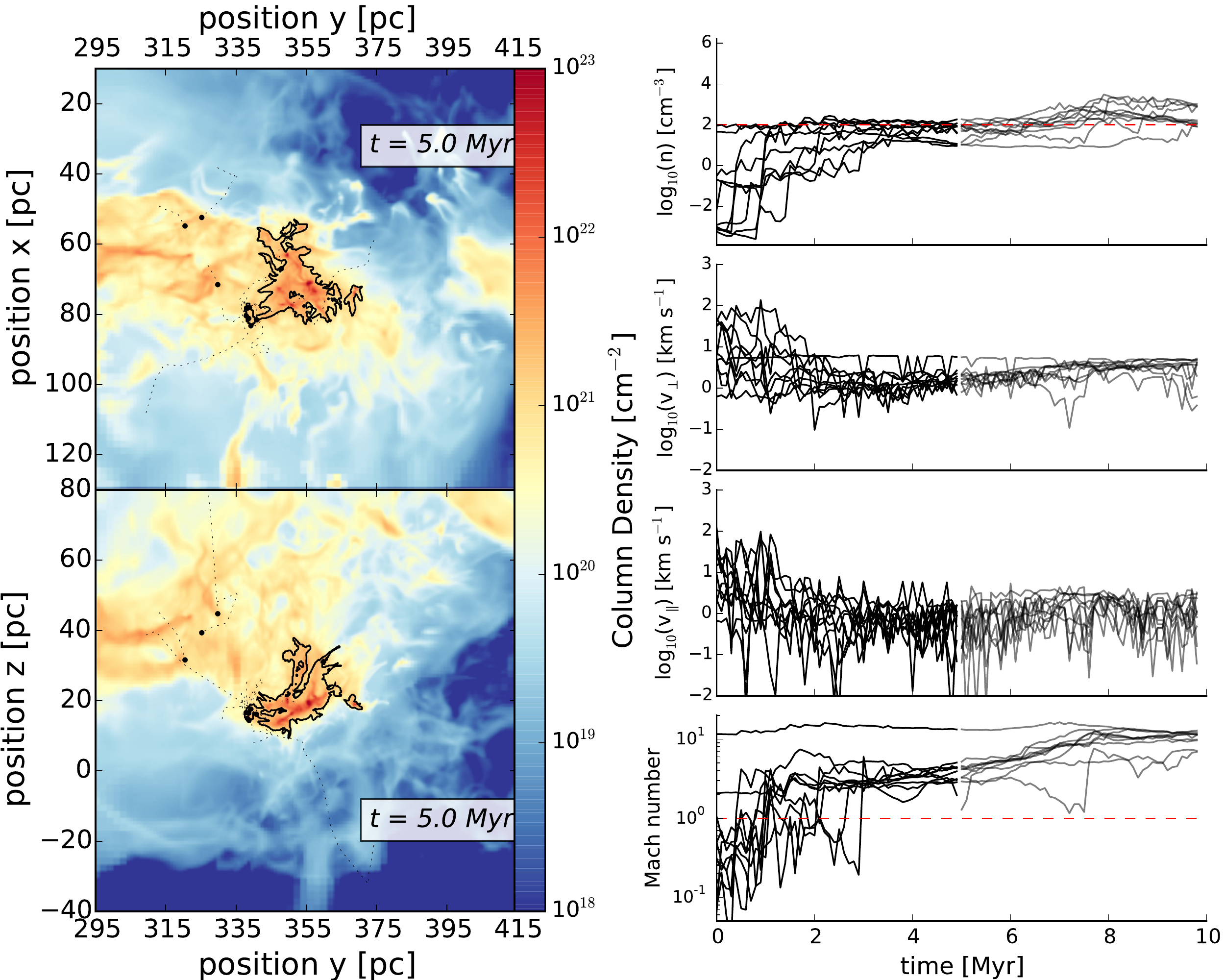} 
\caption{Evolution of randomly selected tracer particles eventually accreted by cloud
  M8 using same diagnostics and notation as
  Figure~\ref{fig:M3_GlobalAcc}. 
    \label{fig:M8_GlobalAcc} } 
\end{figure*}

\paragraph{Cloud M8} Out of the ten particles shown in Figure~\ref{fig:M8_GlobalAcc}, one is already part of the cold envelope, one is part of the warm environment, and the rest of the particles are in the hot environment close to the location of the SN explosion. 
Although most of these particles are moving at velocities of $\approx 10$--100~km~s$^{-1}$, their associated Mach numbers are low, ${\cal  M}<1$, because they are embedded in hot gas.

At the moment the nearby SN explodes at $t=0.62$~Myr, most of the particles in the hot phase are pushed towards the cloud, quickly climbing up the density gradient. 
As these particles move to denser environments, both their parallel and perpendicular velocities drop to lower values, while their Mach numbers climb, reaching values of order unity.  

Some of these particles continue on to quickly join the cold envelope and then become part of the cloud. 
However, as previously discussed, nearby SN explosions have the dual role of compressing the cloud but also eroding the envelope.   
For this reason, some of the particles in the cloud envelope move back to lower density environments. 
As seen in Figure \ref{fig:M8_evol}, following this sudden accretion phase, M8 goes through a low accretion phase that lasts $\approx 4 $~Myr, until two new nearby SN explode, on opposite sides of the cloud. 
These nearby events compress the envelope, increasing the mass accretion rate of the cloud, characterized by the accretion of material at supersonic velocities, reaching Mach numbers of ${\cal M}\approx10$.  

\subsubsection{Cloud Energetics}

\begin{figure}[t]
\centering 
\includegraphics[width=0.5\textwidth]{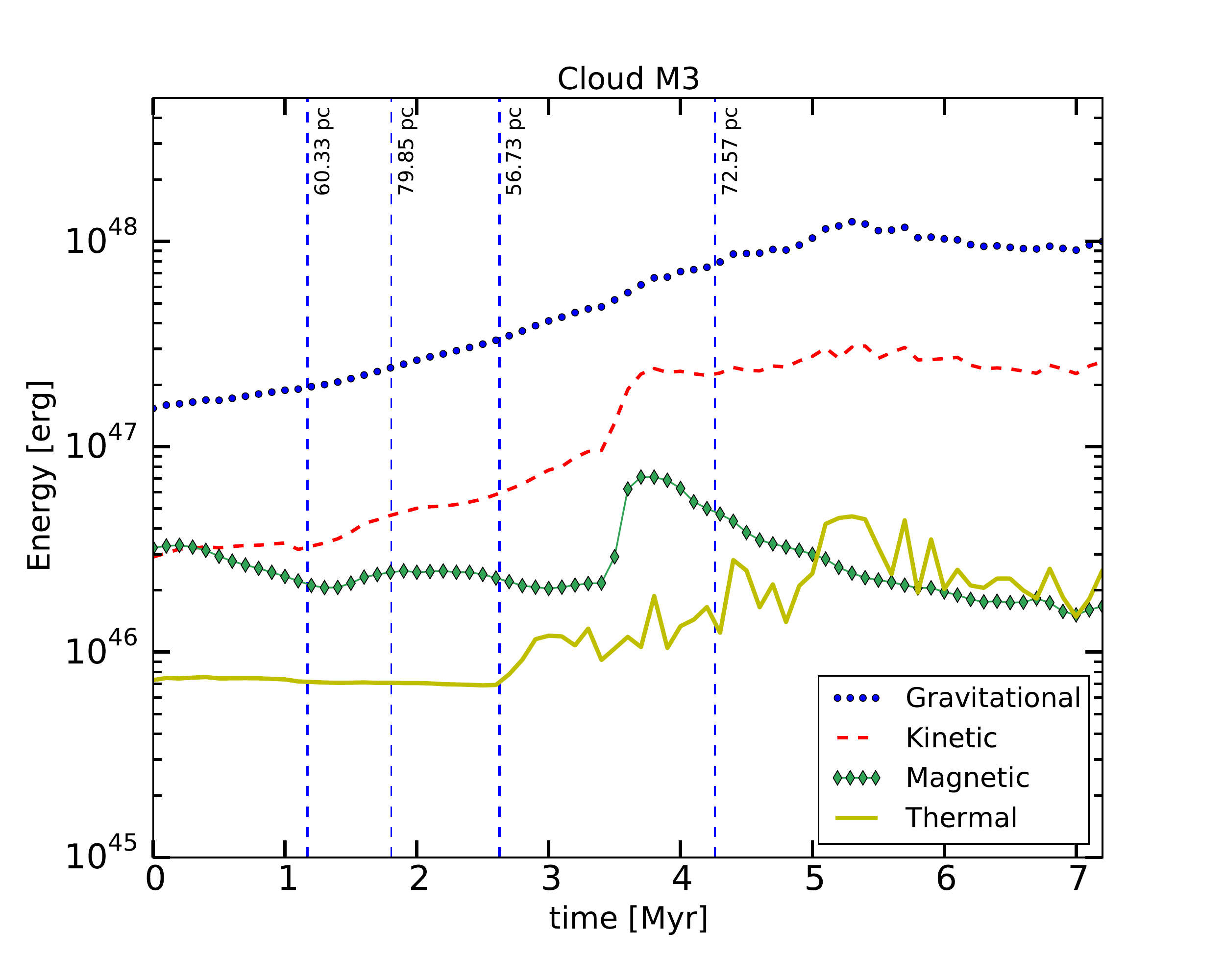} 
\vspace*{-6mm}
\caption{Evolution of the {\em{(dotted blue line)}} gravitational potential,
{\em{(dashed red line)}} kinetic, {\em{(solid green with diamonds line)}} magnetic, and {\em (solid olive line)} thermal energies within cloud M3. Explosion times of nearby SN events are shown {\em{(Vertical dashed blue lines)}}, with their corresponding distance to the cloud center of mass  (line thickness depends on SN distance). 
\label{fig:M3_Energy} } 
\end{figure}

\paragraph{Cloud M3} Figure~\ref{fig:M3_Energy} shows the gravitational, kinetic and magnetic contributions to the total energy of the cloud as a function of time, compared to the gravitational potential energy, which dominates over the entire evolution of the simulation.  
This indicates that the evolution of cloud M3 is dominated by self-gravity, with minor to moderate contributions from the other energy reservoirs. 

The initial kinetic energy is very weak compared to the gravitational potential energy, but it is affected by the gravitational collapse, the inflow of material onto the cloud and shocks from the turbulent environment.
A sudden increase in the cloud's gravitational potential, kinetic, and magnetic energies occurs shortly after the third nearby SN explosion.
This event triggered an enhanced burst of accretion, followed by a phase of erosion and fragmentation of the cloud's surface.
Finally the magnetic energy fluctuates following the general behavior of the other energy reservoirs, but systematically shifted by an order of magnitude below as is characteristic of small-scale dynamos \citet{Balsara2004AmplificationTurbulence}. 
Magnetic fields make little contribution to the fate of the cloud and appear unable to prevent it from collapsing. 

Comparing the evolution of the energies over time, we find that while the cloud is mostly dominated by the gravitational potential energy, sudden events such as the accretion spike at $t\approx3.6$~Myr, can trigger a chain reaction inside the cloud that can potentially have an effect on the global dynamics of the cloud. 
Another interesting feature we see here is that as the cloud contracts it converts some of its gravitational potential energy into kinetic energy.
For this reason Figure~\ref{fig:M3_Energy} shows that the gravitational and kinetic energies are closely related to one another, especially at later times in their evolution. 

\begin{figure}[!t]
\centering 
\includegraphics[width=0.5\textwidth]{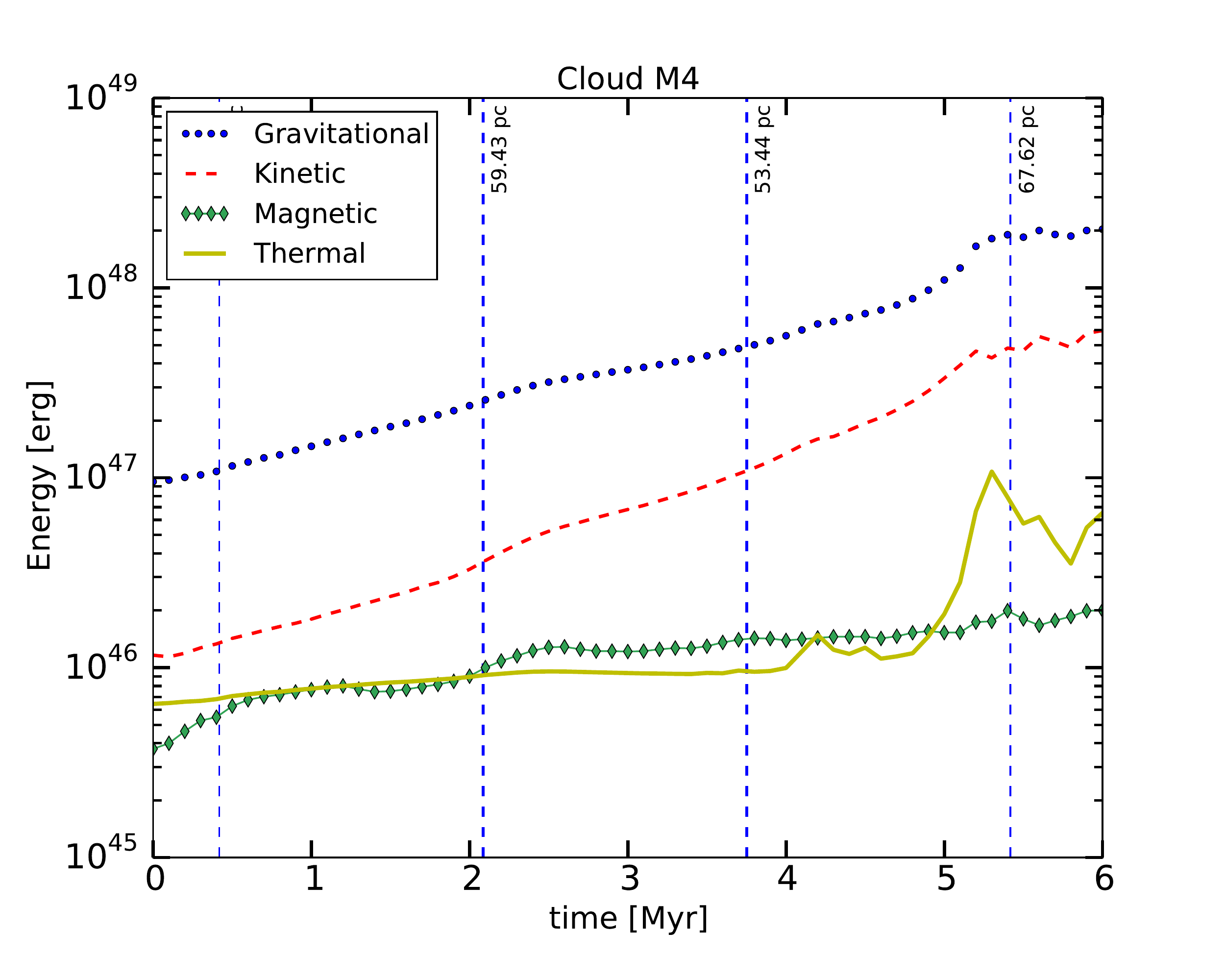} 
\vspace*{-6mm}
\caption{Evolution of the {\em{(dotted blue line)}} gravitational potential,
{\em{(dashed red line)}} kinetic, {\em{(thin green line with diamond markers)}} magnetic, and {\em (solid olive line)} thermal energies within cloud M4. Explosion times of nearby SN events are shown {\em{(Vertical dashed blue lines)}}, with their corresponding distance to the cloud center of mass (line thickness depends on SN distance).
\label{fig:M4_Energy} } 
\end{figure}

\paragraph{Cloud M4} Similarly to cloud M3, the gravitational potential energy dominates the energy budget throughout the resolved evolution of the cloud (Fig.~\ref{fig:M4_Energy}). 
Despite the number of nearby SN explosions, the total energy of the cloud shows no sign of being affected by them.

The kinetic energy is almost an order of magnitude smaller than the gravitational potential energy.
However, as the cloud collapses, it converts gravitational energy into kinetic energy, seen as a steady climb of the kinetic energy.
The magnetic energy is low compared to the other energy reservoirs of
the cloud, and again is unable to prevent collapse.

\begin{figure}[t]
\centering 
\includegraphics[width=0.5\textwidth]{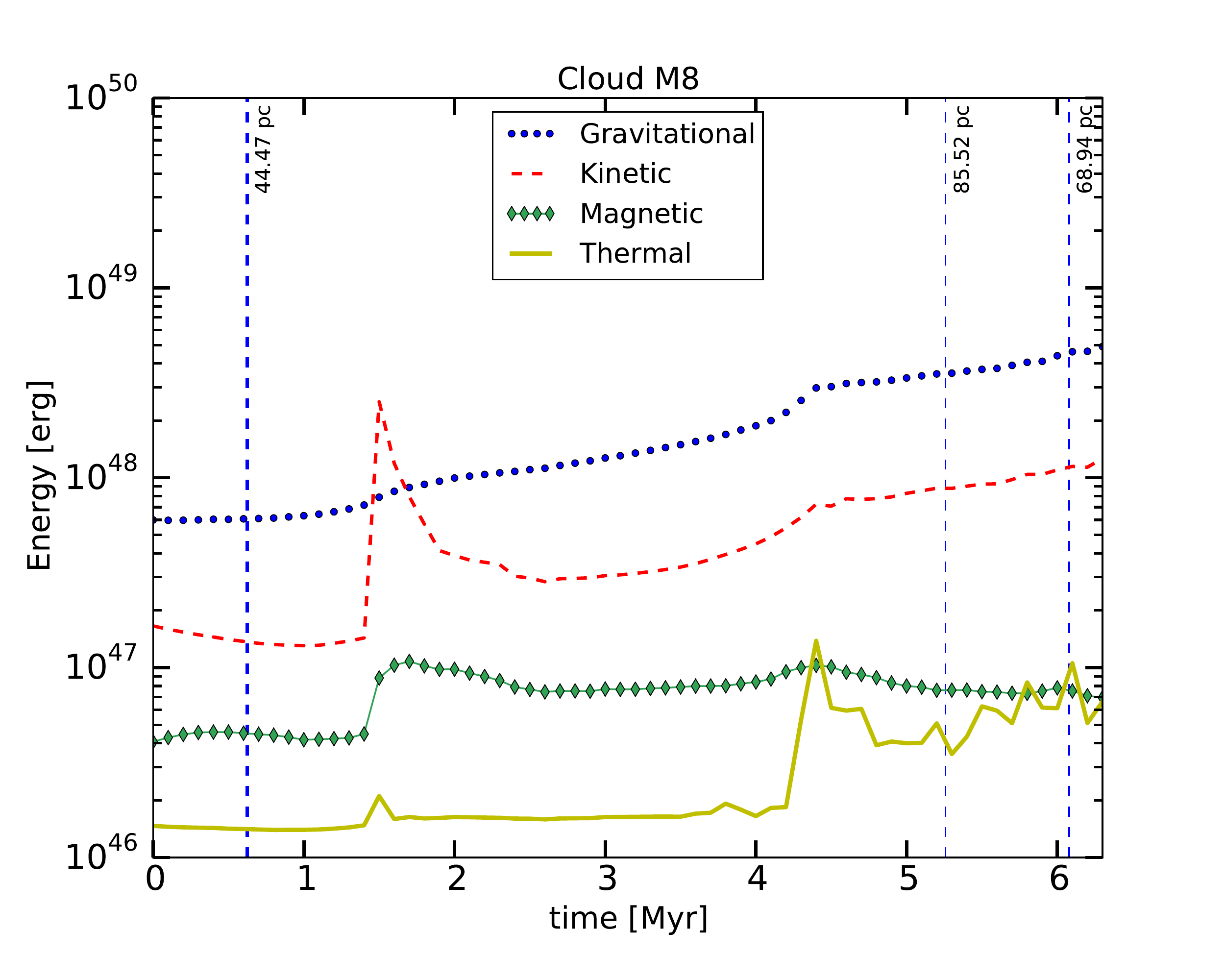} 
\caption{Evolution of the {\em{(dotted blue line)}} gravitational potential,
{\em{(dashed red line)}} kinetic, {\em{(solid green with diamonds line)}} magnetic, and {\em (solid olive line)} thermal energies within cloud M8. Explosion times of nearby SN events are shown {\em{(Vertical dashed blue lines)}}, with their corresponding distance to the cloud center of mass  (line thickness depends on SN distance).
\label{fig:M8_Energy} } 
\end{figure}

\paragraph{Cloud M8} The gravitational potential energy again dominates over most of the evolution of the cloud (Fig.~\ref{fig:M8_Energy}).
However, in comparison to the other clouds, there is a major injection of kinetic energy from the accreted gas after the first nearby SN explosion.
This sudden inflow of mass onto the cloud causes the internal kinetic energy to jump by an order of magnitude, exceeding the total binding energy of the entire cloud by a factor of two.
However, instead of dispersing the entire cloud, the excess kinetic energy breaks up fragments of the cloud and its envelope, and again decays in a crossing time. 
Such behavior also occurs in numerical simulations of isolated clouds with initial virial parameters of $\alpha_{vir} = 0.5-5$ 
\citep{Howard2016SimulatingBoundedness}. 
At later times, the cloud continues contracting gravitationally without significant accretion.
During this time the kinetic energy of the cloud increases steadily as the cloud converts gravitational energy into kinetic energy during contraction.
Two nearby SN explosions do cause enhanced accretion onto the
cloud. 
They reach the cloud at the end, or after, the internal dynamics become unresolved, so their effects on total energy could not be captured.
The magnetic energy remains below both the gravitational potential and
the kinetic energy of the cloud, unable to prevent the cloud from
collapsing.

\subsubsection{Energy Distribution}

\begin{figure*}[!t]
\centering 
\includegraphics[width=1.0\textwidth]{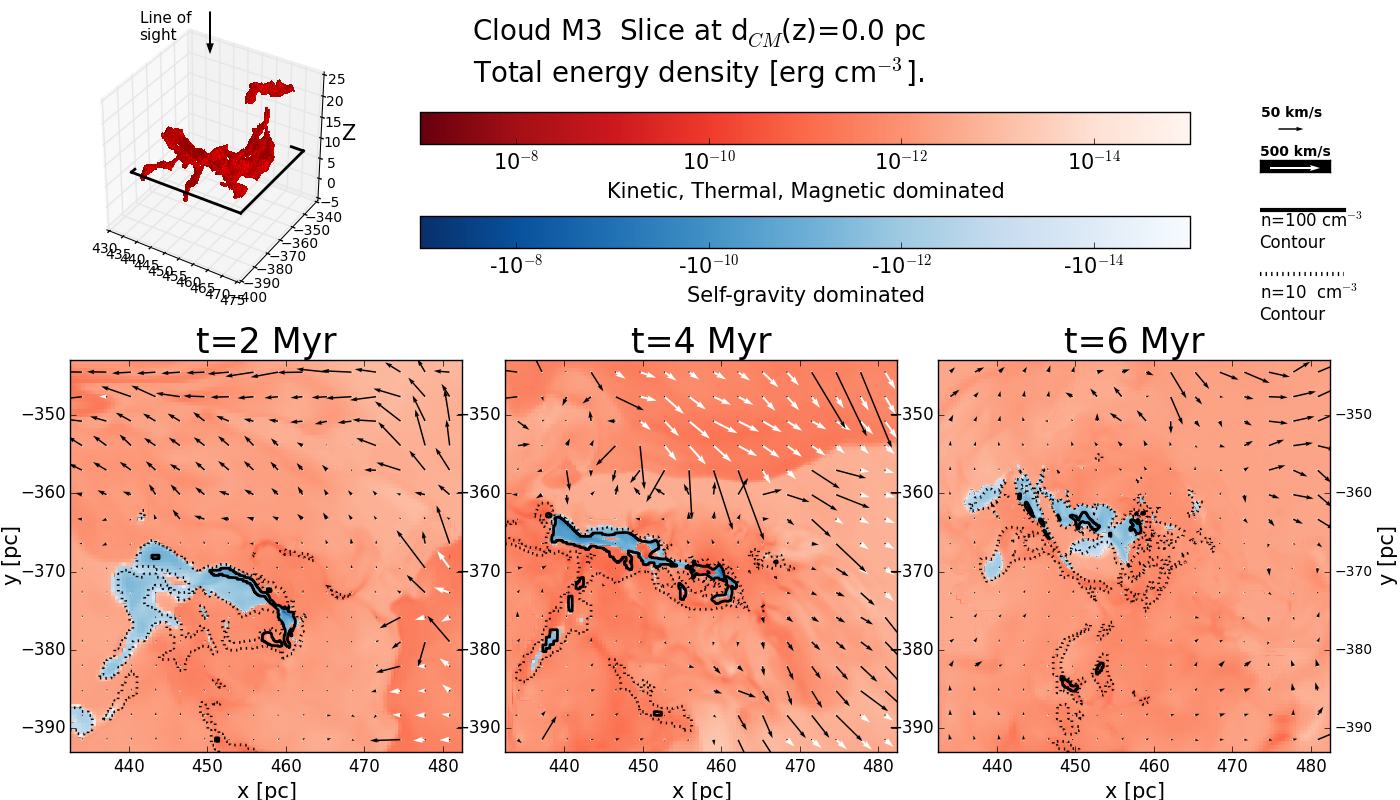}
\vspace*{-3mm}
\caption{Evolution of the total energy density of cloud M3, with blue showing bound regions and red showing unbound regions. 
The figure shows a slice through the center of mass of the cloud, while an online animation shows every slice. {\em{(Top left)}} Cloud surface at a time of  4~Myr (defined as connected structure and gravitationally bound fragments above a number density of $n=100$~cm$^{-3}$). 
On top, an arrow indicates the normal to the slice plots, while the height of the current cut is indicated by the square surrounding the cloud. {\em{(Bottom, left to right)}} Slices of the total energy density at times of 2, 4 and 6~Myr after self-gravity. 
Density contours are shown at {\em (solid line)} $100$~cm$^{-3}$ and {\em (dotted line)} $10$~cm$^{-3}$. 
In-plane velocity vectors are given with scales at upper right. Black arrows show  velocities below $|v|<100$~km~s$^{-1}$, while white arrows show velocities above. 
\label{fig:redvsblueM3}}
\end{figure*}
We are interested in understanding the evolution of not only the material inside the cloud, but also the material surrounding it. 
Thus we determine the total energy density of all gas in the region, and the contributions of different components.  
Figure \ref{fig:redvsblueM3} shows the sum of the components: gas self-gravity, turbulent kinetic energy, magnetic pressure, and thermal pressure, where all but self-gravity make a positive contribution to the total
energy density. 
Regions with positive total energy density are red, while negative regions are blue, again showing that material within the cloud ({\em{solid line contour}}), is mostly gravitationally dominated and unstable to contraction. 
Outside the cloud surface, the envelope is composed of gas at number densities between 10--100~cm$^{-3}$.  
Much of the envelope gas is bound to the cloud and gravitationally dominated, suggesting that this material is collapsing onto the cloud and contributing to its mass growth. 
Gas with densities below $10$~cm$^{-3}$ near the cloud is generally unbound from the cloud, with almost uniformly positive energy densities.  
Very fast flows, with relative velocities of $\approx 200$~km~s$^{-1}$ with respect to the cloud, occur, driven by nearby SN explosions. 

\begin{figure*}[!t]
\centering 
\includegraphics[width=1.0\textwidth]{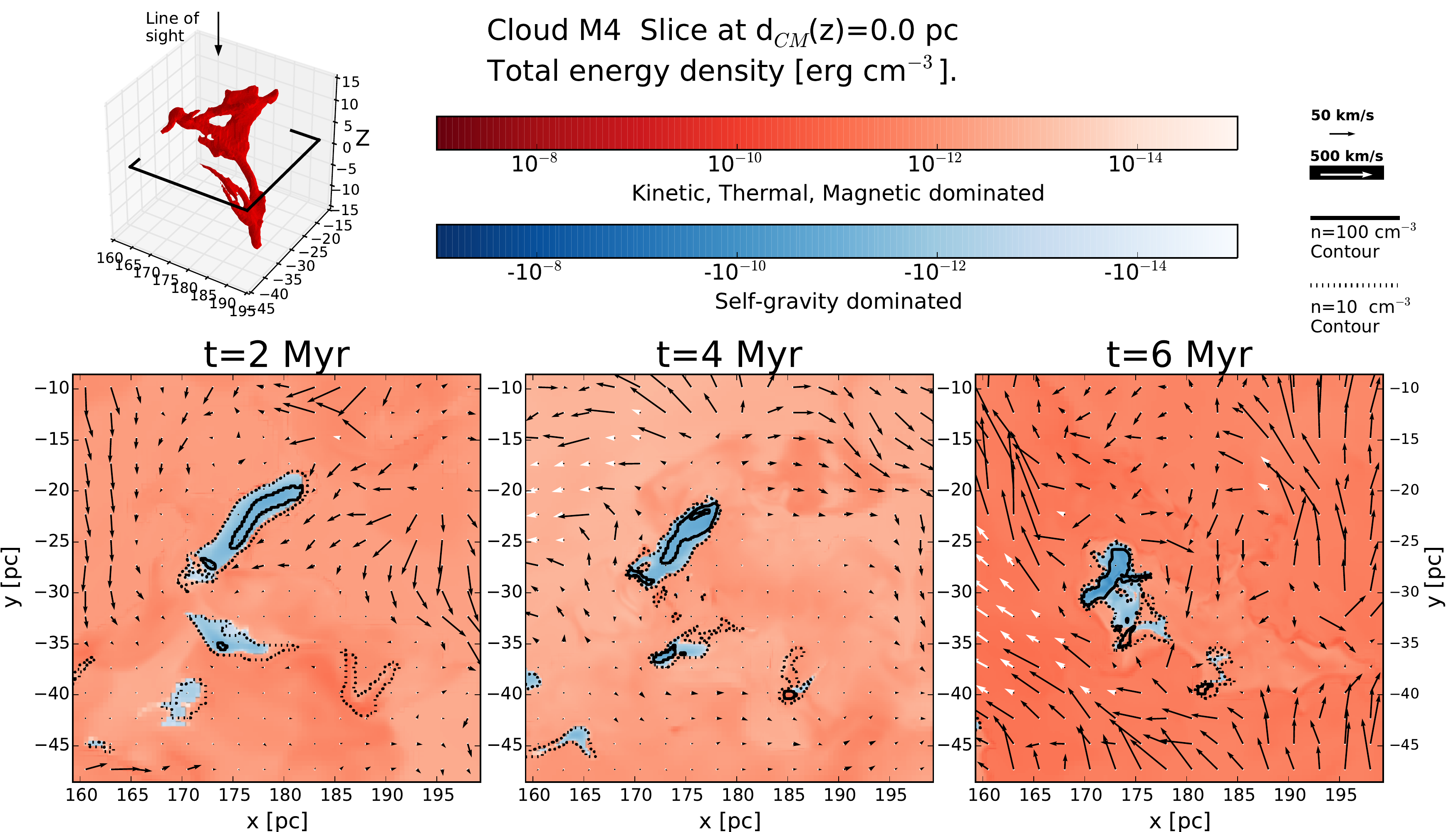}
\vspace*{-3mm}
\caption{Evolution of the total energy density of cloud M4, with blue showing bound regions and red showing unbound regions. 
    The figure shows a slice through the center of mass of the cloud, while an online animation shows every slice. 
    {\em{(Top left)}} Cloud surface ($n=100$~cm$^{-3}$) at a time of 4~Myr. 
    The arrow is normal to slices, while square shows current cut. 
  {\em{(Bottom, left to right)}} Total energy  density at times of 2, 4 and 6~Myr after self-gravity. 
  Density contours are shown at {\em (solid line)} $n = 100$~cm$^{-3}$ and {\em (dotted line)} $10$~cm$^{-3}$. 
  In-plane velocity vectors are given with scales at upper right. Black arrows show  velocities below $|v|<100$~km~s$^{-1}$, while white arrows show velocities above. 
\label{fig:redvsblueM4}}
\end{figure*}
Figure \ref{fig:redvsblueM4} shows the three-dimensional distribution of cloud M4, which clearly has a filamentary structure, elongated in the vertical direction.  
The material inside the cloud is strongly gravitationally dominated and is in runaway collapse.
Right outside the cloud surface, the envelope, with densities between $n=10-100$~cm$^{-3}$, is also mostly gravitationally bound to the cloud and infalling. 
Lower density material has uniformly positive total energy density, dominated by a combination of turbulent, thermal and magnetic energies, and is thus unbound. 

\begin{figure*}[!t]
\centering 
\includegraphics[width=1.0\textwidth]{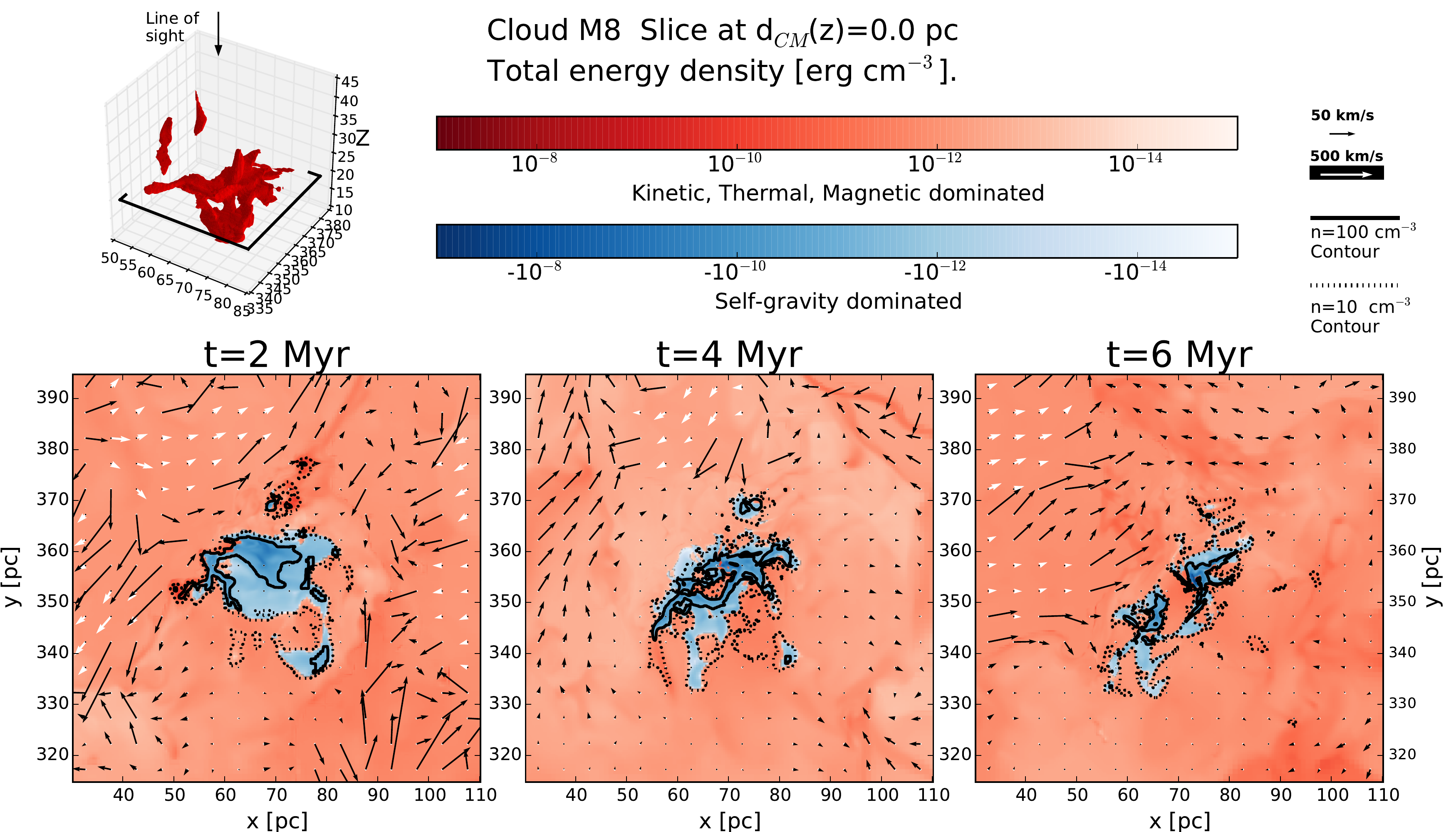}
\vspace*{-3mm}
\caption{Evolution of the total energy density of cloud M8, with blue showing bound regions and red showing unbound regions. 
  The figure shows a slice through the center of mass of the cloud, while an online animation shows every slice. {\em{(Top left)}} Cloud surface ($n=100$~cm$^{-3}$) at a time of 4~Myr. 
  Arrow shows normal to slices, while square shows current cut. {
  \em{(Bottom, left to right)}} Total energy density at times of 2, 4 and
  6~Myr after self-gravity. 
  Density contours are shown at {\em (solid line)} $n = 100$~cm$^{-3}$ and {\em (dotted line)} $10$~cm$^{-3}$. 
  In-plane velocity vectors are given with scales at upper right. Black arrows show  velocities below $|v|<100$~km~s$^{-1}$, while white arrows show velocities above. 
\label{fig:redvsblueM8}}
\end{figure*}
Figure \ref{fig:redvsblueM8} shows that cloud M8 has mostly been compressed into the $x-y$ plane, as it was falling towards the midplane and it got shocked by a SN explosion from below. 
This explosion deposited a significant amount of material onto the cloud, but also fragmented large portions of it. 
Some of these fragments where not accelerated strongly enough to become unbound and thus fall back. 

The material inside the cloud is again gravitationally dominated and it is in the process of collapse. 
As cloud M8 evolves and contracts, the dense envelope surrounding it is either accreted, or eroded by the fast-moving surroundings.
Low density gas, $n<10$~cm$^{-3}$, is unbound and reaches high velocities with respect to the cloud, of the order of $\approx 200$~km/s.

%
%

\section{Energy Budget}
\label{sec:discussion}
The simulations presented here suggest that mass accretion onto MCs subsequent to their formation is an essential part of their life cycle. 
These accretion flows not only increase the mass available in the clouds to form stars, but also carry a significant amount of kinetic energy that could affect the clouds' evolution. 

In Section~\ref{sec:mass_and_energy_influx} we analytically calculated that the mass accretion rate onto molecular clouds should correlate with the cloud mass, not just because of the increase in the gravitational pull resulting from the increase in mass, but also because of the increased surface area for the gas to flow through.
We then measured mass accretion rates for a simulated cloud population
evolving in a realistic Galactic environment and found a correlation of the accretion rates with both the cloud mass and surface area.
Comparing simulations with and without the effects of self-gravity showed that for a wide range of masses, the accretion rates driven by random turbulent flows and by gravitational attraction seem to be indistinguishable, but for the most massive clouds with $M>10^{5}$~M$_{\odot}$, the infall of gravitationally bound gas onto the cloud dominates the accretion rates. 
This is the range where both the calculated and simulated accretion rates come in to agreement with the global average estimates of mass accretion rates for GMCs in the LMC by \citet{Fukui2009MOLECULARI, Kawamura2009THEFORMATION}, and \citet{Fukui2010MolecularGalaxies}.

Simulating the evolutionary history of individual, low-mass clouds with initial masses in the range 3--8$\times 10^3$~M$_{\odot}$, we find that their average accretion rates agree with the estimated rates for a combination of turbulent and gravitationally driven accretion in Section \ref{sec:mass_and_energy_influx}.
However, mass accretion rates fluctuate on timescales of order the crossing time or less because of the disturbance of the gas reservoir by the turbulence in the environment.
As clouds accrete from a non-uniform, dynamic envelope, nearby SN explosions can simultaneously promote and prevent the inflow of mass onto the cloud by compressing and disrupting different parts of its envelope.
After a nearby SN explosion, a period of high accretion rate is observed, carrying a large influx of kinetic energy into the cloud, followed by a period of low accretion as the enveloping material has been either deposited onto the cloud or puffed up and eroded.
It appears that the gravitational collapse of the envelope, gas sweeping, and turbulent accretion have similar influence on the mass accretion rates over the entire lifetime of these low-mass clouds, where gas seems to flow in with velocities close to the free-fall velocities at the edge of the cloud.

In order to understand what controls the global evolution of these clouds, we focus on the time variation of the fractions of energy stored in the form of gravitational potential, kinetic, thermal, and magnetic energy (Fig.\ \ref{fig:M3_Energy}, \ref{fig:M4_Energy}, and
\ref{fig:M8_Energy}). 
We find that the gravitational potential energy is the dominant form of energy in the clouds at almost all times during their evolution, ultimately determining the fates of these clouds.
The kinetic energy for all the clouds starts at low values, presumably due to the lack of gas self-gravity during their initial turbulent assembly, but then climbs from the moment self-gravity is turned on, increasing the internal velocity dispersion of the clouds until it reaches values approaching virial balance (right panels of Fig.\
\ref{fig:M3_evol}, \ref{fig:M4_evol} and \ref{fig:M8_evol}), as expected not just for equilibrium systems, but also for gravitationally dominated clouds far from equilibrium \citep{Vazquez-Semadeni2007MolecularConditions, Vazquez-Semadeni2008TheCase,Ballesteros-Paredes2011GravityRelation,Naranjo-Romero2015HIERARCHICALCLOUDS}.

\begin{figure}[t]
\centering 
\includegraphics[width=0.5\textwidth]{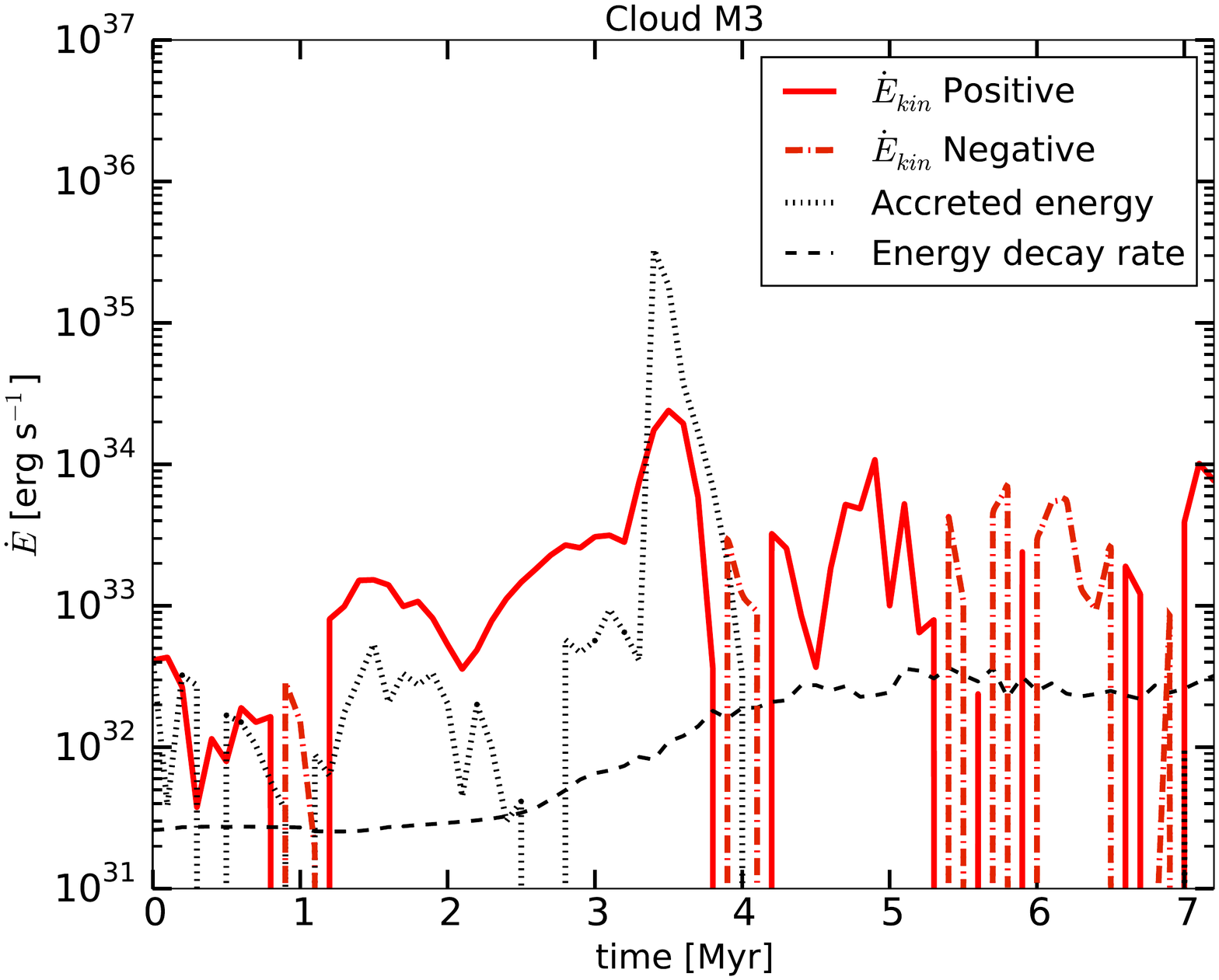} 
\caption{Rate of change for cloud M3 of {\em{(red line)}} internal
  kinetic energy ({\em solid} for positive, {\em dotted} for
  negative), compared to the energy influx rate from the {\em{(dotted
      black line)}} accretion flows, and the predicted {\em{(dashed
      black line)}} energy decay rate of the supersonic turbulence
  given by Equation~(\ref{eq:ekin_decay}).
\label{fig:M3_Edot}} 
\end{figure}

\begin{figure}[t]
\centering 
\includegraphics[width=0.5\textwidth]{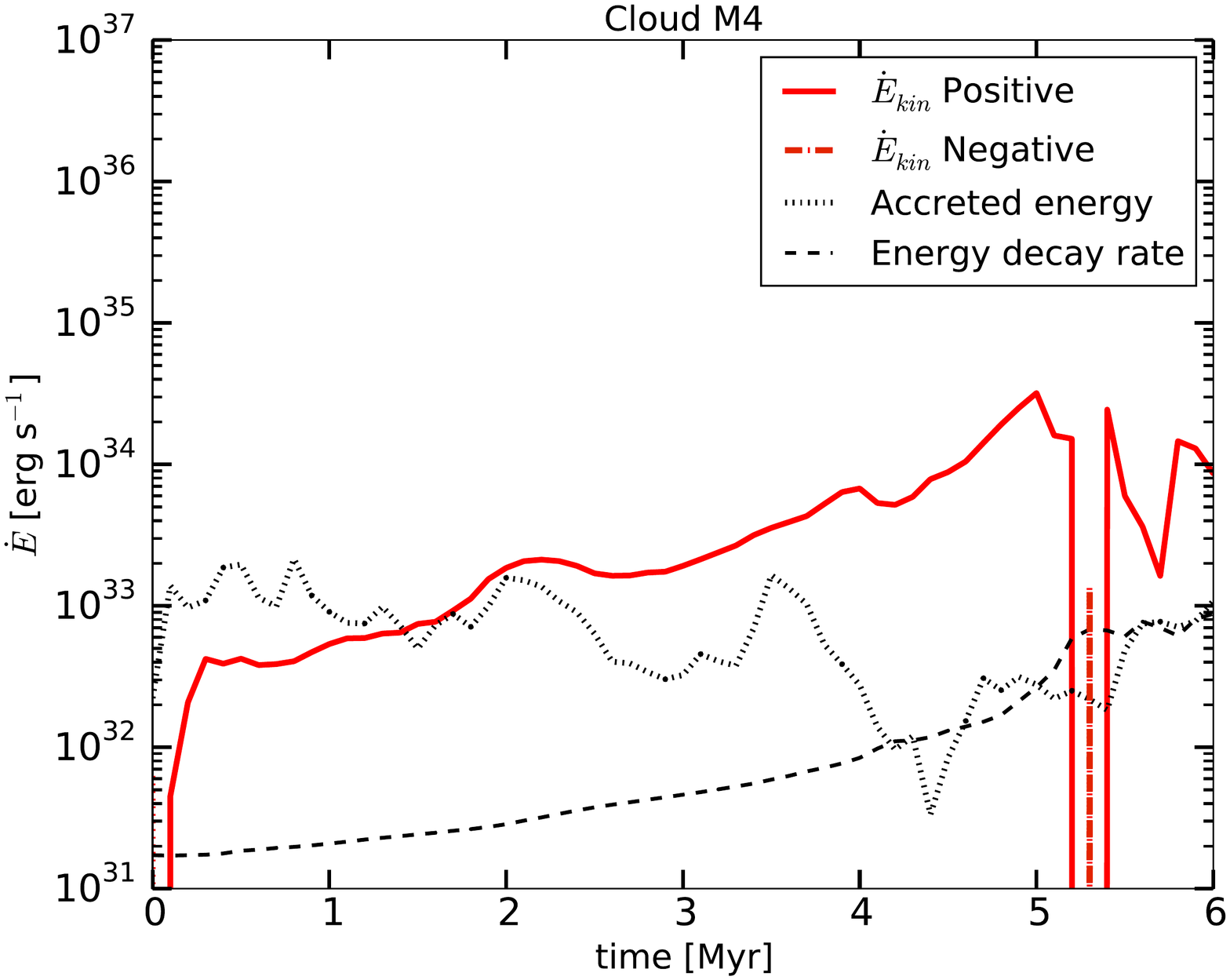} 
\caption{Rate of change for cloud M4 of the {\em{(red
      line)}} internal kinetic energy  ({\em solid} for positive, {\em
    dotted} for negative), compared to the energy influx rate from the
  {\em{(dotted black line)}} accretion flows, and the predicted
  {\em{(dashed black line)}} energy decay rate of the supersonic
  turbulence given by Equation~(\ref{eq:ekin_decay}). 
\label{fig:M4_Edot}} 
\end{figure}

\begin{figure}[t]
\centering 
\includegraphics[width=0.5\textwidth]{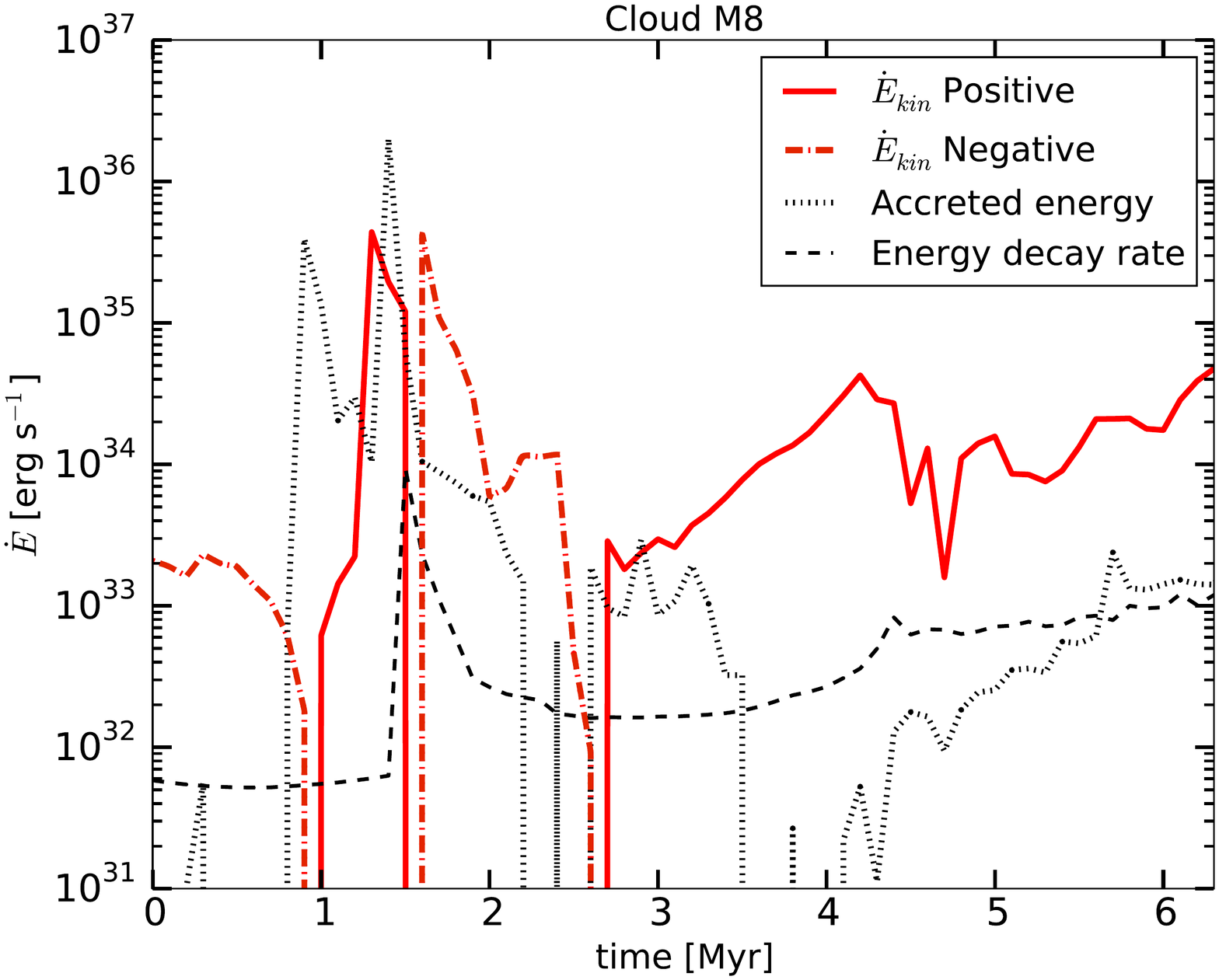} 
\caption{Rate of change for cloud M8 of the {\em (red
      line)} internal kinetic energy ({\em solid} for positive, {\em dotted} for negative), compared to the energy influx rate from the {\em{(dotted black line)}} accretion flows, and the predicted {\em{(dashed black line)}} energy decay rate of the supersonic turbulence given by Equation~(\ref{eq:ekin_decay}).
\label{fig:M8_Edot}} 
\end{figure}
Figures \ref{fig:M3_Edot}--\ref{fig:M8_Edot} show the change over time of kinetic energy in each cloud, compared to the inflow of kinetic energy from the accretion flow, and the predicted decay rate of energy given the cloud mass, size and velocity dispersion from Equation~(\ref{eq:ekin_decay}). 
For all three clouds, the energy deposited onto the cloud by accretion is generally higher than the predicted decay rate of the internal turbulence, making accretion-driven turbulence a reasonable candidate for maintaining the fast motions observed in MCs.
However, once we compare the inflow rate of kinetic energy from accretion to the rate of change of the total internal kinetic energy of the cloud, we find that the internal energy of the cloud increases much faster than the accretion can explain, suggesting that it cannot be the only source driving these motions.
Ultimately, the continued contraction and the final collapsed state of the three clouds suggests that gravitational contraction dominates the evolution of these clouds.

We do not include internal feedback from star formation in collapsing regions in our simulations, which has been argued to constitute one of the main energy sources for preventing runaway cloud collapse.
Internal feedback has been suggested to maintain the clouds in a
quasi-virial equilibrium state \citep{Goldbaum2011THEACCRETION,
  Zamora-Aviles2012ANACTIVITY}, and thus to maintain the observed low star
formation rate. 
In the present work, we stop our simulations once the majority of the mass in our cloud has collapsed to unresolved structures and should undergo vigorous star formation.

%
%

\section{Conclusions}
\label{sec:conclusions}
We present here a set of numerical simulations of the evolution of a cloud population in a Galactic environment. 
We performed detailed zoom-in re-simulations for three moderate-mass clouds from this population, in order to resolve their interaction with their turbulent environment as they accrete gas and collapse.  
We examined the physical processes that drive the observed mass accretion rates onto molecular clouds, and how the accretion scales with cloud mass.
We then investigated the dynamics of the accreting flows, measuring the incoming velocity and the strength of the shocks produced within the cloud. 
Finally we measured the net amount of kinetic energy delivered into the cloud from these accretion flows. 
We compare this with the internal kinetic energy of the cloud and the estimated decay rate of turbulence within molecular clouds, and the net change of kinetic energy from the cloud.
This allows us to understand what physical process drives the non-thermal velocity dispersion observed in molecular clouds.

We find that mass accretion occurs by the combined action of background turbulent flows and the gravitational attraction exerted by the cloud on its environment. 
Figure~\ref{fig:MassAcc_CloudPop} shows that low mass clouds, below a few times $10^{5}$~M$_{\odot}$, primarily accrete mass due to the background turbulence. 
However, it is necessary to account for the gravitational pull of the clouds on their envelopes in order to explain the mass accretion rates for clouds more massive than a few times $10^{5}$~M$_{\odot}$.  
The mass accretion rate appears to be correlated with the cloud mass over four orders of magnitude in mass. 

For the individual zoom-in clouds, we covered an initial cloud mass range of 3--8$ \times 10^{3}$~M$_{\odot}$. 
We found that the mass accretion rates of individual clouds fluctuate by several orders of magnitude over timescales shorter than a crossing time, as shown in the center panels of Figures~\ref{fig:M3_evol}, \ref{fig:M4_evol} and~\ref{fig:M8_evol}.  
These fluctuations are caused by the influence of the background turbulence on the cloud envelopes, particularly shocks from nearby SNe. 
As a consequence of this highly dynamic process, Figures
\ref{fig:M3_InstAcc}, \ref{fig:M4_InstAcc} and \ref{fig:M8_InstAcc}
show that mass accretion often occurs as a supersonic flow. 
We see no evidence for stable envelopes delivering mass at a constant rate in quasi-hydrostatic equilibrium, as is frequently assumed in analytic models. 

Analyzing the detailed evolution of accreted gas shown in Figures
\ref{fig:M3_GlobalAcc}, \ref{fig:M4_GlobalAcc} and
\ref{fig:M8_GlobalAcc}, we see that the accreted gas evolves
dynamically, originating in every phase of the ISM, and jumping between these phases as it is compressed or eroded by background shocks.  
The gas is almost always supersonic during its evolution and does not seem to reach any sort of equilibrium at any point of its trajectory towards becoming part of the cloud and its subsequent collapse. 

Nearby SN explosions drive blast waves that play an important, but two-sided, role in the dynamics of the cloud envelope and the resulting accretion rates
(Figs.~\ref{fig:M3_InstAcc}, \ref{fig:M4_InstAcc},
and~\ref{fig:M8_InstAcc}). 
On the one hand, they compress parts of cloud
envelopes, increasing the instantaneous accretion rates, but on the
other hand they also disrupt parts of the cloud and its envelope,
resulting in extended periods of low or even negative mass accretion
rates (center panels of Figs.~\ref{fig:M3_evol}, \ref{fig:M4_evol},
and~\ref{fig:M8_evol}).  

However, the blast waves seem unable to continuously drive fast
turbulent motions within the dense clouds that can prevent them from
continuing to collapse. As a clear example, the right panel of Figure~\ref{fig:M8_evol} shows the response of the cloud to a blast wave hitting it at $t \approx 2$~Myr: a brief increase in kinetic energy to unbound values, followed by a fast decay and continued gravitational collapse.

Finally, we compare the amount of kinetic energy delivered into the clouds from the accretion flows to the predicted decay rate of internal turbulence and the net rate of change of kinetic energy in the cloud (Figs.~\ref{fig:M3_Edot}, \ref{fig:M4_Edot} and~\ref{fig:M8_Edot}). 
We find that mass accretion does not carry enough kinetic energy into the cloud to power the internal turbulent motions and prevent the cloud from collapsing. 
Instead the conversion of gravitational potential energy into kinetic
energy through the hierarchical contraction of the cloud appears
primarily responsible for driving the observed fast motions within the clouds,
maintaining them in a state of near balance between potential and
kinetic energy, but far from virial equilibrium, as emphasized by
\citet{Ballesteros-Paredes2011GravityRelation}.  

\acknowledgments{Thanks to Javier Ballesteros-Paredes,  Cara
  Battersby, Paul Clark, Simon Glover, Eric Pellegrini, Rowan Smith, Enrique V\'azquez-Semadeni and the anonymous referee for useful discussions and comments
  on the manuscript.   
This work was supported by NSF grant AST11-09395. J. C. I. -M. also acknowledges support by the ISM-SPP, M.-M.M.L. acknowledges support by the Alexander-von-Humboldt Stiftung and R.S.K. also acknowledges support from the DFG via SFB 881 “The Milky Way System” (sub-projects B1, B2 and B8), and SPP 1573 “Physics of the ISM,” as well as funding from the ERC under the European Community’s Seventh Framework Programme via the ERC Advanced Grant “STARLIGHT” (project number 339177). 
Computations were performed on TACC Stampede under grant TG-MCA99S024 from the Extreme Science and Engineering Discovery Environment (XSEDE), which is supported by NSF grant OCI-1053575. 
C.B. acknowledges support from a Kade Fellowship for support of a visit to the AMNH. The FLASH code was in part developed by the DOE-supported ASC/Alliance Center for Astrophysical Thermonuclear Flashes at the University of Chicago.
Visualizations were performed with yt \citep{Turk2010AData}. }

\facilities{XSEDE} 
\software{FLASH \citep{Fryxell2000FLASHFlashes}, yt \citep{Turk2010AData}}

\bibliographystyle{apj}
\bibliography{AccretionPaper}

\begin{thebibliography}{}
\expandafter\ifx\csname natexlab\endcsname\relax\def\natexlab#1{#1}\fi

\bibitem[{{Audit} \& {Hennebelle}(2005)}]{Audit2005Thermalflow}
{Audit}, E., \& {Hennebelle}, P. 2005, \aap, 433, 1

\bibitem[{{Audit} \& {Hennebelle}(2010)}]{Audit2010Onflows}
---. 2010, \aap, 511, A76

\bibitem[{Bakes \& Tielens(1994)}]{Bakes1994TheHydrocarbons}
Bakes, E. L.~O., \& Tielens, A. G. G.~M. 1994, \apj, 427, 822

\bibitem[{Ballesteros-Paredes(2006)}]{Ballesteros-Paredes2006SixClouds}
Ballesteros-Paredes, J. 2006, \mnras, 372, 443

\bibitem[{Ballesteros-Paredes {et~al.}(2011)Ballesteros-Paredes, Hartmann,
  V{\'{a}}zquez-Semadeni, Heitsch, \&
  Zamora-Avil{\'{e}}s}]{Ballesteros-Paredes2011GravityRelation}
Ballesteros-Paredes, J., Hartmann, L.~W., V{\'{a}}zquez-Semadeni, E., Heitsch,
  F., \& Zamora-Avil{\'{e}}s, M.~A. 2011, \mnras, 411, 65

\bibitem[{Balsara {et~al.}(2004)Balsara, Kim, Mac~Low, \&
  Mathews}]{Balsara2004AmplificationTurbulence}
Balsara, D.~S., Kim, J., Mac~Low, M.-M., \& Mathews, G.~J. 2004, \apj, 617, 339

\bibitem[{Banerjee {et~al.}(2009)Banerjee, V{\'{a}}zquez-Semadeni, Hennebelle,
  \& Klessen}]{Banerjee2009ClumpFormation}
Banerjee, R., V{\'{a}}zquez-Semadeni, E., Hennebelle, P., \& Klessen, R.~S.
  2009, \mnras, 398, 1082

\bibitem[{{Beck}(2001)}]{Beck2001GalacticFields}
{Beck}, R. 2001, \ssr, 99, 243

\bibitem[{Bertoldi \& McKee(1992)}]{Bertoldi1992Pressure-confinedClouds}
Bertoldi, F., \& McKee, C.~F. 1992, \apj, 395, 140

\bibitem[{{Col{\'{\i}}n} {et~al.}(2013){Col{\'{\i}}n}, {V{\'a}zquez-Semadeni},
  \& {G{\'o}mez}}]{Colin2013MolecularFeedback}
{Col{\'{\i}}n}, P., {V{\'a}zquez-Semadeni}, E., \& {G{\'o}mez}, G.~C. 2013,
  \mnras, 435, 1701

\bibitem[{Crutcher {et~al.}(2010)Crutcher, Wandelt, Heiles, Falgarone, \&
  Troland}]{Crutcher2010MAGNETICANALYSIS}
Crutcher, R.~M., Wandelt, B., Heiles, C., Falgarone, E., \& Troland, T.~H.
  2010, \apj, 725, 466

\bibitem[{{Daley} {et~al.}(2012){Daley}, {Vanella}, {Dubey}, {Weide}, \&
  {Balaras}}]{Daley2012HybridGravity}
{Daley}, C., {Vanella}, M., {Dubey}, A., {Weide}, K., \& {Balaras}, E. 2012,
  Concurrency and Computation: Practice and Experience, 24, 2346

\bibitem[{Dalgarno \& McCray(1972)}]{Dalgarno1972HeatingRegions}
Dalgarno, A., \& McCray, R.~A. 1972, \araa, 10, 375

\bibitem[{Dehnen \& Binney(1998)}]{Dehnen1998MassWay}
Dehnen, W., \& Binney, J. 1998, \mnras, 294, 429

\bibitem[{Dobbs {et~al.}(2014)Dobbs, Krumholz, Ballesteros-Paredes, Bolatto,
  Heyer, Mac~Low, Ostriker, \&
  V{\'{a}}zquez-Semadeni}]{Dobbs2014FormationFormation}
Dobbs, C.~L., Krumholz, M.~R., Ballesteros-Paredes, J., {et~al.} 2014, in
  Protostars and Planets VI, ed. H.~Beuther, R.~S. Klessen, C.~P. Dullemond, \&
  T.~Henning (Tucson: University of Arizona Press), 3--26

\bibitem[{Draine(1978)}]{Draine1978PhotoelectricGas}
Draine, B.~T. 1978, \apjs, 36, 595

\bibitem[{Falgarone {et~al.}(2004)Falgarone, Hily-Blant, \&
  Levrier}]{Falgarone2004StructureClouds}
Falgarone, E., Hily-Blant, P., \& Levrier, F. 2004, \apss, 292, 89

\bibitem[{Federrath \& Klessen(2012)}]{Federrath2012THEOBSERVATIONS}
Federrath, C., \& Klessen, R.~S. 2012, \apj, 761, 156

\bibitem[{{Fryxell} {et~al.}(2000){Fryxell}, {Olson}, {Ricker}, {Timmes},
  {Zingale}, {Lamb}, {MacNeice}, {Rosner}, {Truran}, \&
  {Tufo}}]{Fryxell2000FLASHFlashes}
{Fryxell}, B., {Olson}, K., {Ricker}, P., {et~al.} 2000, \apjs, 131, 273

\bibitem[{Fukui \& Kawamura(2010)}]{Fukui2010MolecularGalaxies}
Fukui, Y., \& Kawamura, A. 2010, \araa, 48, 547

\bibitem[{Fukui {et~al.}(2009)Fukui, Kawamura, Wong, Murai, Iritani, Mizuno,
  Mizuno, Onishi, Hughes, Ott, Muller, Staveley-Smith, \&
  Kim}]{Fukui2009MOLECULARI}
Fukui, Y., Kawamura, A., Wong, T., {et~al.} 2009, \apj, 705, 144

\bibitem[{{Gatto} {et~al.}(2017){Gatto}, {Walch}, {Naab}, {Girichidis},
  {W{\"u}nsch}, {Glover}, {Klessen}, {Clark}, {Peters}, {Derigs}, {Baczynski},
  \& {Puls}}]{Gatto2017SILCCIII}
{Gatto}, A., {Walch}, S., {Naab}, T., {et~al.} 2017, \mnras, 466, 1903

\bibitem[{Girichidis {et~al.}(2016)Girichidis, Walch, Naab, Gatto,
  W{\"{u}}nsch, Glover, Klessen, Clark, Peters, Derigs, \&
  Baczynski}]{Girichidis2016TheOutflows}
Girichidis, P., Walch, S., Naab, T., {et~al.} 2016, \mnras, 456, 3432

\bibitem[{Goldbaum {et~al.}(2011)Goldbaum, Krumholz, Matzner, \&
  McKee}]{Goldbaum2011THEACCRETION}
Goldbaum, N.~J., Krumholz, M.~R., Matzner, C.~D., \& McKee, C.~F. 2011, \apj,
  738, 101

\bibitem[{Heiles \& Troland(2005)}]{Heiles2005TheTurbulence}
Heiles, C., \& Troland, T.~H. 2005, \apj, 624, 773

\bibitem[{{Heitsch} {et~al.}(2005){Heitsch}, {Burkert}, {Hartmann}, {Slyz}, \&
  {Devriendt}}]{Heitsch2005FormationStudy}
{Heitsch}, F., {Burkert}, A., {Hartmann}, L.~W., {Slyz}, A.~D., \& {Devriendt},
  J.~E.~G. 2005, \apjl, 633, L113

\bibitem[{Heitsch \& Hartmann(2008)}]{Heitsch2008RapidMovies}
Heitsch, F., \& Hartmann, L. 2008, \apj, 689, 290

\bibitem[{Heitsch {et~al.}(2001)Heitsch, Mac~Low, \&
  Klessen}]{Heitsch2001GravitationalTurbulence}
Heitsch, F., Mac~Low, M.-M., \& Klessen, R.~S. 2001, \apj, 547, 280

\bibitem[{{Heitsch} {et~al.}(2011){Heitsch}, {Naab}, \&
  {Walch}}]{Heitsch2011FlowSimulations}
{Heitsch}, F., {Naab}, T., \& {Walch}, S. 2011, \mnras, 415, 271

\bibitem[{Heitsch {et~al.}(2006)Heitsch, Slyz, Devriendt, Hartmann, \&
  Burkert}]{Heitsch2006TheFlows}
Heitsch, F., Slyz, A.~D., Devriendt, J. E.~G., Hartmann, L.~W., \& Burkert, A.
  2006, \apj, 648, 1052

\bibitem[{Heyer {et~al.}(2009)Heyer, Krawczyk, Duval, \&
  Jackson}]{Heyer2009RE-EXAMININGCLOUDS}
Heyer, M., Krawczyk, C., Duval, J., \& Jackson, J.~M. 2009, \apj, 699, 1092

\bibitem[{Hill {et~al.}(2012)Hill, Joung, Mac~Low, Benjamin, Matthew~Haffner,
  Klingenberg, \& Waagan}]{Hill2012VerticalMedium}
Hill, A.~S., Joung, M. K.~R., Mac~Low, M.-M., {et~al.} 2012, \apj, 750, 104

\bibitem[{Howard {et~al.}(2016)Howard, Pudritz, \&
  Harris}]{Howard2016SimulatingBoundedness}
Howard, C., Pudritz, R., \& Harris, W. 2016, \mnras, 461, 2953

\bibitem[{Huang \& Greengard(1999)}]{Huang00Fastmeshes}
Huang, J., \& Greengard, L. 1999, SIAM J. Sci.\ Comp., 21, 1551

\bibitem[{{Ib\'a\~nez-Mej\'{\i}a} {et~al.}(2017){Ib\'a\~nez-Mej\'{\i}a}, {Mac
  Low}, {Klessen}, \& {Baczynski}}]{AccPaperScripts}
{Ib\'a\~nez-Mej\'{\i}a}, J.~C., {Mac Low}, M.-M., {Klessen}, R.~S., \&
  {Baczynski}, C. 2017, doi:https://doi.org/10.5531/sd.astro.1

\bibitem[{Ib{\'{a}}{\~{n}}ez-Mej{\'{i}}a
  {et~al.}(2016)Ib{\'{a}}{\~{n}}ez-Mej{\'{i}}a, Mac~Low, Klessen, \&
  Baczynski}]{Ibanez-Mejia2016GravitationalClouds}
Ib{\'{a}}{\~{n}}ez-Mej{\'{i}}a, J.~C., Mac~Low, M.-M., Klessen, R.~S., \&
  Baczynski, C. 2016, ApJ, 824, 41

\bibitem[{Jeans(1902)}]{Jeans1902TheNebula}
Jeans, J.~H. 1902, Phil.\ Trans.\ Roy.\ Soc.\ A, 199, 1

\bibitem[{Joung \& Mac~Low(2006)}]{Joung2006}
Joung, M. K.~R., \& Mac~Low, M.-M. 2006, \apj, 653, 1266

\bibitem[{Joung {et~al.}(2009)Joung, Mac~Low, \& Bryan}]{Joung2009}
Joung, M. K.~R., Mac~Low, M.-M., \& Bryan, G.~L. 2009, \apj, 704, 137

\bibitem[{Kauffmann {et~al.}(2013)Kauffmann, Pillai, \&
  Goldsmith}]{Kauffmann2013LOWFIELDS}
Kauffmann, J., Pillai, T., \& Goldsmith, P.~F. 2013, \apj, 779, 185

\bibitem[{Kawamura {et~al.}(2009{\natexlab{a}})Kawamura, Minamidani, Mizuno,
  Mizuno, Onishi, Mizuno, \& Fukui}]{Kawamura2009GlobularGalaxies}
Kawamura, A., Minamidani, T., Mizuno, Y., {et~al.} 2009{\natexlab{a}}, in
  Globular Clusters - Guides to Galaxies, ed. T.~Richtler \& S.~Larsen
  (Heidelberg: Springer), 121--122

\bibitem[{Kawamura {et~al.}(2009{\natexlab{b}})Kawamura, Mizuno, Minamidani,
  D.~Fillipovi{\'{c}}, Staveley-Smith, Kim, Mizuno, Onishi, Mizuno, \&
  Fukui}]{Kawamura2009THEFORMATION}
Kawamura, A., Mizuno, Y., Minamidani, T., {et~al.} 2009{\natexlab{b}}, \apjs,
  184, 1

\bibitem[{{Klessen} \& {Glover}(2016)}]{Klessen2014PhysicalMedium}
{Klessen}, R.~S., \& {Glover}, S.~C.~O. 2016, in Star Formation in Galaxy
  Evolution: Connecting Numerical Models to Reality, ed. Y.~Revaz, P.~Jablonka,
  R.~Teyssier, \& L.~Mayer, Vol.~43 (Heidelberg: Springer), 85

\bibitem[{Klessen {et~al.}(2000)Klessen, Heitsch, \&
  Mac~Low}]{Klessen2000GravitationalTurbulence}
Klessen, R.~S., Heitsch, F., \& Mac~Low, M.-M. 2000, \apj, 535, 887

\bibitem[{Klessen \& Hennebelle(2010)}]{Klessen2010Accretion-drivenDisks}
Klessen, R.~S., \& Hennebelle, P. 2010, \aap, 520, A17

\bibitem[{Krumholz \& McKee(2005)}]{Krumholz2005AGalaxies}
Krumholz, M.~R., \& McKee, C.~F. 2005, \apj, 630, 250

\bibitem[{Kuijken \& Gilmore(1989)}]{Kuijken1989TheSun}
Kuijken, K., \& Gilmore, G. 1989, \mnras, 239, 605

\bibitem[{Lada(2005)}]{Lada2005StarOverview}
Lada, C.~J. 2005, Progress of Theoretical Physics Supplement, 158, 1

\bibitem[{{Larson}(1981)}]{Larson1981}
{Larson}, R.~B. 1981, \mnras, 194, 809

\bibitem[{Larson(2003)}]{Larson2003TheFormation}
Larson, R.~B. 2003, Rep.\ Prog.\ Phys., 66, 1651

\bibitem[{Lee {et~al.}(2015)Lee, Chang, \& Murray}]{Lee2015TIME-VARYINGRATE}
Lee, E.~J., Chang, P., \& Murray, N. 2015, \apj, 800, 49

\bibitem[{Mac~Low(1999)}]{MacLow1999TheClouds}
Mac~Low, M.-M. 1999, \apj, 524, 169

\bibitem[{Mac~Low \& Klessen(2004)}]{MacLow2004ControlTurbulence}
Mac~Low, M.-M., \& Klessen, R.~S. 2004, \rmp, 76, 125

\bibitem[{McKee \& Ostriker(2007)}]{McKee2007TheoryFormation}
McKee, C.~F., \& Ostriker, E.~C. 2007, \araa, 45, 565

\bibitem[{Naranjo-Romero {et~al.}(2015)Naranjo-Romero, V{\'{a}}zquez-Semadeni,
  \& Loughnane}]{Naranjo-Romero2015HIERARCHICALCLOUDS}
Naranjo-Romero, R., V{\'{a}}zquez-Semadeni, E., \& Loughnane, R.~M. 2015, \apj,
  814, 48

\bibitem[{Navarro {et~al.}(1996)Navarro, Frenk, \& White}]{Navarro1996TheHalos}
Navarro, J.~F., Frenk, C.~S., \& White, S. D.~M. 1996, \apj, 462, 563

\bibitem[{Padoan {et~al.}(2012)Padoan, Haugb{\o}lle, \&
  Nordlund}]{Padoan2012AFormation}
Padoan, P., Haugb{\o}lle, T., \& Nordlund, Ã. 2012, \apj, 759, L27

\bibitem[{{Peters} {et~al.}(2017){Peters}, {Naab}, {Walch}, {Glover},
  {Girichidis}, {Pellegrini}, {Klessen}, {W{\"u}nsch}, {Gatto}, \&
  {Baczynski}}]{Peters2017SILCCIV}
{Peters}, T., {Naab}, T., {Walch}, S., {et~al.} 2017, \mnras, 466, 3293

\bibitem[{Ricker(2008)}]{Ricker2008AMeshes}
Ricker, P.~M. 2008, \apjs, 176, 293

\bibitem[{{Seifried} {et~al.}(2017){Seifried}, {Walch}, {Girichidis}, {Naab},
  {W{\"u}nsch}, {Klessen}, {Glover}, {Peters}, \&
  {Clark}}]{Seifried2017SILCC-ZOOM}
{Seifried}, D., {Walch}, S., {Girichidis}, P., {et~al.} 2017, ArXiv e-prints,
  arXiv:1704.06487

\bibitem[{Solomon {et~al.}(1987)Solomon, Rivolo, Barrett, \&
  Yahil}]{Solomon1987MassClouds}
Solomon, P.~M., Rivolo, A.~R., Barrett, J., \& Yahil, A. 1987, \apj, 319, 730

\bibitem[{Sutherland \& Dopita(1993)}]{Sutherland1993CoolingPlasmas}
Sutherland, R.~S., \& Dopita, M.~A. 1993, \apjs, 88, 253

\bibitem[{Tammann {et~al.}(1994)Tammann, Loeffler, \&
  Schroeder}]{Tammann1994TheRate}
Tammann, G.~A., Loeffler, W., \& Schroeder, A. 1994, \apjs, 92, 487

\bibitem[{Truelove {et~al.}(1997)Truelove, Klein, McKee, Holliman~II, Howell,
  \& Greenough}]{Truelove1997TheHydrodynamics}
Truelove, J.~K., Klein, R.~I., McKee, C.~F., {et~al.} 1997, \apj, 489, L179

\bibitem[{Truelove {et~al.}(1998)Truelove, Klein, McKee, Holliman~II, Howell,
  Greenough, \& Woods}]{Truelove1998SelfgravitationalFragmentation}
---. 1998, \apj, 495, 821

\bibitem[{Turk {et~al.}(2010)Turk, Smith, Oishi, Skory, Skillman, Abel, \&
  Norman}]{Turk2010AData}
Turk, M.~J., Smith, B.~D., Oishi, J.~S., {et~al.} 2010, \apjs, 192, 9

\bibitem[{V{\'{a}}zquez-Semadeni {et~al.}(2010)V{\'{a}}zquez-Semadeni,
  Col{\'{i}}n, G{\'{o}}mez, Ballesteros-Paredes, \&
  Watson}]{Vazquez-Semadeni2010MOLECULARFEEDBACK}
V{\'{a}}zquez-Semadeni, E., Col{\'{i}}n, P., G{\'{o}}mez, G.~C.,
  Ballesteros-Paredes, J., \& Watson, A.~W. 2010, \apj, 715, 1302

\bibitem[{V{\'{a}}zquez-Semadeni {et~al.}(2007)V{\'{a}}zquez-Semadeni, Gomez,
  Jappsen, Ballesteros-Paredes, Gonzalez, \&
  Klessen}]{Vazquez-Semadeni2007MolecularConditions}
V{\'{a}}zquez-Semadeni, E., Gomez, G.~C., Jappsen, A.~K., {et~al.} 2007, \apj,
  657, 870

\bibitem[{V{\'{a}}zquez-Semadeni {et~al.}(2009)V{\'{a}}zquez-Semadeni,
  G{\'{o}}mez, Jappsen, Ballesteros-Paredes, \&
  Klessen}]{Vazquez-Semadeni2009HIGH-EFFICIENCIES}
V{\'{a}}zquez-Semadeni, E., G{\'{o}}mez, G.~C., Jappsen, A.-K.,
  Ballesteros-Paredes, J., \& Klessen, R.~S. 2009, \apj, 707, 1023

\bibitem[{V{\'{a}}zquez-Semadeni {et~al.}(2008)V{\'{a}}zquez-Semadeni,
  Gonz{\'{a}}lez, Ballesteros-Paredes, Gazol, \&
  Kim}]{Vazquez-Semadeni2008TheCase}
V{\'{a}}zquez-Semadeni, E., Gonz{\'{a}}lez, R.~F., Ballesteros-Paredes, J.,
  Gazol, A., \& Kim, J. 2008, \mnras, 390, 769

\bibitem[{Waagan {et~al.}(2011)Waagan, Federrath, \&
  Klingenberg}]{Waagan2011ATests}
Waagan, K., Federrath, C., \& Klingenberg, C. 2011, J. Comput.\ Phys., 230,
  3331

\bibitem[{Walch {et~al.}(2015)Walch, Girichidis, Naab, Gatto, Glover,
  W{\"{u}}nsch, Klessen, Clark, Peters, Derigs, \& Baczynski}]{Walch2015TheISM}
Walch, S., Girichidis, P., Naab, T., {et~al.} 2015, \mnras, 454, 246

\bibitem[{Williams {et~al.}(1999)Williams, Blitz, \&
  McKee}]{Williams1999TheIMF}
Williams, J.~P., Blitz, L., \& McKee, C.~F. 1999, in Protostars and Planets IV,
  ed. V.~Mannings, A.~P. Boss, \& S.~S. Russell (Tucson: U. of Arizona Press),
  97

\bibitem[{Wolfire {et~al.}(1995)Wolfire, Hollenbach, McKee, Tielens, \&
  Bakes}]{Wolfire1995TheMedium}
Wolfire, M.~G., Hollenbach, D., McKee, C.~F., Tielens, A. G. G.~M., \& Bakes,
  E. L.~O. 1995, \apj, 443, 152

\bibitem[{Zamora-Avil{\'{e}}s {et~al.}(2012)Zamora-Avil{\'{e}}s,
  V{\'{a}}zquez-Semadeni, \& Col{\'{i}}n}]{Zamora-Aviles2012ANACTIVITY}
Zamora-Avil{\'{e}}s, M., V{\'{a}}zquez-Semadeni, E., \& Col{\'{i}}n, P. 2012,
  \apj, 751, 77

\bibitem[{Zuckerman \& Palmer(1974)}]{Zuckerman1974RadioMolecules}
Zuckerman, B., \& Palmer, P. 1974, \araa, 12, 279

\end{thebibliography}

\appendix
\section{Resolution Study}

\begin{figure*}[!h]
\centering 
\includegraphics[width=1\textwidth]{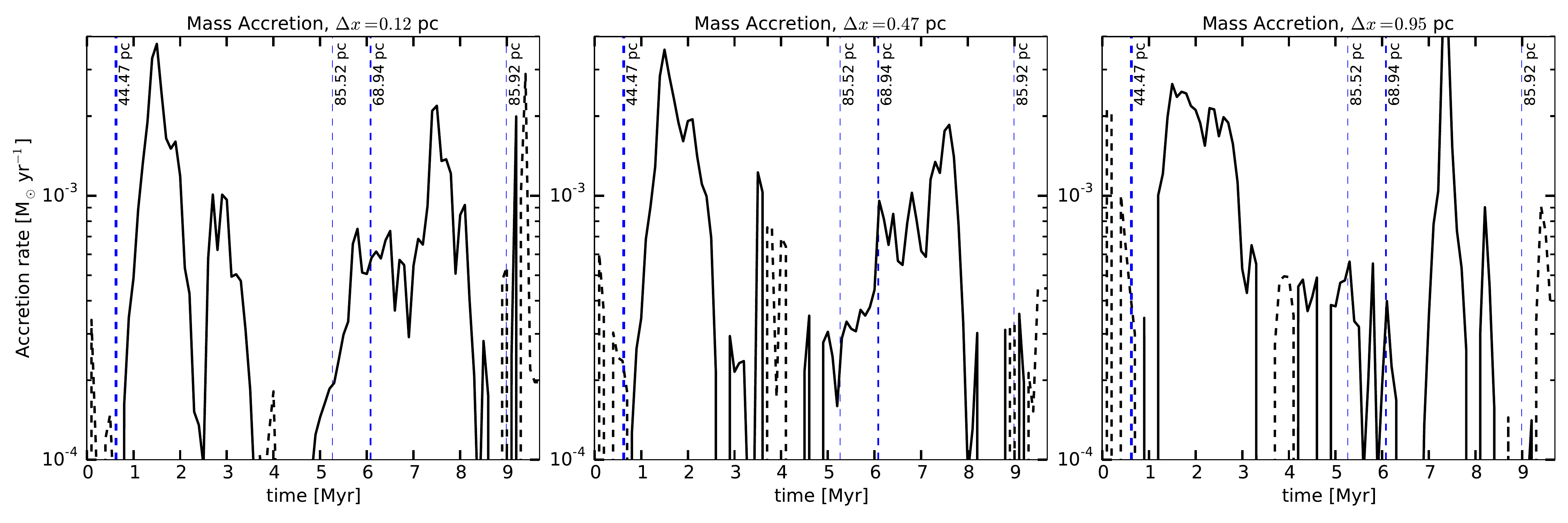} 
\caption{Mass accretion rate of cloud M8 as a function of time for the ({\em{left}}) high resolution, $\Delta x = 0.12$~pc, ({\em {center}}) intermediate resolution, $\Delta x = 0.47$~pc, and ({\em {right}}) low resolution, $\Delta x = 0.95$~pc, simulations. {\em Solid lines} denote mass gain while {\em dashed lines} denote mass loss.
Vertical {\em dashed blue lines} denote nearby SN events, with the thickness of the line showing the distance to the explosion.
\label{fig:M8_Macc_resStudy}}
\end{figure*}      

\begin{figure*}[t]
\centering 
\includegraphics[width=1\textwidth]{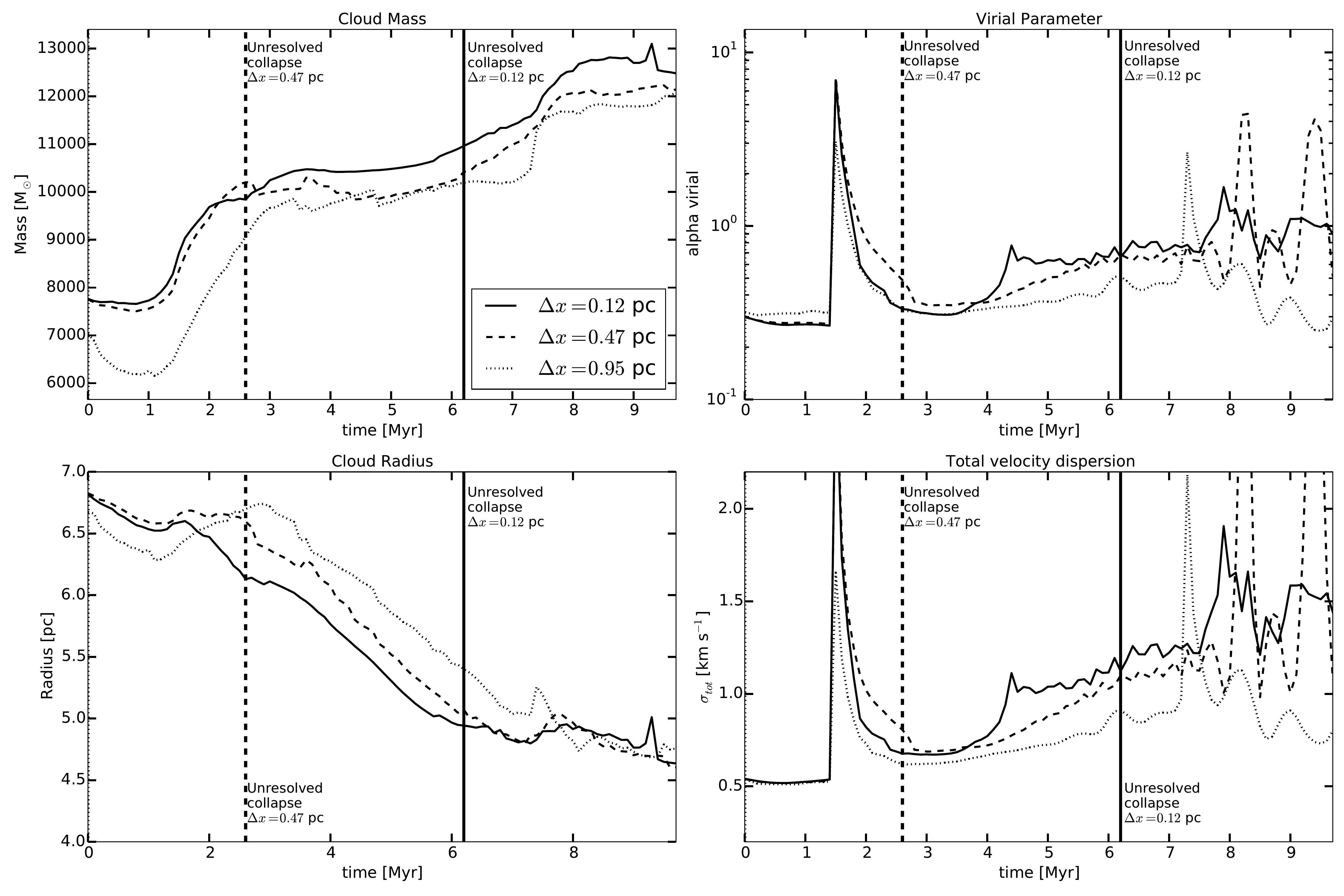} 
\caption{Evolution of the properties of cloud M8 at three different numerical resolutions starting with the inclusion of self-gravity and tracer particles. 
Resolutions shown are {\em{(solid line)}} $\Delta x=0.12$ pc,  {\em{(dashed line)}} $\Delta x=0.47$ pc, and {\em{(dotted line)}} $\Delta x=0.95$~pc.
Histories are shown of the {\em{(top left)}} cloud mass, 
{\em{(top right)}} virial parameter, 
{\em{(bottom left)}} cloud radius, and
{\em{(bottom right)}} total velocity dispersion. 
Vertical lines show the point at which collapse becomes unresolved for the different resolutions. 
\label{fig:M8_resStudy}} 
\end{figure*}     

\begin{figure*}[t]
\centering 
\includegraphics[width=1\textwidth]{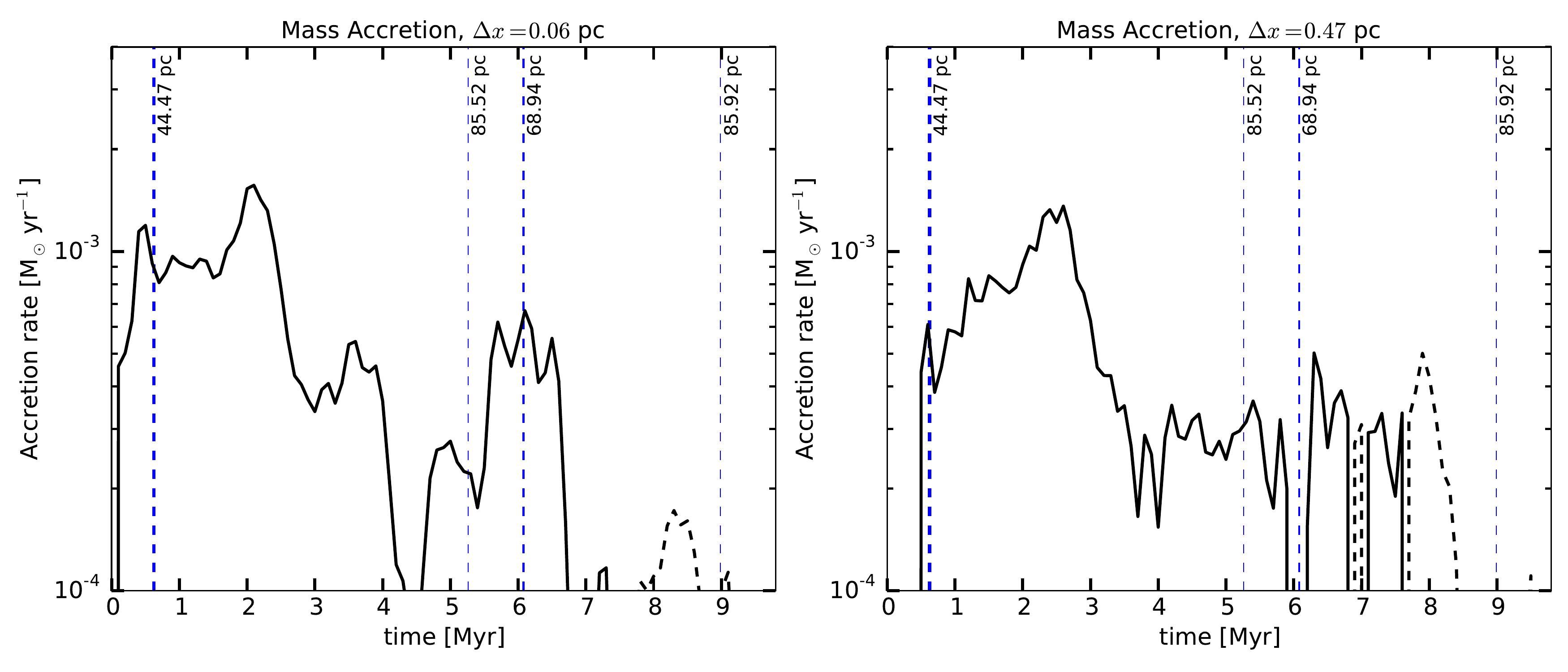} 
\caption{Mass accretion rates of cloud M4 as a function of time for the ({\em{left}}) high resolution, $\Delta x = 0.06$~pc, simulation and the ({\em {right}}) low resolution, $\Delta x = 0.47$~pc, simulation. {\em Solid lines} denote mass gain while {\em dashed lines} denote mass loss.
Vertical {\em dashed blue lines} denote nearby SN events, with the thickness of the line correlated with the distance to the explosion.
\label{fig:Macc_M4_resStudy}}
\end{figure*}

\begin{figure*}[t]
\centering 
\includegraphics[width=1\textwidth]{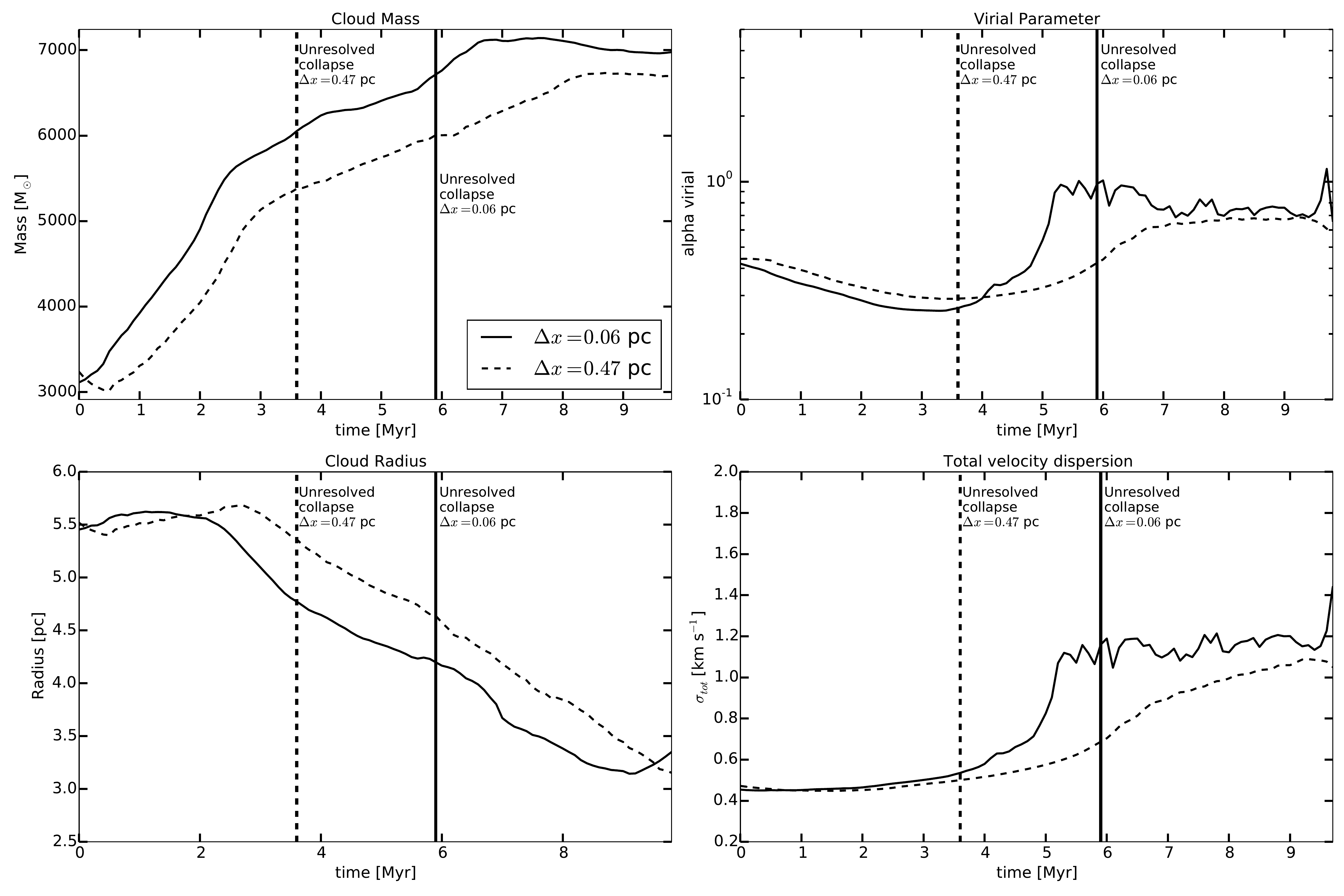} 
\caption{\label{fig:M4_resStudy} Evolution of the properties of cloud M4 at two different numerical resolutions starting with the inclusion of self-gravity and tracer particles. 
Resolutions shown are {\em{(solid line)}} $\Delta x=0.06$ pc and  {\em{(dashed line)}} $\Delta x=0.47$ pc.
Histories are shown of the {\em{(top left)}} cloud mass, 
{\em{(top right)}} virial parameter, 
{\em{(bottom left)}} cloud radius, and
{\em{(bottom right)}} total velocity dispersion. 
Vertical lines show the point at which collapse becomes unresolved for the different resolutions. 
}
\end{figure*}      

In this paper we study mass accretion onto molecular clouds and determine its influence on cloud dynamics. 
For this purpose, it is important to understand how accurately our simulations resolve the mass accretion rates onto the clouds as well as their internal dynamics. 
In this Appendix we examine the mass accretion rates and history of the properties of clouds M4 and M8 for simulations with numerical resolutions $\Delta~x=0.12, 0.47$, and $0.95$~pc for cloud M8, and $\Delta~x=0.06$ and $0.47$~pc for cloud M4. 

Figure \ref{fig:M8_Macc_resStudy} shows the mass accretion and loss rates of cloud M8 as a function of time for the three resolutions.
The global behavior of the mass accretion is conserved at the different resolutions, with the peaks of major accretion events occurring at roughly the same evolutionary time and with similar strengths.
This shows that the major accretion events due to the compression of the envelope by nearby SN explosions are well resolved. 
Note that SN events always occur at the same position, in time and space, for all re-simulations of the same cloud.

There is one prominent accretion event occurring at $\sim7.5$~Myr in the $0.95$~pc resolution simulation that differs from its counterpart simulations at $0.47$ and $0.12$~pc resolution. 
In this event, cloud M8 accretes a large fragment of gas that is gravitationally bound to the cloud, but in the low resolution case, this fragment is not big enough to be considered a resolved structure, and therefore is not counted as part of the cloud before this time.  
The moment this fragment connects with the cloud, the mass accretion rate artificially spikes up to $5\times10^{-3}$~M$_{\odot}$yr$^{-1}$.

We then compare the mean mass accretion rates of these simulations obtaining $\dot{M}_{0.12pc}=4.8\times10^{-3}$~M$_{\odot}$yr$^{-1}$, $\dot{M}_{0.47pc}=4.4\times10^{-3}$~M$_{\odot}$yr$^{-1}$ and $\dot{M}_{0.95pc}=5.2\times10^{-3}$~M$_{\odot}$yr$^{-1}$, in good agreement with each other, and showing no clear sign of further convergence. 

Figure \ref{fig:M8_resStudy} shows the evolution of the mass, size, internal velocity dispersion and virial parameter of the cloud as a function of time for the three different numerical resolutions.
The integrated mass evolution of the cloud is similar for the three resolutions, even though the internal structure of cloud M8 at the lowest resolution is already formally unresolved at the moment self-gravity is turned on.
From the Figure, it is clear that the cloud gains more mass at higher resolutions, but the level of convergence is already quite good.
Examining the shape of the cloud at different resolutions at the same evolutionary time, at higher resolutions the cloud shows a more complex and filamentary structure, while at $\Delta x=0.95$~pc, the M8 surface looks more smooth and uniform.
The evolution of the cloud size is relatively similar for the three resolutions, but it appears that the cloud begins contracting earlier at higher resolutions. 
This is because blast waves can compress to higher densities in the better resolved model, so the boundary of the cloud shrinks earlier at higher resolution.
Afterwards, contraction is mainly due to gravitational collapse of the cloud, which seems to occur at a similar rate at the three different resolutions despite the lack of interior resolution at lower resolutions.

The cloud velocity dispersion and virial parameter show that the major SN impact at $1.8$~Myr is captured by all of the simulations and reaches a similar maximum velocity and virial parameter with well converged decay at the higher resolutions.
Later collapse becomes the main driver of the internal velocity dispersion of the cloud. 
Here the internal resolution of the cloud becomes important, and the higher resolution simulations can reach higher velocities as in them collapsing regions reach higher densities before becoming unresolved.  

Well after the collapse for all of the simulations is unresolved, there are a number of spikes in the velocity dispersion, with similar counterparts on the cloud's virial parameter, which differ between the different resolutions.
This spikes are provoked by the collision between two unresolved dense cores.
This unresolved cores carry a large amount of mass and accelerate toward each other, appearing as high velocity dispersion spikes.
At higher resolutions fragmentation and collapse are better resolved, delaying the appearance of the spikes and reducing their magnitudes as now the unresolved cores carry less mass and accelerate less dramatically toward each other.

Figure \ref{fig:Macc_M4_resStudy} shows the mass accretion rate as a function of time for cloud M4 at two different numerical resolutions, $\Delta x= 0.06$~pc and $\Delta x= 0.47$~pc.
The global behavior of the mass accretion rate is preserved between the two simulations, with an initial high mass accretion rate of $\sim10^{-3}$~M$_{\odot}$yr$^{-1}$ for 2~Myr, later followed by a decrease in the accretion rates until accretion is finally shut down around 7--8~Myr.
The mean mass accretion rate measured is $\dot{M}_{0.06pc} = 3.88\times 10^{-4} $~M$_{\odot}$~yr$^{-1}$, and $\dot{M}_{0.47pc} = 3.58\times 10^{-4} $~M$_{\odot}$~yr$^{-1}$.

Figure~\ref{fig:M4_resStudy} shows the evolution of the mass, radius, virial parameter, and velocity dispersion of cloud M4 as a function of time.
The high resolution cloud starts growing in mass faster than its low resolution counterpart. 
Cloud M4 is shocked by a nearby SN that compresses it and causes mass accretion at $t\sim2$~Myr.
This shock compresses the high resolution cloud more efficiently than the one at lower resolution. 
Then, shortly after, gravitational collapse takes over the highest density peaks at the same time that the simulation with $\Delta x = 0.47$~pc resolution becomes unresolved.
For this reason the high resolution cloud increases its internal velocity dispersion much faster, until this simulation too becomes unresolved.

For the questions that we address in this paper, our simulations appear to be sufficiently well-converged to support the conclusions we have reached.  
A more detailed analysis of the structure of cloud interiors will require additional computational effort.

\section{Jeans Refinement Criterion}

\begin{figure*}[t]
\centering 
\includegraphics[width=0.4\textwidth]{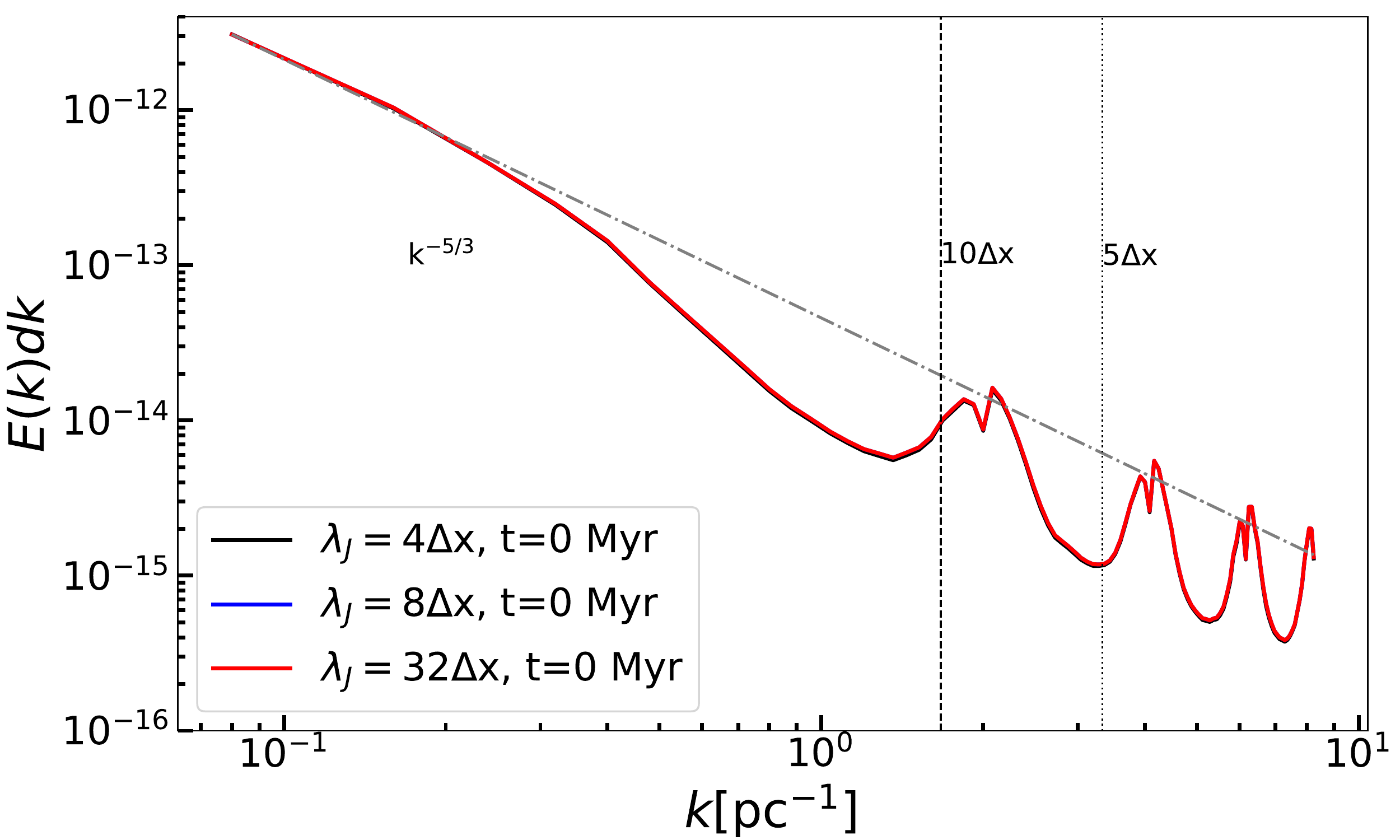} 
\includegraphics[width=0.4\textwidth]{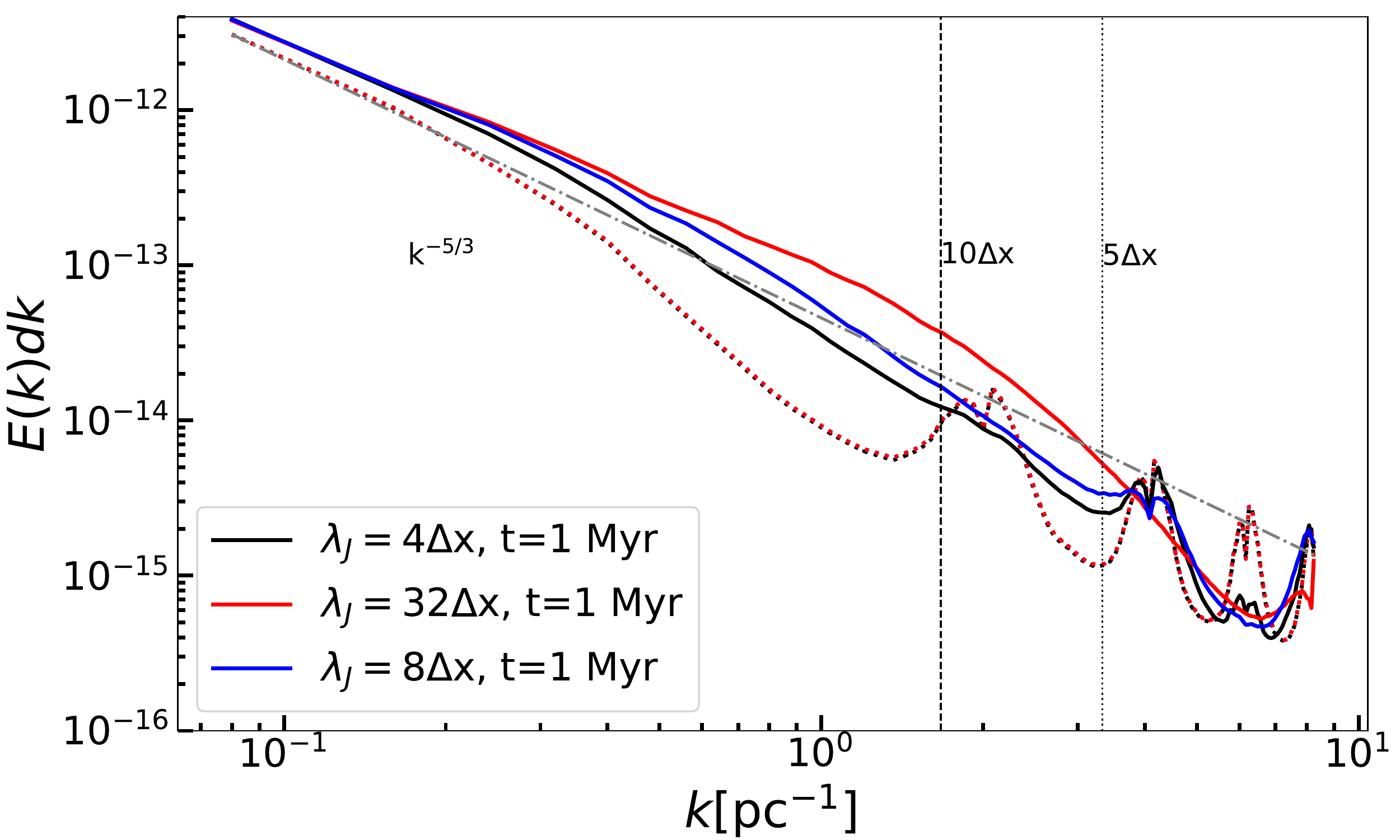}
\includegraphics[width=0.4\textwidth]{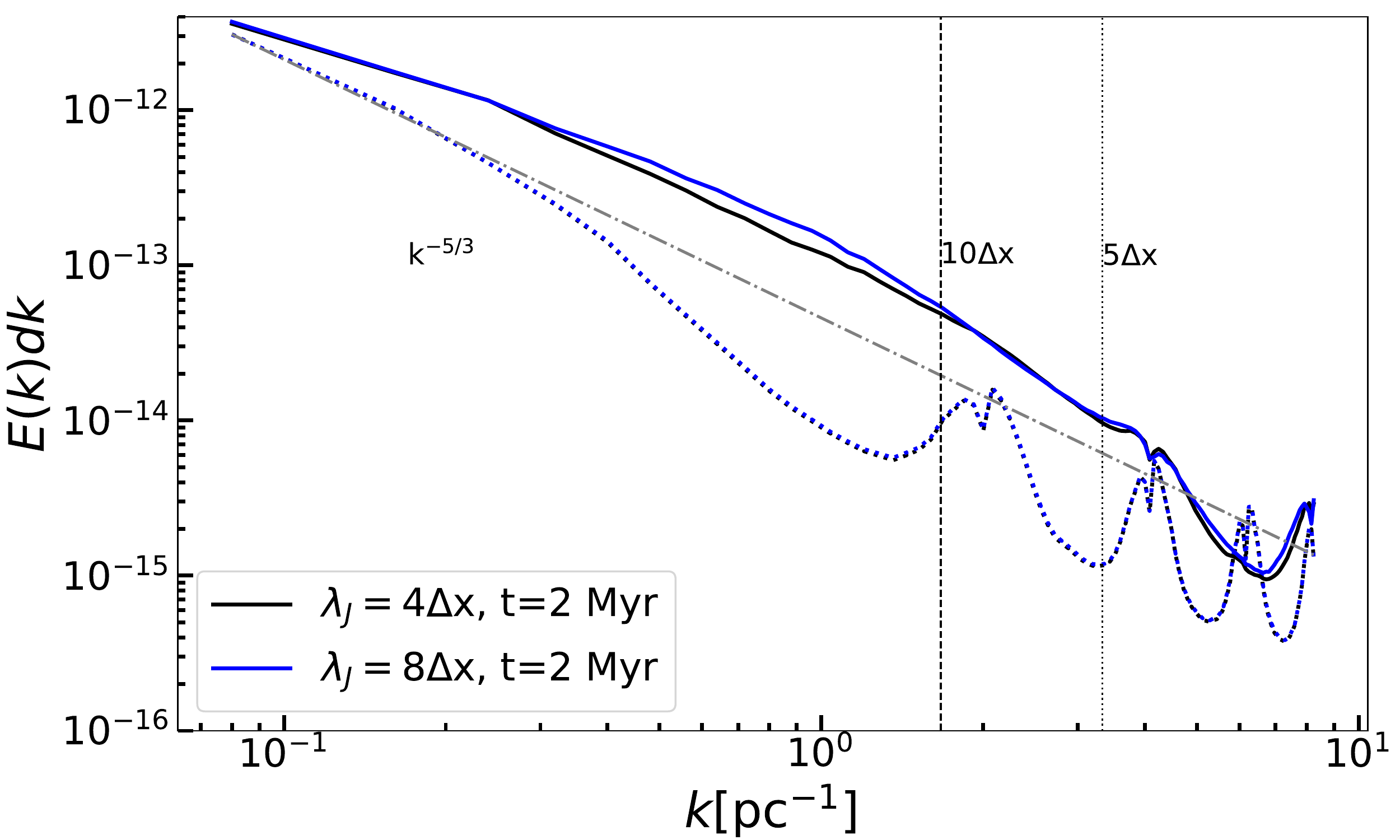} 
\includegraphics[width=0.4\textwidth]{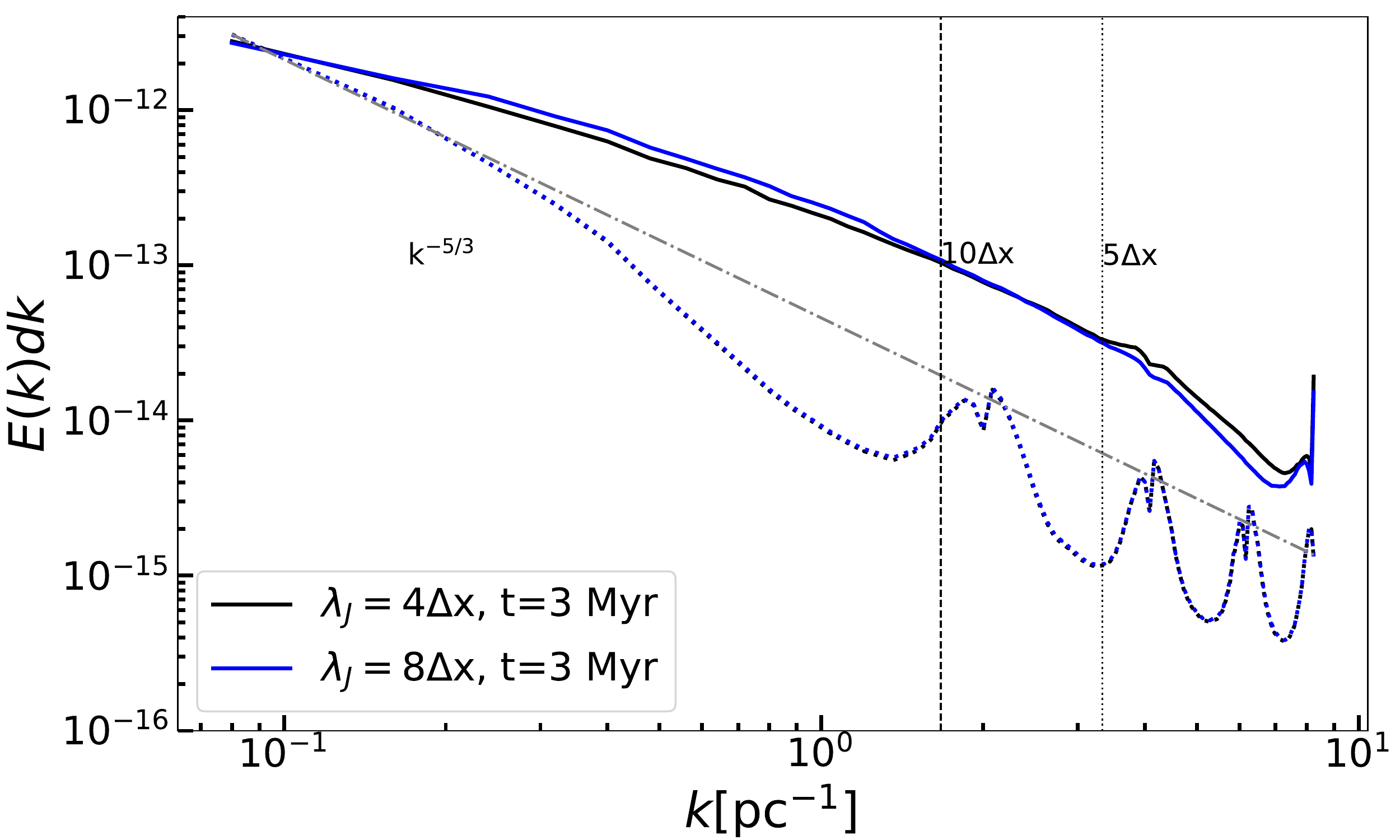}
\vspace*{-2mm}
\caption{
Turbulent kinetic energy power spectra of a 25~pc$^{3}$ box, centered in the cloud center of mass, for the simulations resolving the Jeans length with 4, 8 and 32 zones, at times
from the moment self-gravity is turned on shown in the legend.
Times with $t > 0$ also contain the turbulent spectrum at $t=0$ (\emph{dotted lines}) for comparison. Vertical (\emph{dashed} and \emph{dotted}) lines, show the scales corresponding to 10 and 5 $\Delta x$. A dashed-dotted line shows $k^{-5/3}$ scaling for comparison.
\label{fig:M3_powerspectra} }
\end{figure*}      

Resolving the Jeans length with only four zones is the minimum required to avoid artificial fragmentation in simulations of self-gravitating gas \citep{Truelove1998SelfgravitationalFragmentation}. 
Here, we examine the validity of our results by comparing the kinetic turbulent energy of cloud M3 during the first 3~Myr of evolution to simulations resolving the Jeans length with four and eight zones, and to a simulation for 1~Myr resolving the Jeans length with 32~zones.
We measure the energy using the Fourier power spectrum of $\mathbf{V}=\rho^{1/2} \mathbf{u}$, the density-weighted velocity field,
\begin{eqnarray}
	E_{\ell}(k) = \int \frac{1}{2} \hat{\mathbf{V}}(k)\cdot \hat{\mathbf{V}}^{*}(k) 4 \pi k^{2} dk
\end{eqnarray}
where the index $\ell$ corresponds to the number of cells used to resolve the Jeans length, $\hat{\mathbf{V}}$ is the Fourier-transformed field and $\hat{\mathbf{V}}^{*}$ its complex conjugate.
Figure~B5 shows the turbulent kinetic energy spectrum for a 25~pc$^{3}$ box, centered on the cloud center of mass at times 0, 1, 2 and 3~Myr after self-gravity has been activated.
It is known that as collapse proceeds, power moves towards small scales in regions around local centers of gravitational collapse \citep{Lee2015TIME-VARYINGRATE}, also seen at the small scales in the power spectra in Figure B5. 

Our zoom-in procedure produces the bumps in the power spectrum seen at scales of $8 \Delta x$, $4 \Delta x$ and $2 \Delta x$. 
The first bump, at $8\Delta x$, corresponds to the accumulation of gas at a resolution of $0.48$~pc, which is the resolution of the box before the zoom-in was started and self-gravity turned on. 
We then turn on self-gravity and refine on unstable regions by factors of two, filling in the smaller scales in the power spectrum.
The simulations resolving the Jeans length with more cells initially fill in the turbulent cascade faster, but as the cascade proceeds, and Jeans unstable regions get better resolved, the bumps in the power spectrum disappear.
At this point, the power spectrum is dominated by the turbulent energy cascade and the flattening of it due to global gravitational collapse.

In order to calculate the loss of kinetic energy as a function of time, we compare the total kinetic energy of the simulations resolving the Jeans length with four zones to those with eight and 32 zones. 
We calculate the energy difference, $\Delta E_{8, 32}(t) = E_{8,32}(t) - E_{4}(t)$, and present the fractional difference, $\Delta E_{8, 32} / E_{8, 32}$ as a function of time for the first 3~Myr of evolution.
\begin{figure}[t]
\centering 
\includegraphics[width=0.6\textwidth]{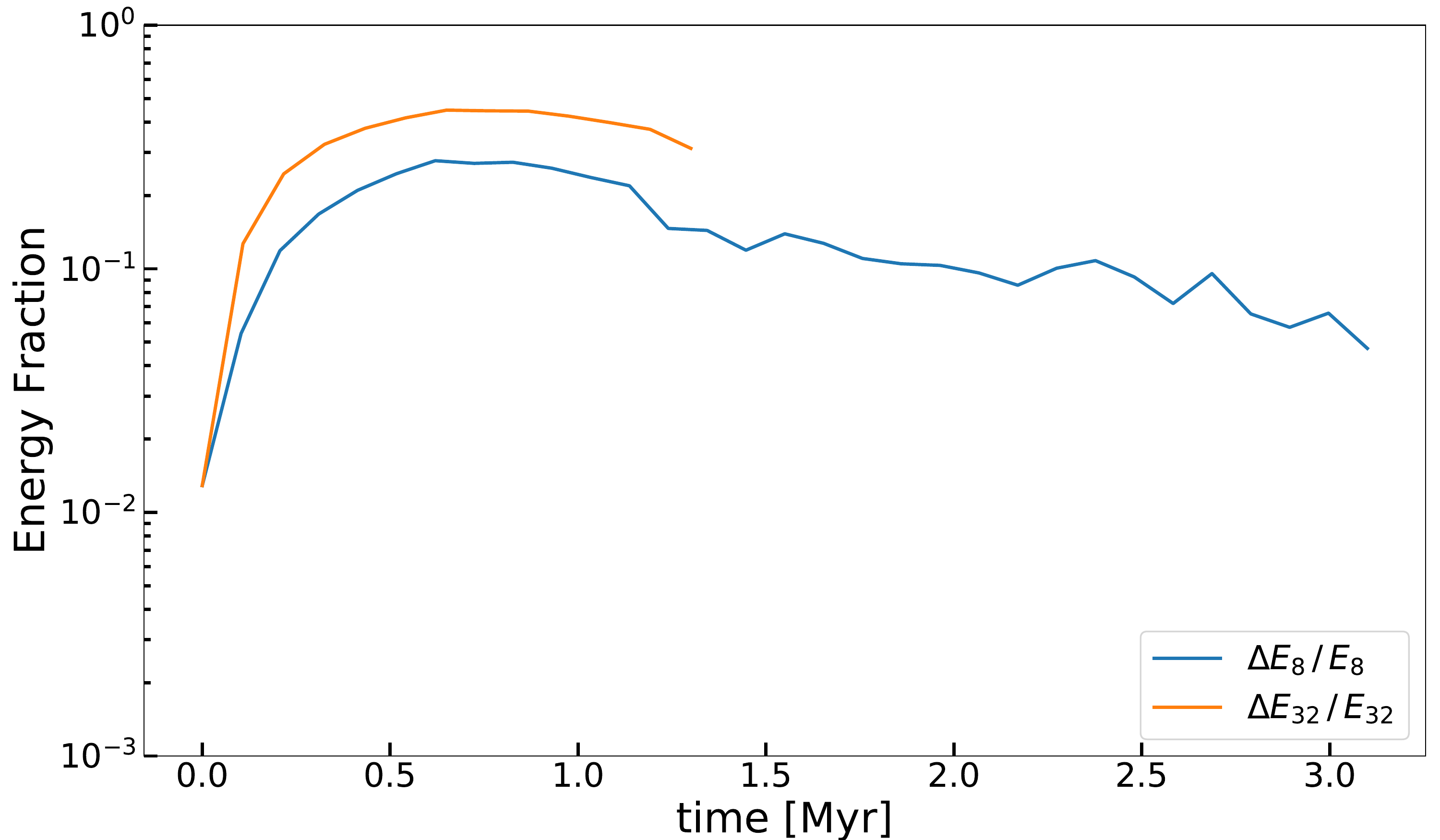} 
\caption{Fractional turbulent kinetic energy difference as a function of time between the simulations resolving the Jeans length with four and eight zones, and simulations with four and 32 zones.
\label{fig:M3_energydifference}}
\end{figure}      
%
Figure~B6 shows that increasing the number of zones resolving the Jeans length better resolves the turbulent cascade and gravitational collapse, resulting in higher turbulent kinetic energies.
However, the energy difference does not grow with time, but rather appears to decrease. 
The mean fractional difference fluctuates around a mean of  $\approx 13$\% between the simulations resolving the Jeans length with four and eight zones, and a mean of  $\approx 33$\% between the simulations resolving the Jeans length with four and 32 zones. 
We do not expect that this increase in total turbulent kinetic energy will have a major impact on the global evolution of the clouds as they are dominated by gravitational potential energy by almost an order of magnitude compared to total turbulent kinetic energy as seen in Figures 12, 13, and 14.

\end{document}